\documentclass[12pt]{article}
\usepackage{epsfig}
\usepackage{amsmath}
\usepackage{hhline}
\usepackage{amssymb}
\usepackage{times}
\usepackage{cite}
\usepackage{lineno}
\usepackage{xcolor}
\usepackage{xspace}
\usepackage{upgreek}
\usepackage{units}
\usepackage{placeins}
\usepackage{booktabs}
\usepackage{dcolumn}

\newlength{\dinwidth}
\newlength{\dinmargin}	
\renewcommand{\Theta}{\theta}

\newcommand{\Cascade}{{\sc Cascade}\xspace}
\newcommand{\Pythia}{{\sc Pythia}\xspace}
\newcommand{\DiffVM}{{\sc DIFFVM}\xspace}

\newcommand{\be}{\begin{eqnarray}}
\newcommand{\ee}{\end{eqnarray}}

\newcommand{\mr}[1]{{{#1}}}

\newcommand{\kT}{\ensuremath{{k_{T}}}\xspace}

\newcommand{\bbb}{\ensuremath{b\bar{b}}\xspace}
\newcommand{\ccb}{\ensuremath{c\bar{c}}\xspace}
\newcommand{\ra}{\rightarrow}

\newcommand{\Mll}{\ensuremath{{m_{\ell\ell}}}\xspace}

\newcommand{\JPsi}{\ensuremath{{J/\psi}}\xspace}
\newcommand{\iPsi}{\ensuremath{{\psi}}\xspace}
\newcommand{\PsiPrime}{\ensuremath{{\psi(2S)}}\xspace}

\newcommand{\PtMu}{\ensuremath{{P_{T,\mu}}}\xspace}
\newcommand{\PtLepton}{\ensuremath{{P_{T,\ell}}}\xspace}

\newcommand{\ThetaElectron}{\ensuremath{{\Theta_{\mr{e}}}}\xspace}
\newcommand{\ThetaMuon}{\ensuremath{{\Theta_{\mu}}}\xspace}

\newcommand{\Qsquared}{\ensuremath{{Q^{2}}}\xspace}
\newcommand{\Wgp}{\ensuremath{{W_{\gamma p}}}\xspace}

\newcommand{\ThetaTrk}{\ensuremath{\Theta}\xspace}
\newcommand{\ThetaJPsi}{\ensuremath{\Theta_{\iPsi}}\xspace}
\newcommand{\ZJPsi}{\ensuremath{{z}}\xspace}
\newcommand{\PtJPsi}{\ensuremath{{P_{T,\iPsi}}}\xspace}
\newcommand{\PtJPsiSquare}{\ensuremath{{P^2_{T,\iPsi}}}\xspace}

\newcommand{\PtStarJPsi}{\ensuremath{{P^{*}_{T,\iPsi}}}\xspace}
\newcommand{\PtStarJPsiSquare}{\ensuremath{{P^{*2}_{T,\iPsi}}}\xspace}

\newcommand{\PtPtStarJPsi}{\ensuremath{{P_{T,\iPsi}(P^{*}	_{T,\iPsi})}}\xspace}

\newcommand{\CosThetaStar}{\ensuremath{\cos(\Theta^{*})}\xspace}

\newcommand{\PhiStar}{\ensuremath{{\phi^{*}}}\xspace}

\DeclareMathAlphabet{\mathitbf}{OML}{cmm}{b}{it}
\DeclareMathSymbol{\psi}{\mathalpha}{letters}{"20}

\begin{document}

% Journal macro
\def\Journal#1#2#3#4{{#1} {\bf #2} (#3) #4}
\def\NCA{\em Nuovo Cimento}
\def\NIM{\em Nucl. Instrum. Methods}
\def\NIMA{{\em Nucl. Instrum. Methods} {\bf A}}
\def\NPB{{\em Nucl. Phys.}   {\bf B}}
\def\PLB{{\em Phys. Lett.}   {\bf B}}
\def\PRL{\em Phys. Rev. Lett.}
\def\PRD{{\em Phys. Rev.}    {\bf D}}
\def\ZPC{{\em Z. Phys.}      {\bf C}}
\def\EJC{{\em Eur. Phys. J.} {\bf C}}
\def\CPC{\em Comp. Phys. Commun.}

\pagestyle{empty}

\begin{titlepage}

\noindent
\begin{flushleft}
{\tt DESY 09-225    \hfill    ISSN 0418-9833} \\
{\tt December 2009}                  \\
\end{flushleft}

\noindent
%Version:     3.0 (after final reading, 17.12.2009) \\
%Editors:     M.~Steder (michael.steder@desy.de), A.~Meyer (andreas.meyer@desy.de) \\
%Referees:    H.~Jung (hannes.jung@desy.de), D.~Wegener (wegener@physik.uni-dortmund.de) \\

\vspace*{2cm}

\begin{center}
  \Large
  {\bf 
   Inelastic Production of $\mathitbf  J\mathitbf/\mathitbf\psi$ Mesons\\
in Photoproduction and Deep Inelastic Scattering\\
at HERA
  }	

  \vspace*{2cm}
    {\Large H1 Collaboration} 
\end{center}
\vspace*{2cm}

%\linenumbers

\begin{abstract}
\noindent
A measurement is presented of inelastic photo- and electroproduction of \JPsi 
mesons in
$ep$ scattering at HERA. The data were recorded with the H1 detector 
in the period from 2004 to 2007. 
Single and double differential cross sections are determined and the 
\mbox{helicity} distributions of the \JPsi mesons are analysed. 
The results are compared to theoretical predictions in the colour singlet model and in the framework of non-relativistic QCD.
Calculations in the colour singlet model using a \kT factorisation ansatz are able to give a good description of the data, 
while colour singlet model calculations to next-to-leading order in collinear factorisation underestimate the data.

\end{abstract}

\vspace{1cm}
\centerline{Submitted to \EJC}

\end{titlepage}

\renewcommand{\thepage}{\arabic{page}}
\setcounter{page}{2}
\noindent
%\include{h1auts}
%-- H1AUTS Author list by names 
%-- Status: Tue Dec  1 09:22:13 CET 2009  Number of authors = 228 

F.D.~Aaron$^{5,49}$,           %BUCH-PD        11/06           Aaron               
C.~Alexa$^{5}$,                %BUCH-PD        06/06           Alexa               
%K.~Alimujiang$^{11,51}$,       %DESY-PD        07/08           Alimujiang          
V.~Andreev$^{25}$,             %LPI -PD        8/88            Andreev             
B.~Antunovic$^{11}$,           %DESY-LEFT      12/08           Antunovic           
S.~Backovic$^{30}$,            %PODG-PD        03/2            Backovic            
A.~Baghdasaryan$^{38}$,        %YERE-PD        09/03           Baghdasaryana       
E.~Barrelet$^{29}$,            %PARI-PD        11/99           Barrelet            
W.~Bartel$^{11}$,              %DESY-PD        8/88            Bartel              
K.~Begzsuren$^{35}$,           %ULBA-PD        04/06           Begzsuren           
A.~Belousov$^{25}$,            %LPI -PD        8/88            Belousov            
J.C.~Bizot$^{27}$,             %ORSA-PD        8/88            Bizot               
V.~Boudry$^{28}$
% ...

%-- H1 Institutes 
\bigskip{\it
% ...
$ ^{5}$ National Institute for Physics and Nuclear Engineering (NIPNE) ,
          Bucharest, Romania \\
% ...
$ ^{11}$ DESY, Hamburg, Germany \\
% ...
$ ^{25}$ Lebedev Physical Institute, Moscow, Russia$^{ e}$ \\
% ...
$ ^{27}$ LAL, Universit\'e Paris-Sud, CNRS/IN2P3, Orsay, France \\
 $ ^{28}$ LLR, Ecole Polytechnique, CNRS/IN2P3, Palaiseau, France \\
 $ ^{29}$ LPNHE, Universit\'e Pierre et Marie Curie Paris 6,
          Universit\'e Denis Diderot Paris 7, CNRS/IN2P3, Paris, France \\
 $ ^{30}$ Faculty of Science, University of Montenegro,
          Podgorica, Montenegro$^{ e}$ \\
% ...
$ ^{35}$ Institute of Physics and Technology of the Mongolian
          Academy of Sciences, Ulaanbaatar, Mongolia \\
% ...
$ ^{38}$ Yerevan Physics Institute, Yerevan, Armenia \\
% ...
$ ^{49}$ Also at Faculty of Physics, University of Bucharest,
          Bucharest, Romania \\
% ...
$ ^e$ Supported by the Deutsche Forschungsgemeinschaft \\
% ...
}

\newpage

%\linenumbers

\pagestyle{plain}
%
%%%%%%%%%%%%%%%%%%%%%%%%%%%%%%%%%%%%%%%%%%%%%%%%%%%%%%%%%%%%
\section{Introduction}
\label{intro}

The description of the process 
of charmonium production in interactions of photons and hadrons is a challenge to theory,
since it involves both the production of the heavy quark system and the formation of the bound state. 
Charmonium production in electron\footnote{In this paper "electron" is used to denote both electron and positron.}-proton collisions at HERA is dominated by photon-gluon fusion: 
a photon emitted from the incoming electron interacts with a gluon from 
the proton to produce a $c \bar c$ pair that evolves into a charmonium state.
In the colour singlet model, only those states with the same quantum numbers as the 
resulting charmonium contribute to the formation of a bound \ccb state.
This is achieved by radiating a hard gluon in a perturbative process.
In the factorisation ansatz of non-relativistic quantum chromodynamics, also colour octet \ccb states
contribute to the charmonium production cross section via soft gluon radiation.

Previous measurements in electroproduction ($ep$) and photoproduction ($\gamma p$) at HERA~\cite{Aid:1996dn,Adloff:1999zs,Adloff:2002ex,Adloff:2002ey,Breitweg:1997we,Chekanov:2002at,ZEUS05} are not described by predictions in the colour singlet model to leading order.
In contrast, the calculation of photoproduction cross sections to next-to-leading order (NLO)~\cite{Kraemer} showed a reasonable description of the photoproduction cross sections.
The calculation proved that the corrections with respect to leading order results are very large, 
increasing towards large transverse momentum of the \JPsi meson. 
The same calculation, repeated recently with an up-to-date set of theoretical parameters~\cite{Maltoni}, results in a prediction which is about a factor of three below the measured cross sections, 
indicating that corrections beyond NLO are needed and/or that contributions 
from colour octet states may be sizable.

In this paper a measurement is presented of inelastic \JPsi meson production at HERA.
The measurement uses a larger data sample than previous results~\cite{Aid:1996dn,Adloff:1999zs,Adloff:2002ex,Adloff:2002ey} and benefits from improved systematics.
The data sets were collected in the years 2004 to 2007 with the H1 detector.
The \JPsi meson candidates are identified by the leptonic decay into two 
muons or electrons.
The cross sections are measured for both electroproduction and photoproduction. 
For the photoproduction sample \JPsi meson polarisation variables are determined. 
The data samples are restricted to the region of phase space where contributions from 
diffractive charmonium production are suppressed.

%%%%%%%%%%%%%%%%%%%%%%%%%%%%%%%%%%%%%%%%%%%%%%%%%%%%%%%%%%%%
\section{Theoretical Models}
\label{theory}

In order to describe inelastic charmonium production in the framework of 
perturbative QCD different models have been proposed, such as the colour-eva\-po\-ration model~\cite{Halzen:1977rs,Eboli:1998xx}, the colour-singlet model 
(CSM)~\cite{CSM_Chang,Berger:1980ni,Baier:1981uk,Baier:1981zz,Baier:1983va}, the factorisation ansatz in non-relativistic quantum chromodynamics (NRQCD)~\cite{Caswell:1985ui,Thacker:1990bm,Bodwin:1994jh} and soft 
colour interactions~\cite{Edin:1997zb}. 
In this paper the most recent calculations using the CSM or NRQCD are compared to the data.

In the CSM, only charm quark pairs in a colour singlet state with the same quantum numbers as the resulting charmonium contribute to the formation of a bound \ccb state.
This is achieved by radiating a hard gluon in the perturbative process.
The factorisation ansatz in NRQCD includes also colour octet $c\bar{c}$ states 
in the charmonium production cross section. 
The size of these colour octet contributions, described by long distance 
matrix elements (LDME), is defined by additional free parameters which
were determined in fits to the Tevatron data~\cite{Braaten:1999qk}.
The NRQCD factorisation approach contains also the colour singlet model
which is recovered in the limit in which the colour-octet LDME tend to zero.

The following calculations are compared to the measurements presented in this paper:
\begin{itemize}
\item A calculation of \JPsi meson photoproduction via a colour singlet mechanism \cite{Maltoni} 
provides predictions for both cross sections and helicity distributions to next-to-leading order. 
The uncertainty of this calculation is estimated by variations of the charm quark mass  
and the factorisation and renormalisation scales.
\item A calculation at NLO for photoproduction cross sections includes the full framework 
of NRQCD~\cite{Kniehl}. 
The uncertainty of this calculation is dominated by the limited knowledge of 
the LDMEs.
\item CSM predictions in the \kT factorisation approach are employed as implemented in the MC generator \Cascade \cite{Jung:2000hk}.
Higher order parton emissions based on the CCFM evolution equations~\cite{ccfm} are matched to ${\cal O}(\alpha_s)$ matrix elements in which the incoming parton can be off-shell.
The uncertainty on the calculation is estimated by varying the renormalisation scale by a factor of two.
In addition polarisation variables in the \kT factorisation approach are calculated analytically~\cite{Baranov}. 
\end{itemize}
Parameters and variations used in the theoretical calculations are given in table~\ref{tab:theoryparam}.

%%%%%%%%%%%%%%%%%%%%%%%%%%%%%%%%%%%%%%%%%%%%%%%%%%%%%%%%%%%%%%%%%%%%%%%%%%%%%%%%%%%%%%%%%%%%%%%%%%

\begin{table}[tb]
\begin{center}
\begin{tabular}{ l p{8cm}}
%
%
%% \hline
%%  CSM (NLO), M.~Kr\"amer~\cite{Kraemer} &  \\
%% \hline
%% PDF                                                        & MRST 
%% renormalisation and factorisation scale                    & $\sqrt{2}m_c, \mr{max}\left[\sqrt{2}m_c,1/2\cdot\sqrt{m_c^2 + \PtJPsiSquare}~\right] (\mathrm{for}~\PtJPsiSquare) $\\
%% CS LDME  & $\left<\mathcal{O}\left[\underline{1},^3S_1\right]\right> = \unit[1.16]{GeV^3}$\\
%% $m_c$                                                      & $1.3 < m_c < \unit[1.5]{GeV}$\\
%% $\alpha_s(M_Z)$                                            & $0.1200 \pm 0.0025$\\

%
%
\toprule
 CSM (NLO), P.~Artoisenet et al.~\cite{Maltoni} &  \\
\midrule
PDF                                                        & CTEQ6M~\cite{CTEQ6M}\\
renormalisation and factorisation scale                    & $\mu_0 = 4 m_c$\\
scale variation                                            & $0.5 \mu_0 < \mu_\mr{f}$, $\mu_\mr{r} < 2 \mu_0$ and $0.5 < \mu_\mr{r}/\mu_\mr{f} < 2$\\
CS LDME  & $\left<\mathcal{O}\left[\underline{1},^3S_1\right]\right> = \unit[1.16]{GeV^3}$\\
$m_c$                                                      & $1.4 < m_c < \unit[1.6]{GeV}$\\
$\alpha_s(M_Z)$                                            & $0.118$ (+ running at 2 loops)\\
\midrule
 NRQCD (NLO), M.~Butensch\"on et al.~\cite{Kniehl} &  \\
\midrule
PDF                                                        & CTEQ6M~\cite{CTEQ6M}\\
renormalisation and factorisation scale                    & $\mu_0 = \sqrt{4m_c^2 + \PtJPsiSquare}$\\
NRQCD scale                                                & $\mu_\Lambda = \mr{m_c}$\\
%CS LDME  & $\left<\mathcal{O}\left[\underline{1},^3S_1\right]\right> = \unit[1.16]{GeV^3}$\\
$m_c$                                                      & $m_\JPsi / 2 \approx \unit[1.55]{GeV}$\\
$\alpha_s(M_Z)$                                            & $0.1176 \pm 0.002$\\
\midrule
 CSM (\kT factorisation), \Cascade~\cite{Jung:2000hk} &  \\
\midrule
PDF                                                        & CCFM set A0~\cite{SetA0}\\
                                                           &  (\lq set A0$^\pm$\rq\xspace for $\mu_r$ uncertainties)\\
renormalisation scale                                      & $\mu_0 = \sqrt{m_\psi^2 + \PtJPsiSquare}$\\
renormalisation scale variation                            & $0.5 \mu_0 < \mu_\mr{r} < 2 \mu_0$\\
factorisation scale                                        & $\sqrt{\hat{s} + Q^2_\perp}$\\
%CS LDME  & $$\left<\mathcal{O}\left[\underline{1},^3S_1\right]\right>$ = 1.16 \unit[]{GeV^3}$\\
$m_c$                                                      & $\unit[1.5]{GeV}$\\
$\Lambda_\mathrm{QCD}^{(3)}$                                  & $\unit[200]{MeV}$\\
\midrule
 CSM (\kT factorisation), S.~Baranov~\cite{Baranov} &  \\
\midrule
PDF                                                        & CCFM set A0~\cite{SetA0}\\
renormalisation and factorisation scale                    & $\mu_0 = \sqrt{m_\psi^2 + \PtJPsiSquare}$\\
%$\left<\mathcal{O}\left[\underline{1},^3S_1\right]\right>$ & $\unit[]{GeV^3}$\\
$m_c$                                                      & $\unit[1.5]{GeV}$\\
$\Lambda_\mathrm{QCD}^{(3)}$                                  & $\unit[200]{MeV}$\\
\bottomrule
\end{tabular}
\end{center}
\caption{Summary of the parameters employed in the CSM and NRQCD calculations 
used to compare to the measurements in this paper. In this table PDF means 
parton distribution function of the proton, $\hat{s}$ denotes the invariant mass square of the hard subprocess and $Q_\perp$ the initial transverse momentum of the partonic system ($\gamma$g).}
\label{tab:theoryparam}
\end{table}

%%%%%%%%%%%%%%%%%%%%%%%%%%%%%%%%%%%%%%%%%%%%%%%%%%%%%%%%%%%%
\section{H1 Detector}
\label{sec:h1}

The H1 detector is described in detail elsewhere~\cite{Abt:1997xv}. Here only
the components essential to the present analysis are briefly described.
A right handed Cartesian  coordinate system is used with the origin at the nominal primary $ep$ interaction vertex. 
The proton beam direction defines the $z$ axis. The polar angles $\theta$ and transverse momenta $P_T$ of all particles 
are defined with respect to this axis.  The azimuthal angle $\phi$ defines the particle direction in the transverse plane. The pseudorapidity  is defined as $\eta=-\ln {\tan {\frac{\theta}{2}}}$.

Charged particles emerging from the $ep$ interaction region
are measured by the central tracking detector (CTD) in the pseudo-rapidity
range $|\eta| < 1.74$.
The CTD consists of two large cylindrical central jet drift chambers (CJC) which
are interleaved by a $z$-chamber and arranged concentrically 
around the beam-line in a magnetic field of $1.16~\mathrm{T}$. 
The CTD provides triggering information based on track segments from the CJC~\cite{FTT, FTT2},
and on the \ZJPsi-position of the vertex from the 5-layer multi-wire proportional chamber~\cite{CIP} which
is situated inside the inner CJC.
To provide the best possible spatial track reconstruction, CTD tracks are linked to 
hits in the vertex detector, the central silicon tracker CST~\cite{CST}.
The CST is installed close to the interaction point, surrounding the beam pipe in the 
pseudo-rapidity range $|\eta| < 1.3$ and consists of two layers of double sided silicon 
strip sensors. 

Charged and neutral particles are measured in the liquid argon calorimeter (LAr)~\cite{LAr} which surrounds the tracking chambers and covers the range 
$-1.5 < \eta < 3.4$ and a lead/scintillating-fibre calorimeter SpaCal~\cite{Nicholls:1996di}, 
covering the backward region $-4.0 < \eta < -1.4$.
The calorimeters are surrounded by the solenoidal 
magnet and the iron return yoke. The yoke is instrumented with
16 layers of limited streamer tubes, forming the central muon detector (CMD) 
in the range $-2.5 < \eta < 3.4$.

The luminosity determination is based on the measurement of the 
Bethe-Heitler process $ep \to ep\gamma$, where the photon is detected 
in a calorimeter located downstream of the interaction 
point in the electron beam direction at $z=-\unit[104]{m}$.

%%%%%%%%%%%%%%%%%%%%%%%%%%%%%%%%%%%%%%%%%%%%%%%%%%%%%%%%%%%%
\section{Data Analysis}
\label{sec:anal}

The kinematics of inelastic charmonium production at HERA are  
described using the following variables: the square of the $ep$ centre 
of mass energy $s = (p+k)^2$, where $p$ and $k$ denote the four vectors of electron and proton respectively; 
the negative squared four momentum transfer $\Qsquared = -q^2$, where $q$ is the
four vector of the virtual photon; and the mass of the hadronic final state $\Wgp = \sqrt{(p+q)^2}$. 
\Wgp is related to the scaled energy transfer $y = (p\cdot q) / (p \cdot k)$ via $\Wgp^2 = ys-\Qsquared$.  
In addition, the elasticity of the \JPsi meson production process is defined as
$\ZJPsi = (p_\psi\cdot p)/(q\cdot p)$, where $p_\psi$ is the four momentum of the \JPsi meson.
The elasticity denotes the fractional energy of the photon 
transferred to the \JPsi meson in the proton rest system.

Events are selected separately in the photoproduction and electroproduction regimes.
Photoproduction events are selected 
by requiring that no isolated high energy electromagnetic cluster, 
consistent with a signal from a scattered electron, is detected in 
the calorimeters. This limits the virtuality to values of 
$\Qsquared \lesssim \unit[2.5]{GeV^2}$, resulting in a mean value of $\langle \Qsquared\rangle \approx \unit[0.085]{GeV^2}$. 
Conversely, for the electroproduction sample,
a scattered electron with energy of more than $\unit[10]{GeV}$
 is required to be 
reconstructed in the backward calorimeter (SpaCal), corresponding to a range
in photon virtuality $3.6 < \Qsquared < \unit[100]{GeV^2}$.

In this analysis the photon virtuality \Qsquared is reconstructed from the
scattered electron energy $E_e'$ and polar angle $\Theta_e'$ as $\Qsquared=4E_eE_e'\cos^2(\Theta_e'/2)$, where $E_e$ denotes the energy of the beam electron.
The variable $y$ is reconstructed using 
the relation $y=\sum_h (E-p_z)/2E_e$ for photoproduction~\cite{jb} 
and $y=\sum_h (E-p_z)/\sum (E-p_z)$ for electroproduction~\cite{esigma}.
The sums in the numerator include all particles of the hadronic final state without the scattered electron, which is only included in the sum of the denominator for electroproduction.
The elasticity \ZJPsi is then obtained from 
$z = (E-p_z)_{\JPsi}/\sum_h (E-p_z)$, where $(E-p_z)_{\JPsi}$ is calculated from the decay particles of the \JPsi meson.
The kinematics of the final state particles are obtained from charged particle
tracks reconstructed in the CTD and energy depositions in the LAr and 
SpaCal calorimeters~\cite{hadrooII,peez}. 
\begin{figure}[t]
\unitlength1cm
%\frame{
\begin{picture}(16.,4.6)
\put( 0.,0.2){\includegraphics*[width=5.cm]{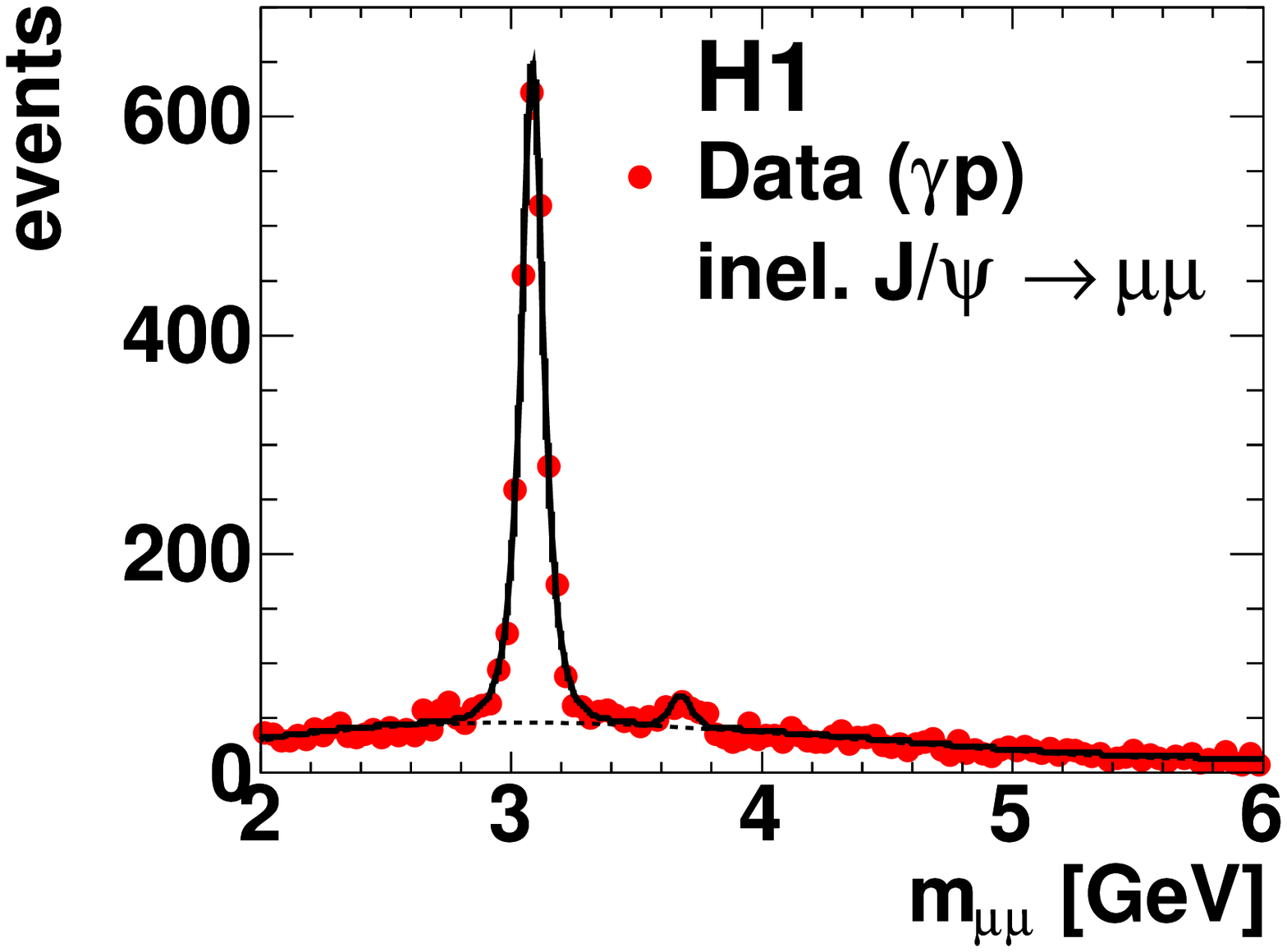}}
\put( 5.5,0.2){\includegraphics*[width=5.cm]{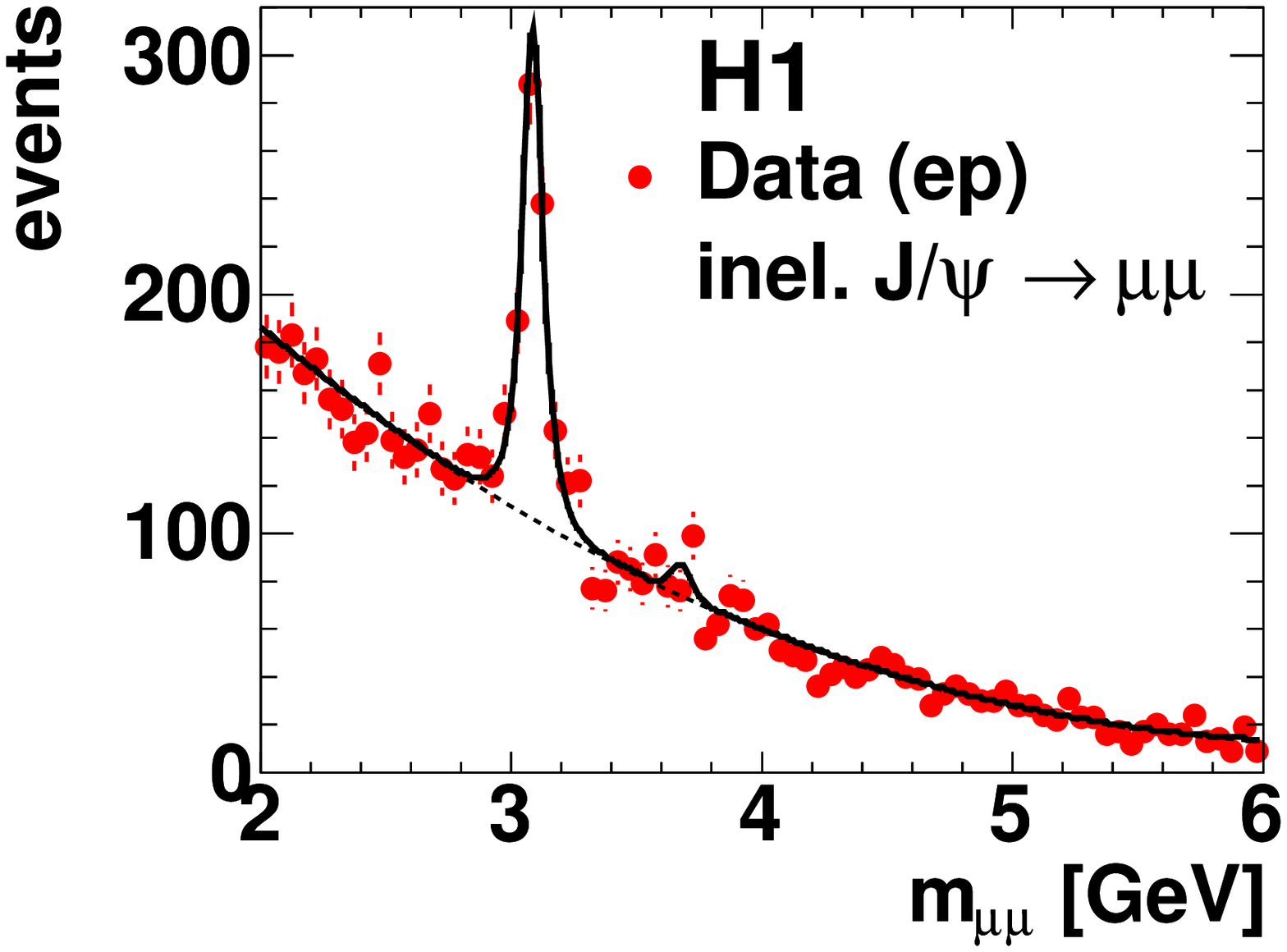}}
\put(11.,0.2){\includegraphics*[width=5.cm]{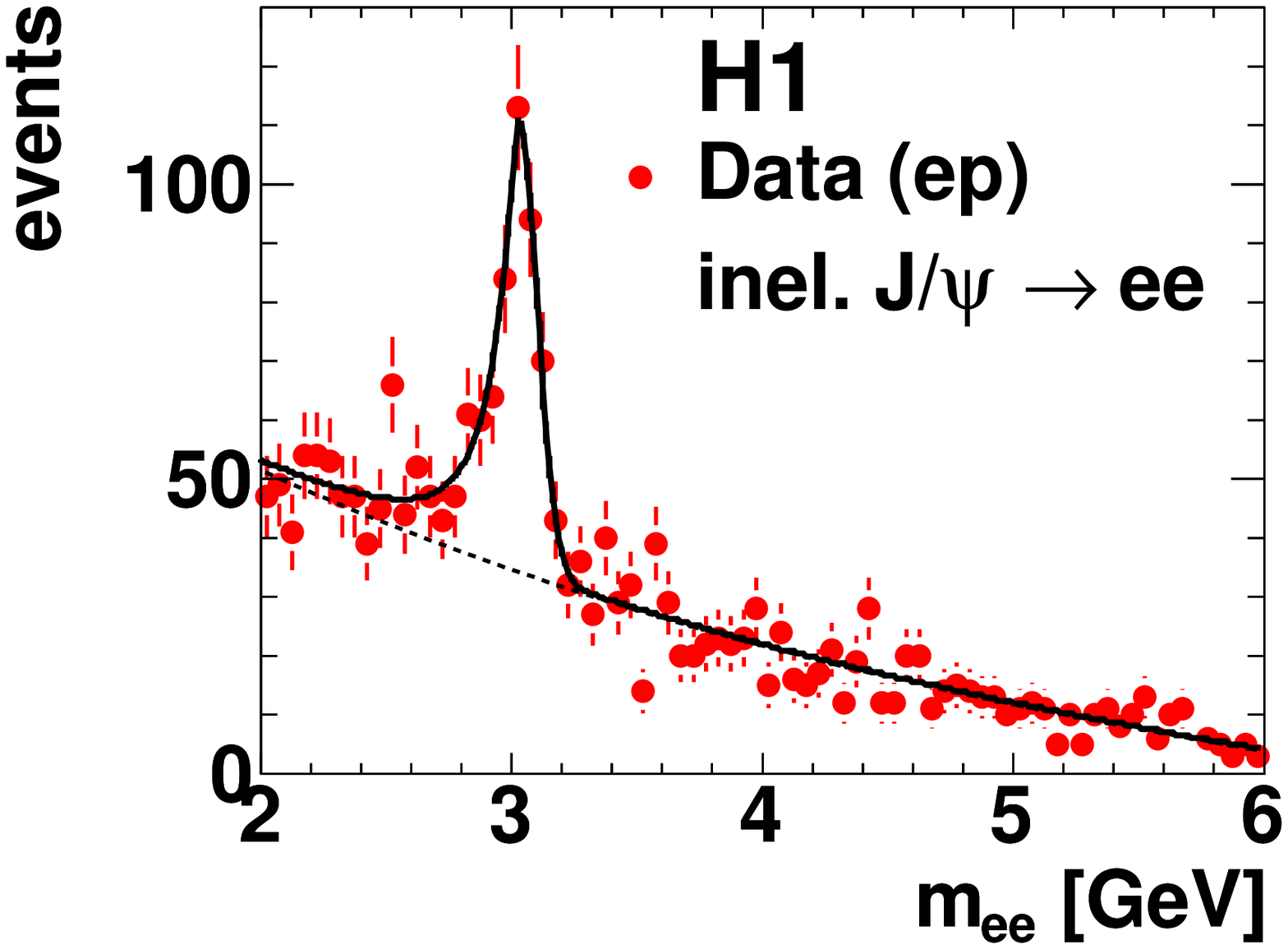}}
\put(1.,4.2){a)}
\put(6.5,4.2){b)}
\put(12.,4.2){c)}
\end{picture}
%}
\caption{Invariant mass spectra of two oppositely charged leptons after all selection
cuts for a) the photoproduction ($\gamma p$) sample and b,c) the 
electroproduction ($ep$) samples as listed in table~\ref{tab:selection}. 
The lines show the results of fits to signals and backgrounds.}
\label{fig:ctrl:mass} 
\end{figure}

\begin{table}[t]
\begin{center}
\begin{tabular}{ rccc }
\toprule
 & Photoproduction & \multicolumn{2}{c}{Electroproduction} \\
 & $\JPsi \rightarrow \mu^+\mu^-$ & $\JPsi \rightarrow \mu^+\mu^-$ & $\JPsi \rightarrow e^+e^-$ \\
 \midrule
 \multicolumn{4}{c}{kinematic range} \\
\midrule
              & $ \Qsquared< \unit[2.5]{GeV^2}$      & \multicolumn{2}{c}{$3.6 < \Qsquared < \unit[100]{GeV^2}$} \\
                & \multicolumn{3}{c}{$ \unit[60]{GeV} < \Wgp < \unit[240]{GeV}$} \\ 
                & \multicolumn{1}{c}{$\PtJPsi > \unit[1]{GeV}$} &\multicolumn{2}{c}{$\PtStarJPsi > \unit[1]{GeV}$} \\ 
                & \multicolumn{3}{c}{$0.3 < \ZJPsi < 0.9$} \\ 
 \midrule
 \multicolumn{4}{c}{event selection} \\
 \midrule
           & \multicolumn{3}{c}{$\PtLepton >\unit[800]{MeV}$} \\ 
        & $20^{\circ} < \ThetaMuon < 160^{\circ}$ & $20^{\circ} < \ThetaMuon < 160^{\circ}$ & $20^{\circ} < \ThetaElectron < 150^{\circ}$ \\
      & \multicolumn{3}{c}{$\mathrm{N_{Trk}} \geq 5$ ~~(in the range $20^\circ < \ThetaTrk < 160^\circ$)} \\ 
 \midrule
 \multicolumn{4}{c}{event samples} \\
 \midrule
  $N_{\JPsi}$ & $2320 \pm 54$ & $501\pm34$  &  $290\pm24$ \\ 
%  $\mathcal{BR}$\cite{PDG} & $(5.93 \pm 0.06)\%$ & $(5.93 \pm 0.06)\%$ & $(5.94 \pm 0.06)\%$  \\
  $\mathcal{L}_\mr{int}$     & $\unit[165]{pb^{-1}}$ & $\unit[315]{pb^{-1}}$ & $\unit[315]{pb^{-1}}$ \\
\bottomrule
\end{tabular}
\end{center}
\caption{List of selection cuts and event yields for each of the three data samples.}
\label{tab:selection}
\end{table}

The \JPsi meson candidates are reconstructed through their decays
into two oppositely charged muons or electrons. These decay leptons
are reconstructed as charged particles in the CTD with a transverse momentum of at least $\unit[800]{MeV}$.
Muon candidates are identified as minimum ionising particles in the 
LAr calorimeter or through track segments in the CMD ($20^\circ < \ThetaMuon < 160^\circ$)~\cite{msteder}. 
Electron candidates are identified through their energy deposit 
in the central calorimeter ($20^\circ < \ThetaElectron < 150^\circ$)~\cite{msauter}.
For trigger reasons the photoproduction sample of \JPsi meson events is  
restricted to decays into $\mu^+\mu^-$, while the electroproduction sample includes
both leptonic decay channels. 
The photoproduction sample was recorded in the years 2006 and 2007 and
corresponds to an integrated luminosity of $\mathcal{L} = \unit[165]{pb^{-1}}$, while the electroproduction sample was recorded in the years 2004 to 2007 and corresponds to an integrated luminosity of $\mathcal{L} = \unit[315]{pb^{-1}}$.

The measurement is performed in the kinematic range 
$60 < \Wgp < \unit[240]{GeV}$, $0.3 < \ZJPsi < 0.9$ and $\PtPtStarJPsi > \unit[1]{GeV}$.
In photoproduction the transverse momentum \PtJPsi is measured in the lab frame, while in electroproduction the transverse momentum \PtStarJPsi is calculated in the $\gamma^
*p$ rest frame.
To suppress contributions from diffractive 
production of \JPsi and \PsiPrime mesons, selected events are required to contain at least 
five reconstructed tracks in the central region of the detector 
($20^\circ < \ThetaTrk < 160^\circ$). 
The reconstruction efficiency accounts for this experimental cut and the measured cross sections are corrected for this track multiplicity cut.

Figure~\ref{fig:ctrl:mass} shows the invariant mass spectra of the leptons in the selected event samples. 
The number of signal events, $N_\JPsi$, is obtained in all bins of the cross section measurements from a fit to the mass distributions in the interval $2 < \Mll < 6$ GeV.
For the decay into muons the signal peak is described using a modified Gaussian~\cite{ZEUS05}. 
In the case of a decay into two electrons an exponential is added to the lower mass region of the signal Gaussian in order to account for the radiative tail~\cite{msteder}.
For the differential cross section measurements, the width and asymmetry term of
the mass peak in each bin are fixed to the values obtained from the full samples. 
For both decay channels, the background is parametrised by a polynomial of third order. 
At $m_{\ell\ell} \approx \unit[3.7]{GeV}$, the nominal mass of \PsiPrime mesons, an additional Gaussian with fixed position and width is allowed in all analysis bins.

The selection criteria and the obtained event samples are summarised in table~\ref{tab:selection}.

%%%%%%%%%%%%%%%%%%%%%%%%%%%%%%%%%%%%%%%%%%%%%%%%%%%%%%%%%%%%
\section{Monte Carlo Simulations}
\label{sec:mc}

Cross sections and polarisation parameters are derived by correcting the measured number of events
and angular distributions for detector effects, such as detector 
resolutions and inefficiencies.
Several Monte Carlo generator programs are used to determine
the corrections. 
All samples are passed through a detailed simulation 
of the H1 detector response based on the GEANT program~\cite{geant} and through the
same reconstruction and analysis algorithms as used for the data.

Signal events are generated using the Monte 
Carlo generator \Cascade \cite{Jung:2000hk}. 
Elastic and proton-dissociative production of \PsiPrime mesons is simulated using  
\DiffVM~\cite{diffvm} with parameters tuned to describe the results of previous H1 measurements~\cite{psi2s97,psi2s02}.
% Psi(2s) 2002: pd: 0.59+/-0.13+/-0.12 / elas: 4.31+-0.57+-0.46
% \DiffVM:    : pd: 1.07                       4.57
The Monte Carlo generator \Pythia~\cite{pythia} is used for the description of 
the contribution from $b$ hadron decays as described in section~\ref{sec:bg}. 
All generators use the JETSET part of the \Pythia program~\cite{pythia} 
to simulate the hadronisation and decay processes.

\begin{figure}[p]
\centering
\unitlength1cm
\begin{picture}(16,19.5)
\put( 0. ,13.){\includegraphics*[width=7.5cm]{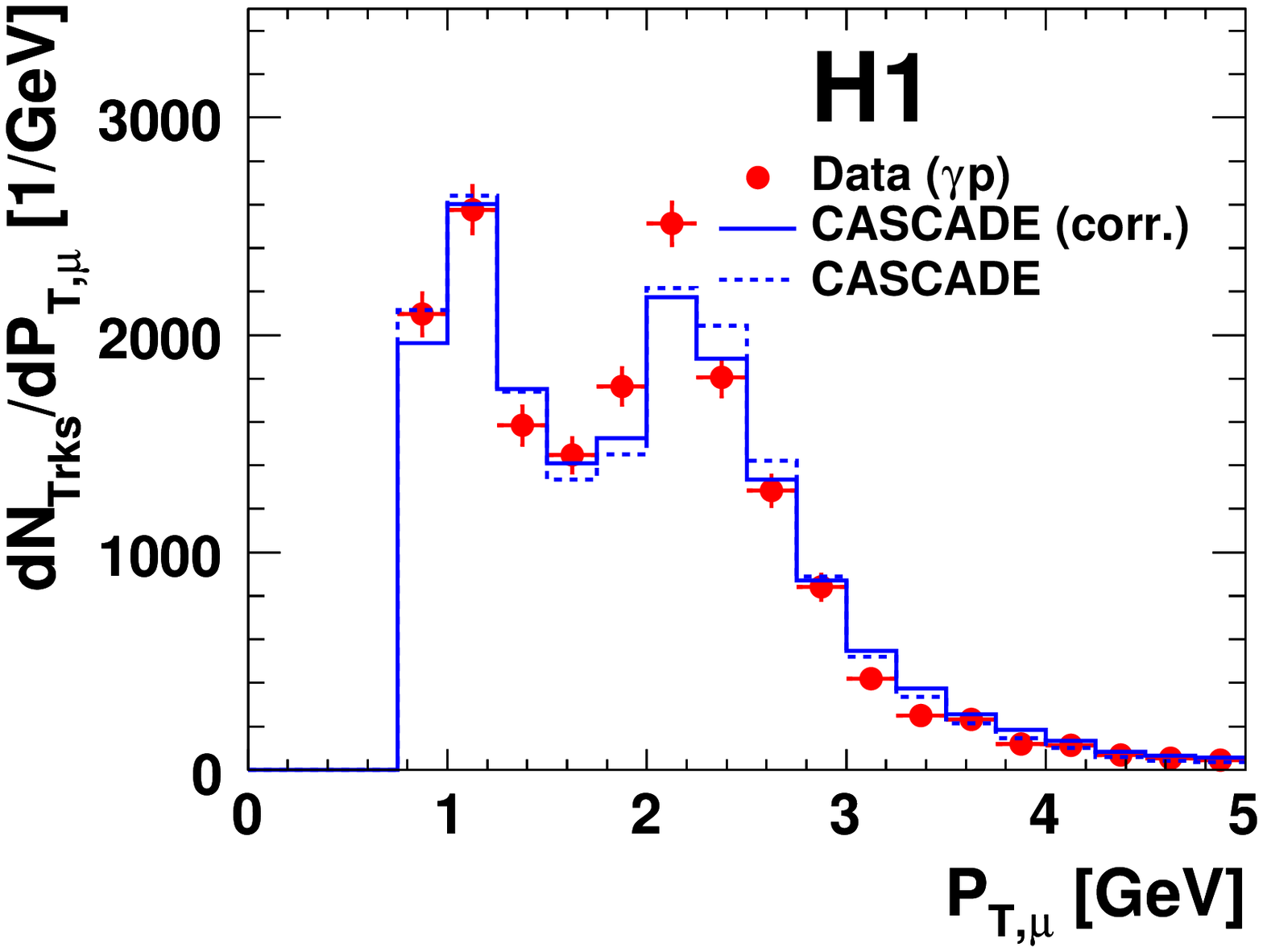}}%fin_control_PtTracks.eps}}
\put( 8.5,13.){\includegraphics*[width=7.5cm]{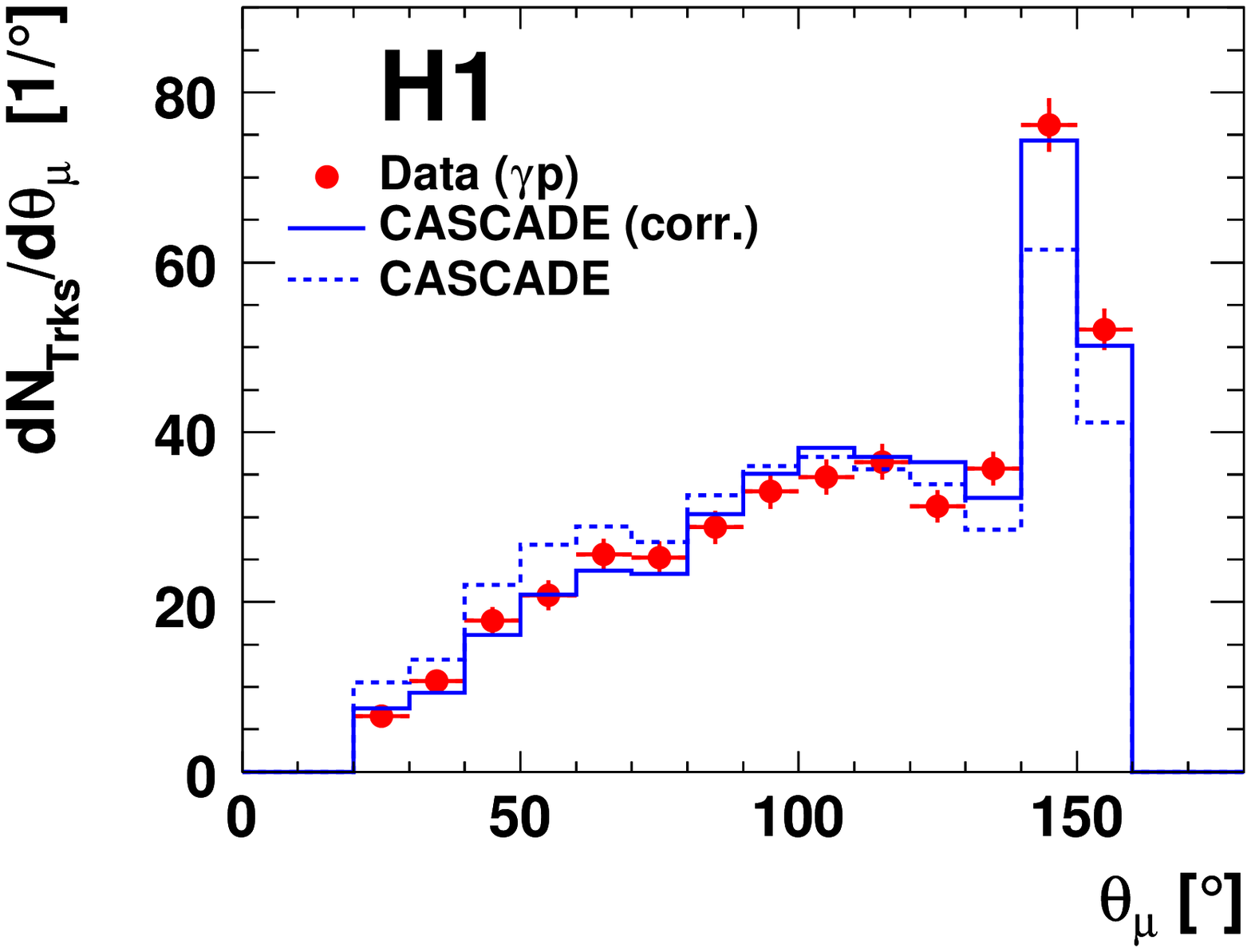}}%fin_control_ThetaTracks.eps}} 
\put( 0. ,6.5){\includegraphics*[width=7.5cm]{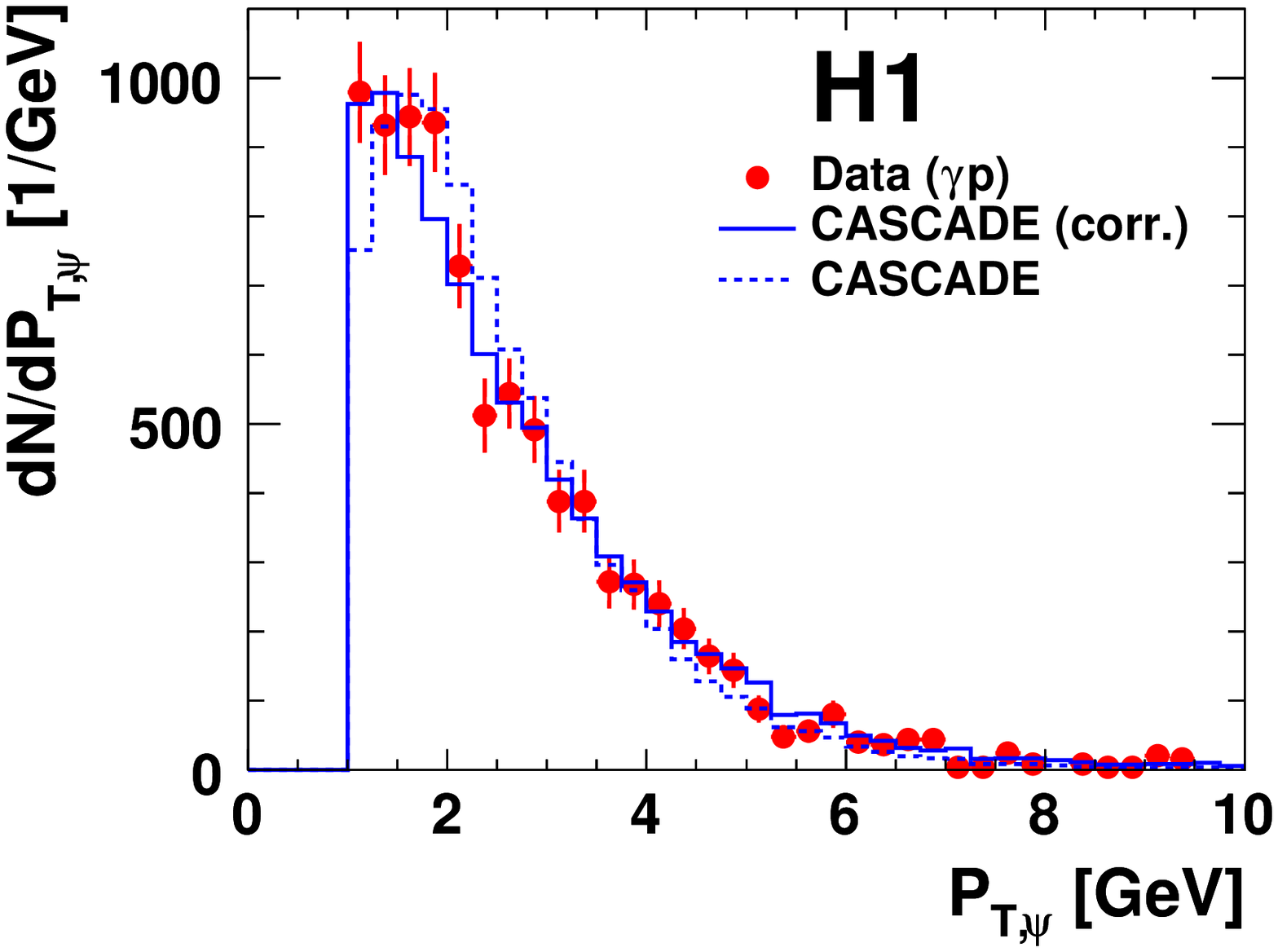}}%fin_control_PtJPsi.eps}} 
\put( 8.5,6.5){\includegraphics*[width=7.5cm]{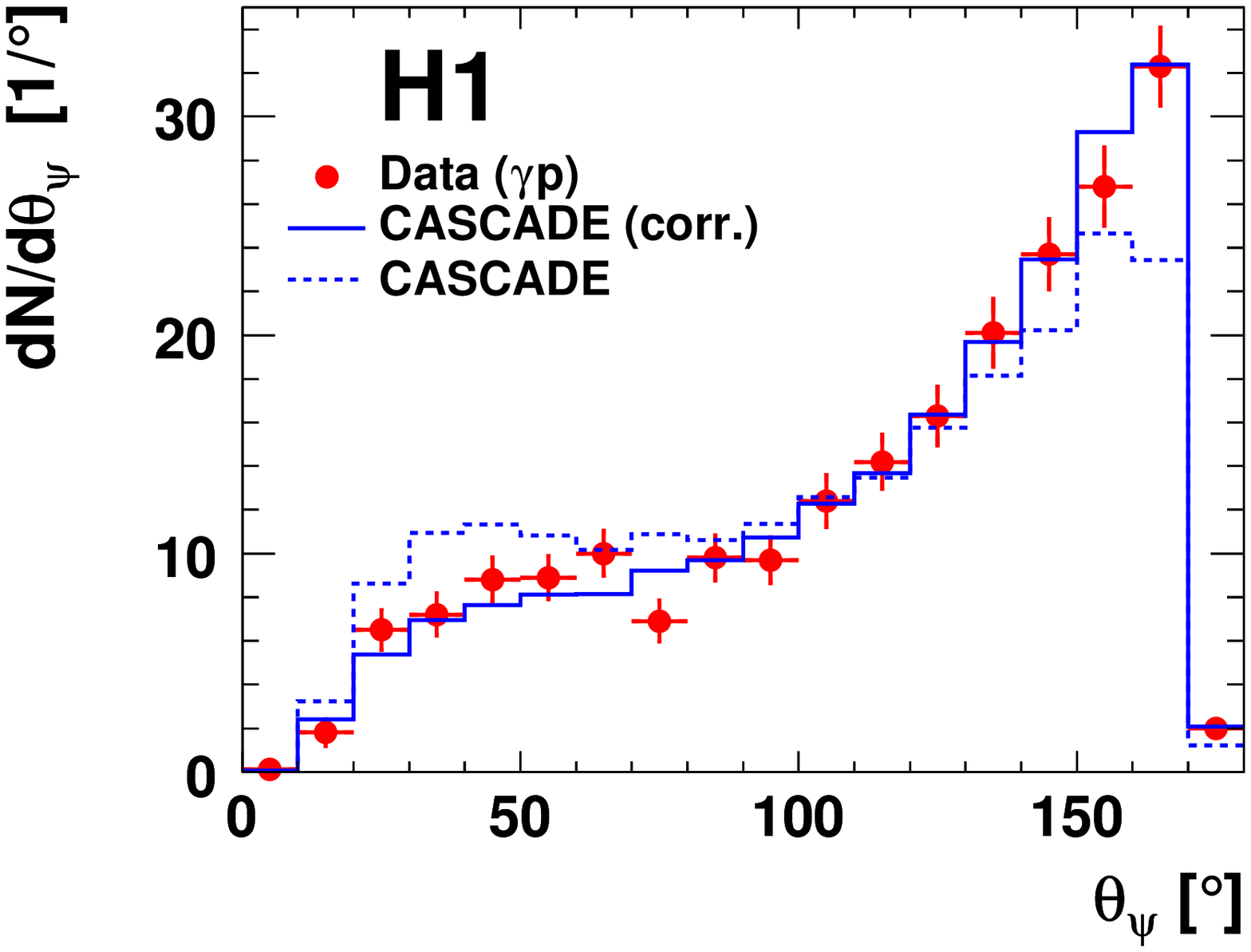}}%fin_control_ThetaJPsi.eps}}
\put( 0. , 0.){\includegraphics*[width=7.5cm]{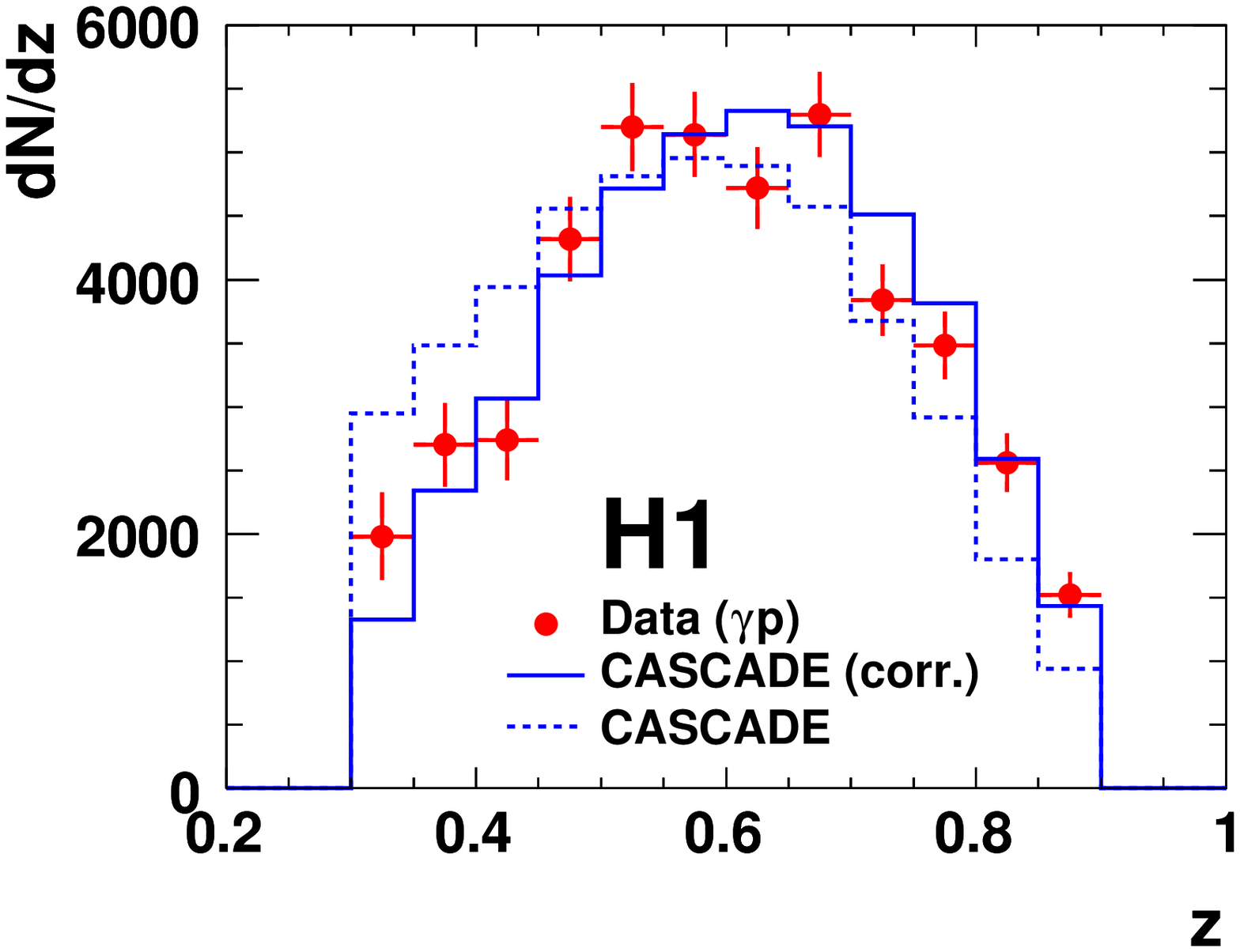}}%fin_control_ZJPsi.eps}}
\put( 8.5, 0.){\includegraphics*[width=7.5cm]{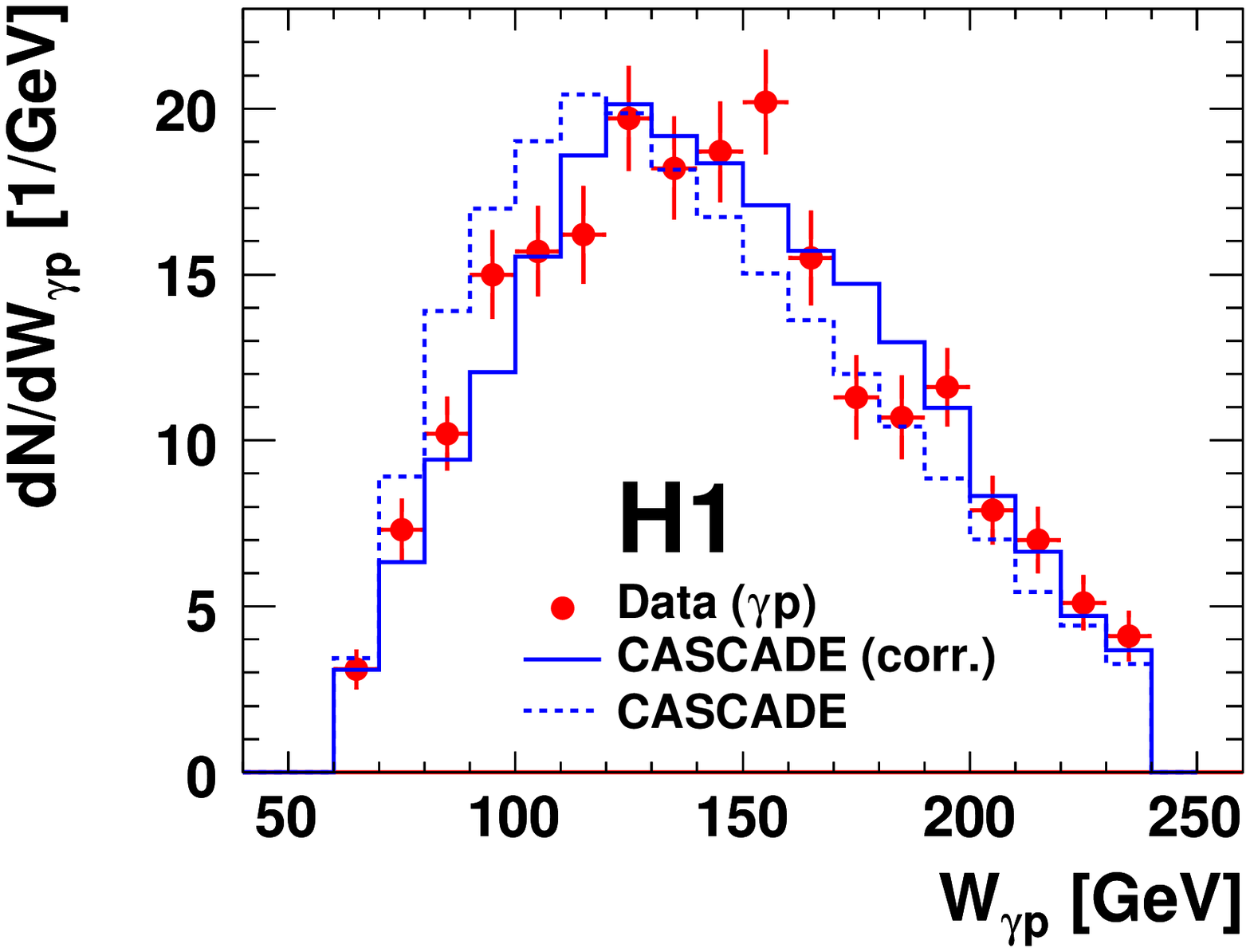}}%fin_control_Wgp.eps}}
\put(4.7,19.5){\large \sf Inelastic \JPsi Photoproduction}

\put(1.1,18.9){a)}
\put(9.6,18.9){b)}
\put(1.1,12.4){c)}
\put(9.6,12.4){d)}
\put(1.1,5.9){e)}
\put(9.6,5.9){f)}
\end{picture}
\caption{Control distributions of the photoproduction sample: 
a) the transverse momentum \PtMu of the muon tracks, 
b) the polar angle \ThetaMuon of the muon tracks, 
c) the transverse momentum \PtJPsi of the \JPsi meson, 
d) the polar angle \ThetaJPsi of the \JPsi meson, 
e) the elasticity \ZJPsi and
f) the photon proton centre-of-mass energy \Wgp.
The data are compared with predictions
from the corrected \Cascade Monte Carlo simulation (solid lines), normalised to
the number of entries in the data. The uncorrected \Cascade Monte Carlo prediction 
is shown as dashed line.}
\label{fig:ctrl:muon} 
\end{figure}

\begin{figure}[p]
\centering
\unitlength1cm
\begin{picture}(16,19.5)
\put( 0. ,13.){\includegraphics*[width=7.5cm]{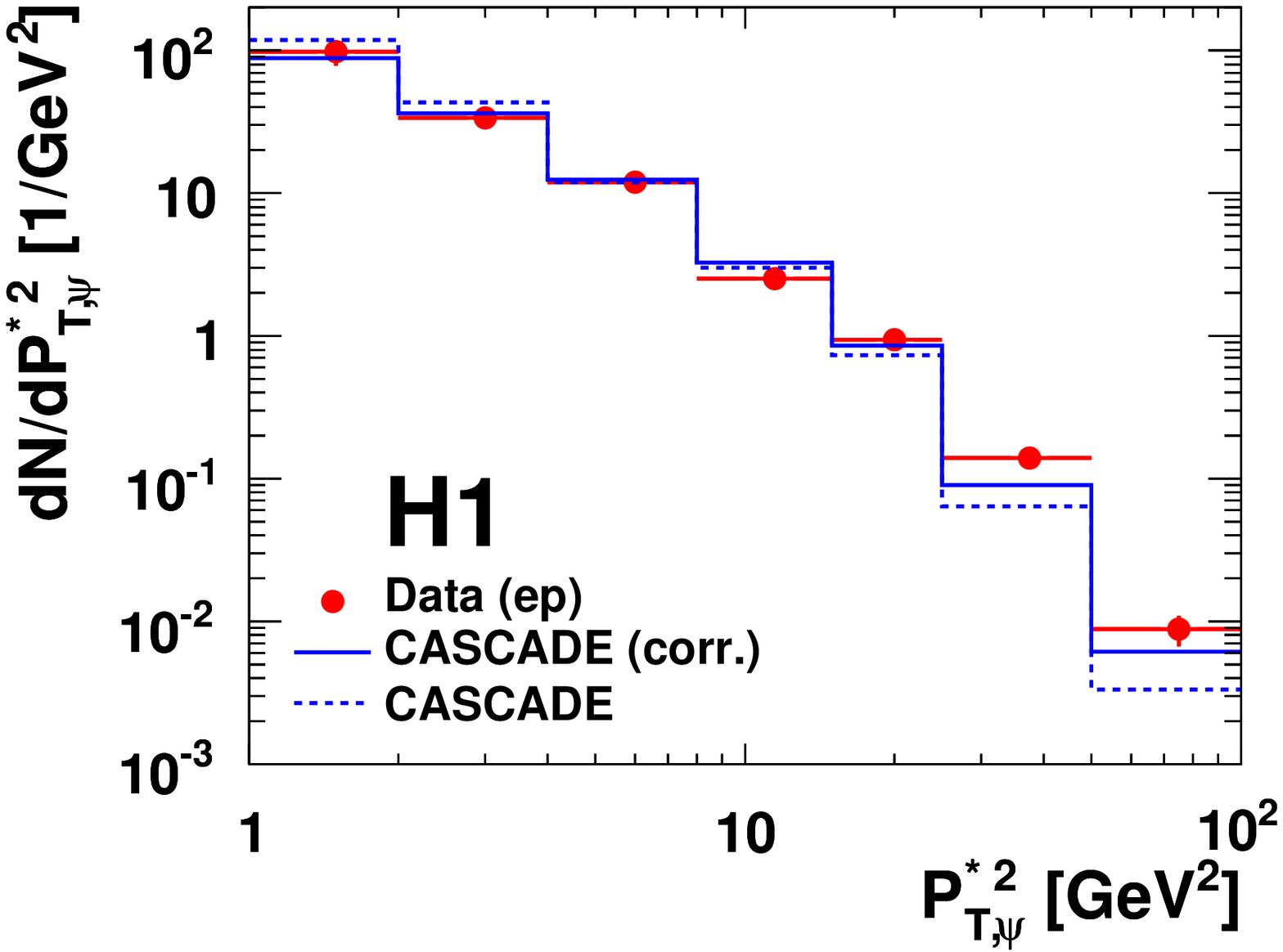}}%fin_control_PtStar2.eps}}
\put( 8.5,13.){\includegraphics*[width=7.5cm]{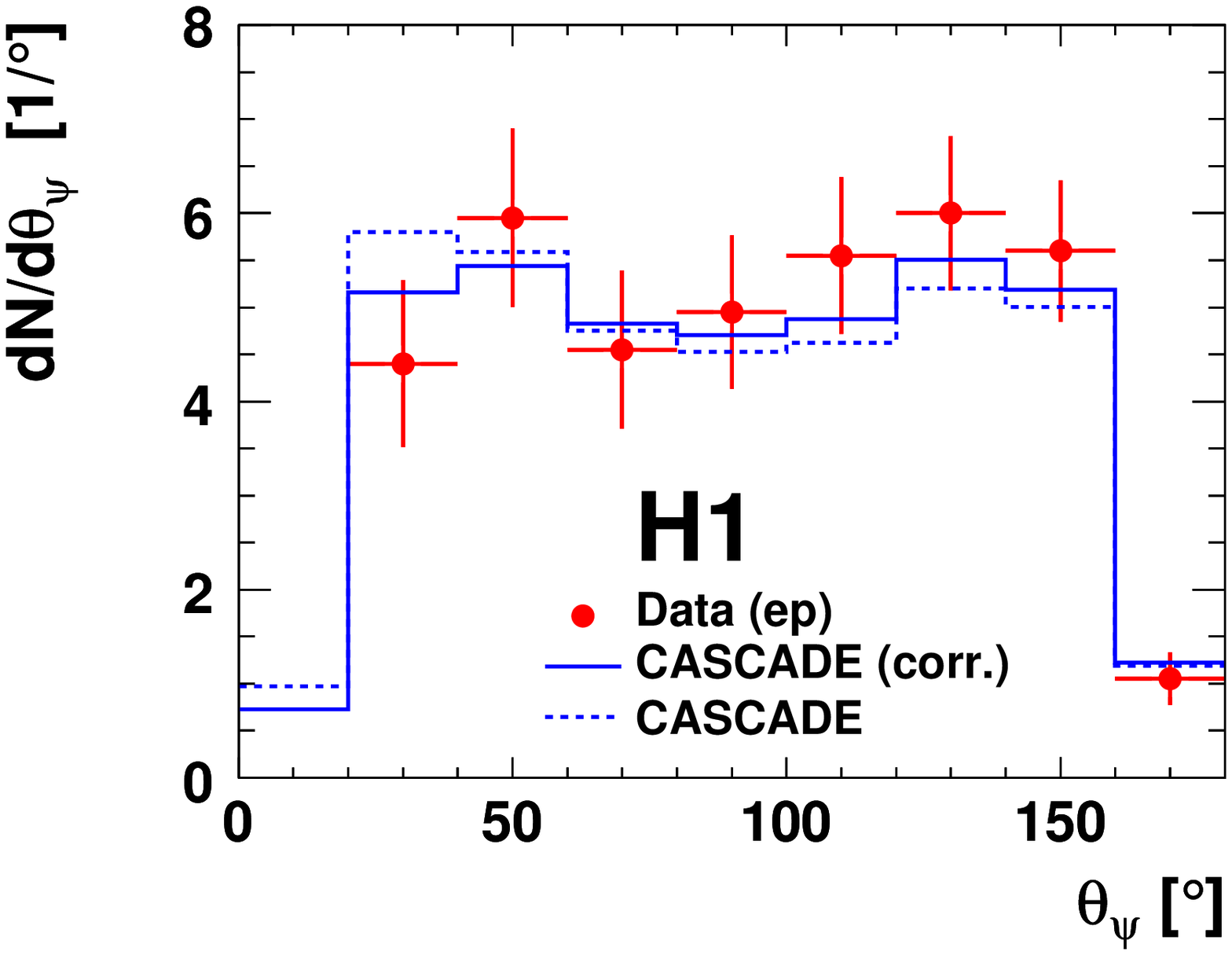}}%fin_control_ThetaJPsi.eps}} 
\put( 0. ,6.5){\includegraphics*[width=7.5cm]{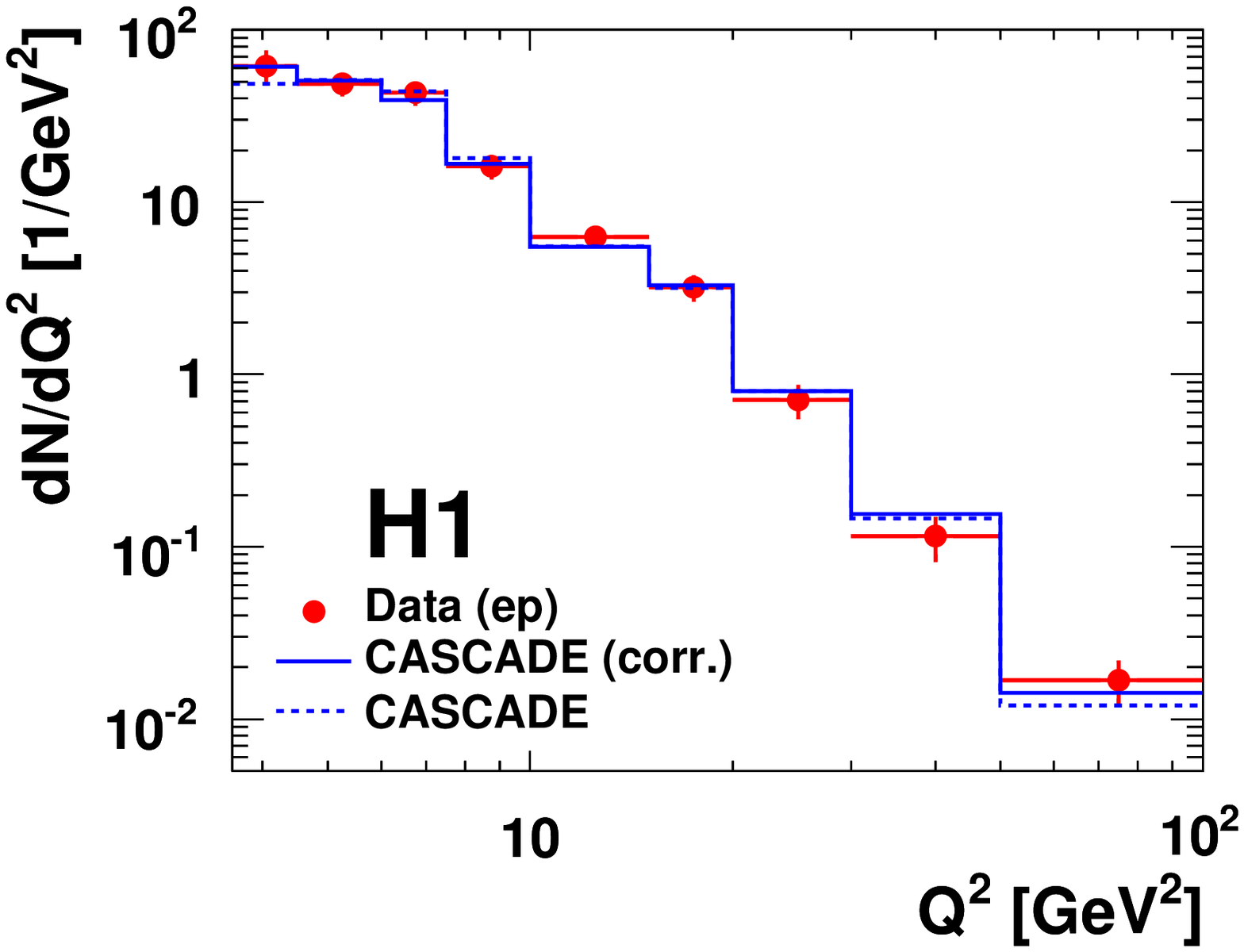}}%fin_control_Q2e.eps}}
\put( 8.5,6.5){\includegraphics*[width=7.5cm]{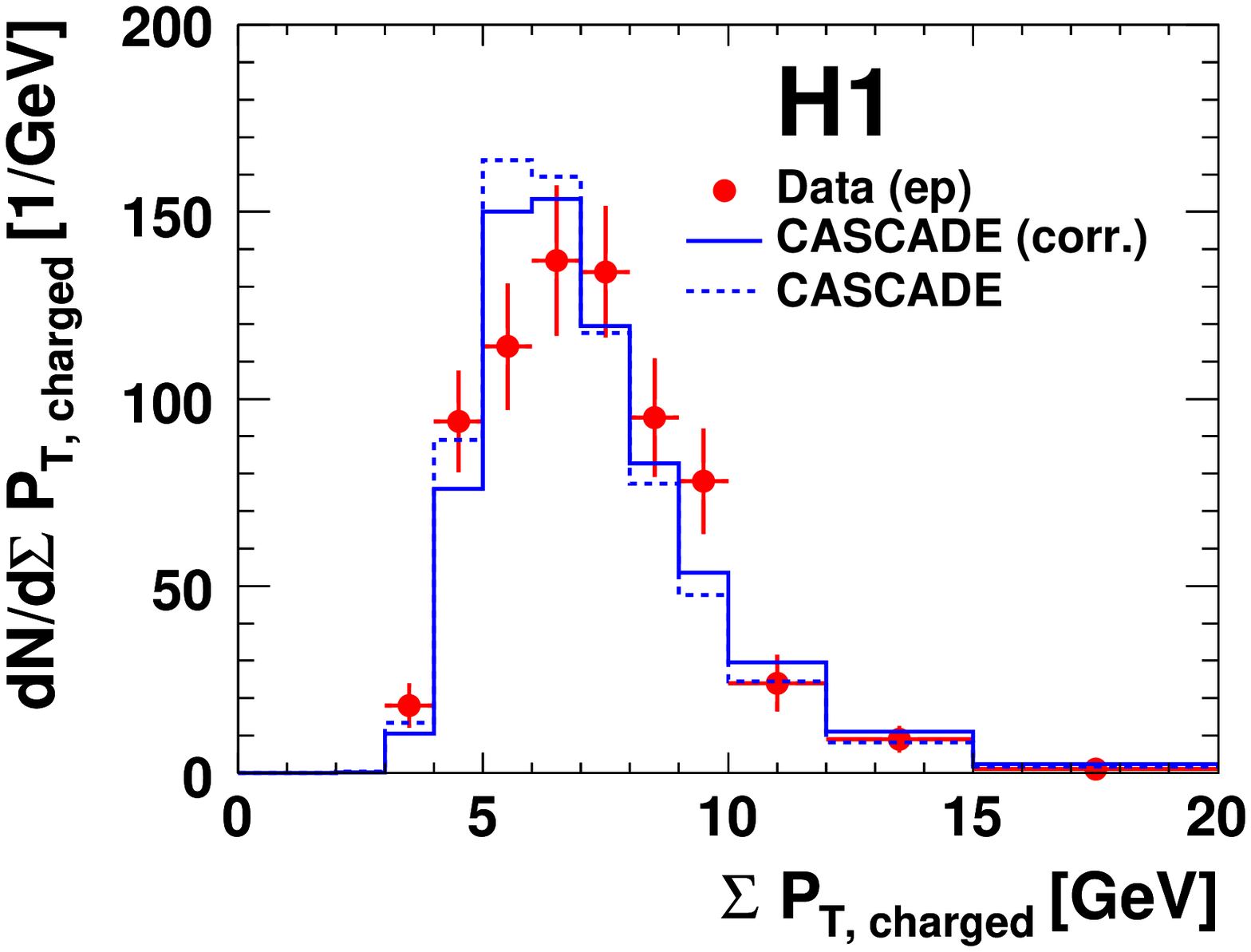}}%fin_control_PtSum_Other.eps}}
\put( 0. , 0.){\includegraphics*[width=7.5cm]{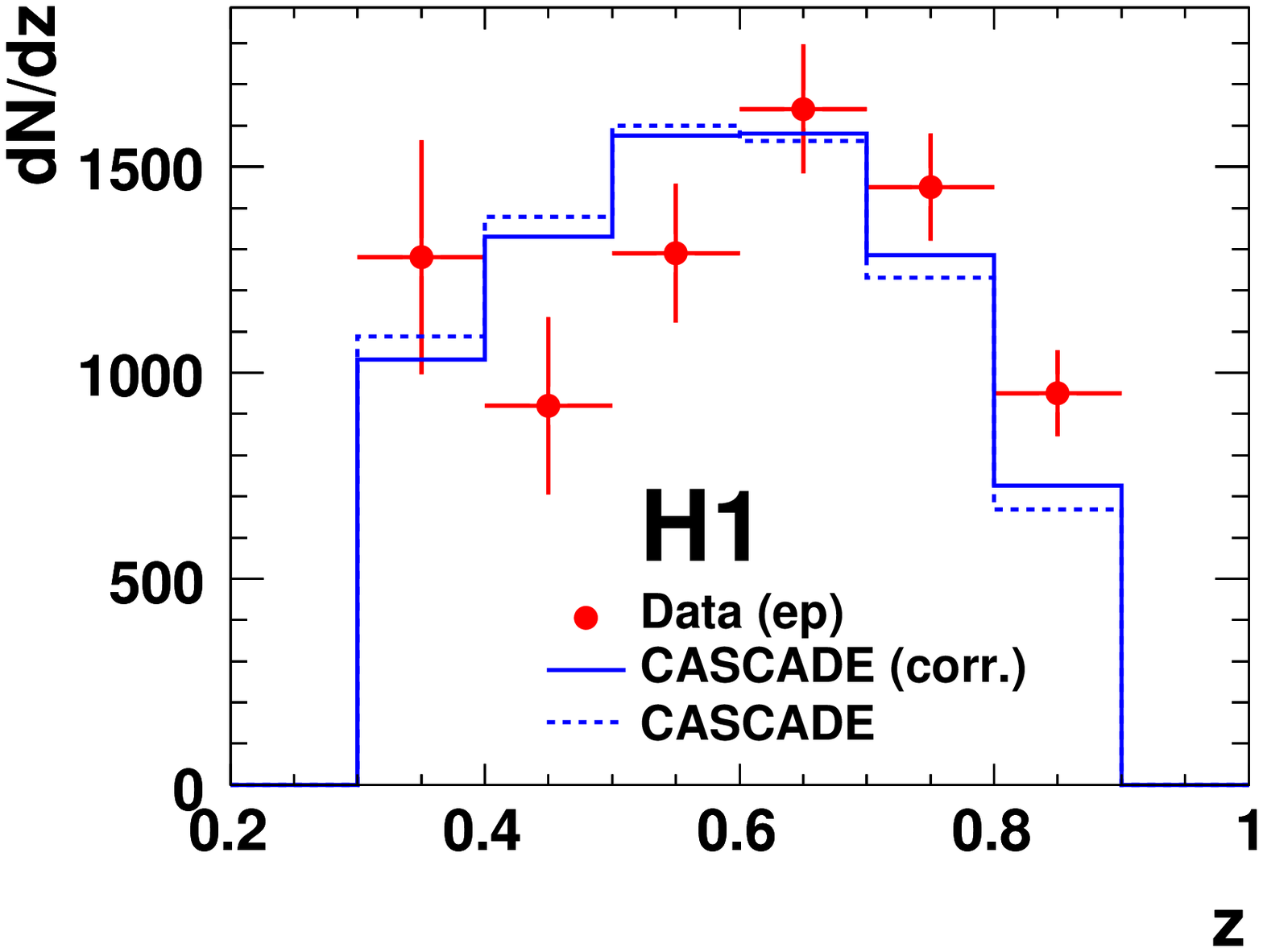}}%fin_control_Z.eps}} 
\put( 8.5, 0.){\includegraphics*[width=7.5cm]{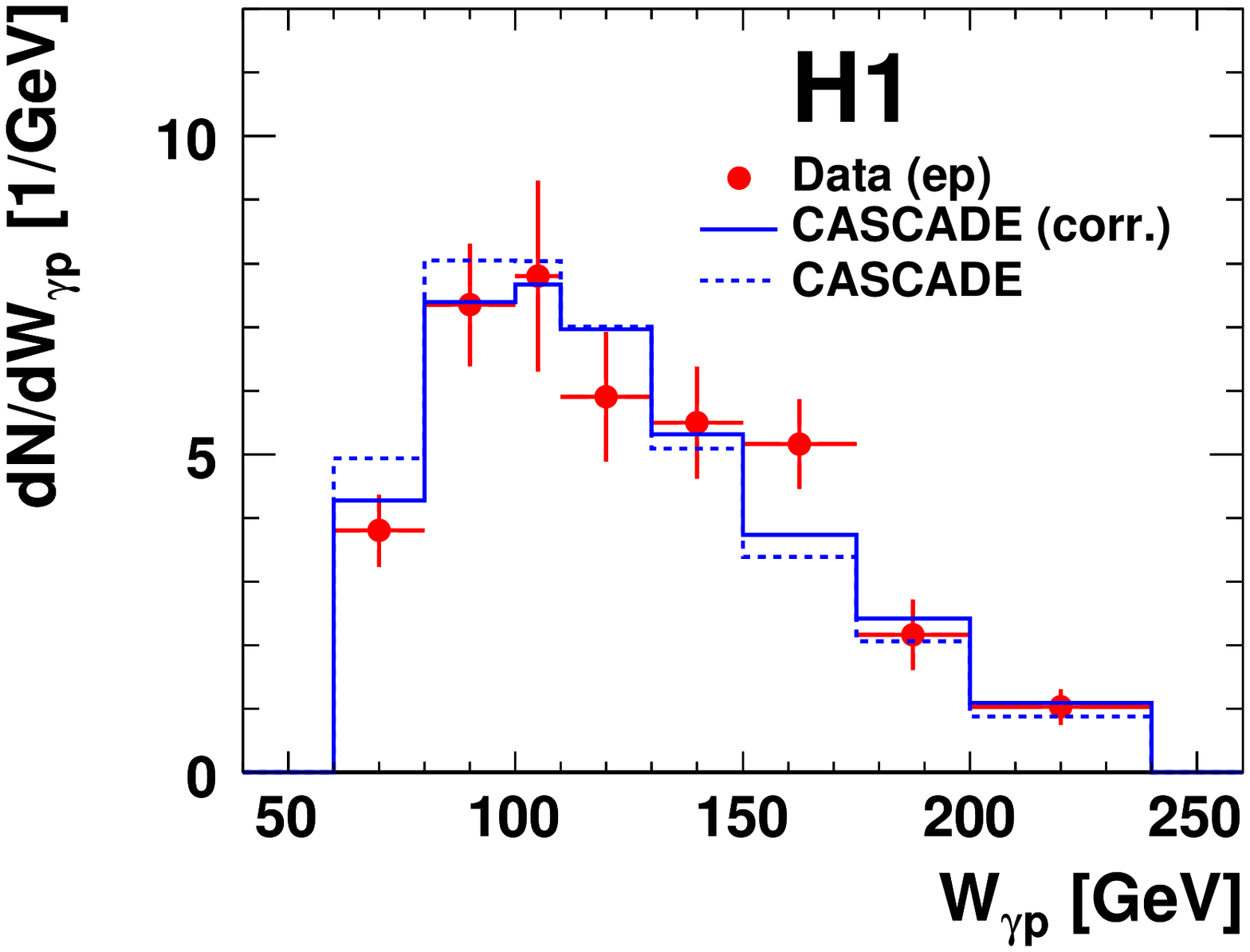}}%fin_control_Wgp.eps}}
\put(4.55,19.5){\large \sf Inelastic \JPsi Electroproduction}

\put(1.1,18.9){a)}
\put(9.6,18.9){b)}
\put(1.1,12.4){c)}
\put(9.6,12.4){d)}
\put(1.1,5.9){e)}
\put(9.6,5.9){f)}
\end{picture}
\caption{Control distributions of the electroproduction sample:
a) The squared transverse momentum of the \JPsi meson in the photon proton rest 
frame \PtStarJPsiSquare, 
b) the polar angle of the \JPsi meson \ThetaJPsi,
c) the photon virtuality \Qsquared, 
d) the scalar transverse sum $\Sigma P_{T,\rm charged}$,
e) the elasticity \ZJPsi and
f) the photon proton centre of mass energy \Wgp.
The data are compared with predictions from the corrected \Cascade Monte Carlo simulation (solid lines), normalised to the number of entries in the data. The uncorrected \Cascade Monte Carlo prediction 
is shown as dashed line.}
\label{fig:ctrl:dis} 
\end{figure}

Signal events as simulated with the Monte Carlo generator \Cascade are compared with the 
data after final selection in figures~\ref{fig:ctrl:muon} and \ref{fig:ctrl:dis}.
All data distributions in these figures are corrected for contributions from non-resonant 
background events using a sideband method described in~\cite{msteder}.

Corrections as a function of \Wgp and \PtJPsi in bins of the elasticity \ZJPsi 
are applied to the \Cascade Monte Carlo simulation in order to describe the data. 
Details of the procedure are described in~\cite{msteder}. 
In figure~\ref{fig:ctrl:muon} distributions for the photoproduction sample are compared to \Cascade Monte Carlo predictions before and after correction for the observables \PtMu, \ThetaMuon, \PtJPsi, \ThetaJPsi, \Wgp and \ZJPsi.
Similarly, in figure~\ref{fig:ctrl:dis}, the summed distributions for the two 
electroproduction samples ($\JPsi \ra \mu^+\mu^-$ and $\JPsi \ra e^+e^-$) are shown for the observables \PtStarJPsi, \ThetaJPsi, 
\Qsquared, $\Sigma P_{T,\rm charged}$, \Wgp and \ZJPsi. Here, $\Sigma P_{T,\rm charged}$ 
is the scalar sum over the transverse momenta of all measured charged particles except for  
the scattered electron and the \JPsi meson decay leptons.
The corrected \Cascade simulation gives a good description of all aspects of the data and
is used to correct the data for losses due to limited acceptance and efficiency
of the detector.

%%%%%%%%%%%%%%%%%%%%%%%%%%%%%%%%%%%%%%%%%%%%%%%%%%%%%%%%%%%%
\section{Backgrounds}
\label{sec:bg}
Remaining backgrounds to prompt \JPsi meson production in the selected sample originate 
from feed-down processes, i.e.\,\JPsi mesons produced in decays of diffractively or 
inelastically produced \PsiPrime mesons and $\chi_c$ mesons or of $b$ hadrons. 

Inelastic production of \PsiPrime mesons with a subsequent decay  into \JPsi mesons is expected
to contribute about $15-\unit[20]{\%}$ to the selected \JPsi meson samples~\cite{ZEUSPsi2S, Kraemer}. Since the production processes are the same, the inelastic \PsiPrime mesons show similar dependences on the kinematic variables.

\begin{figure}[p]
\centering
\unitlength1cm
\begin{picture}(16,13)
\put(3.8 ,6.5){\includegraphics*[width=7.5cm]{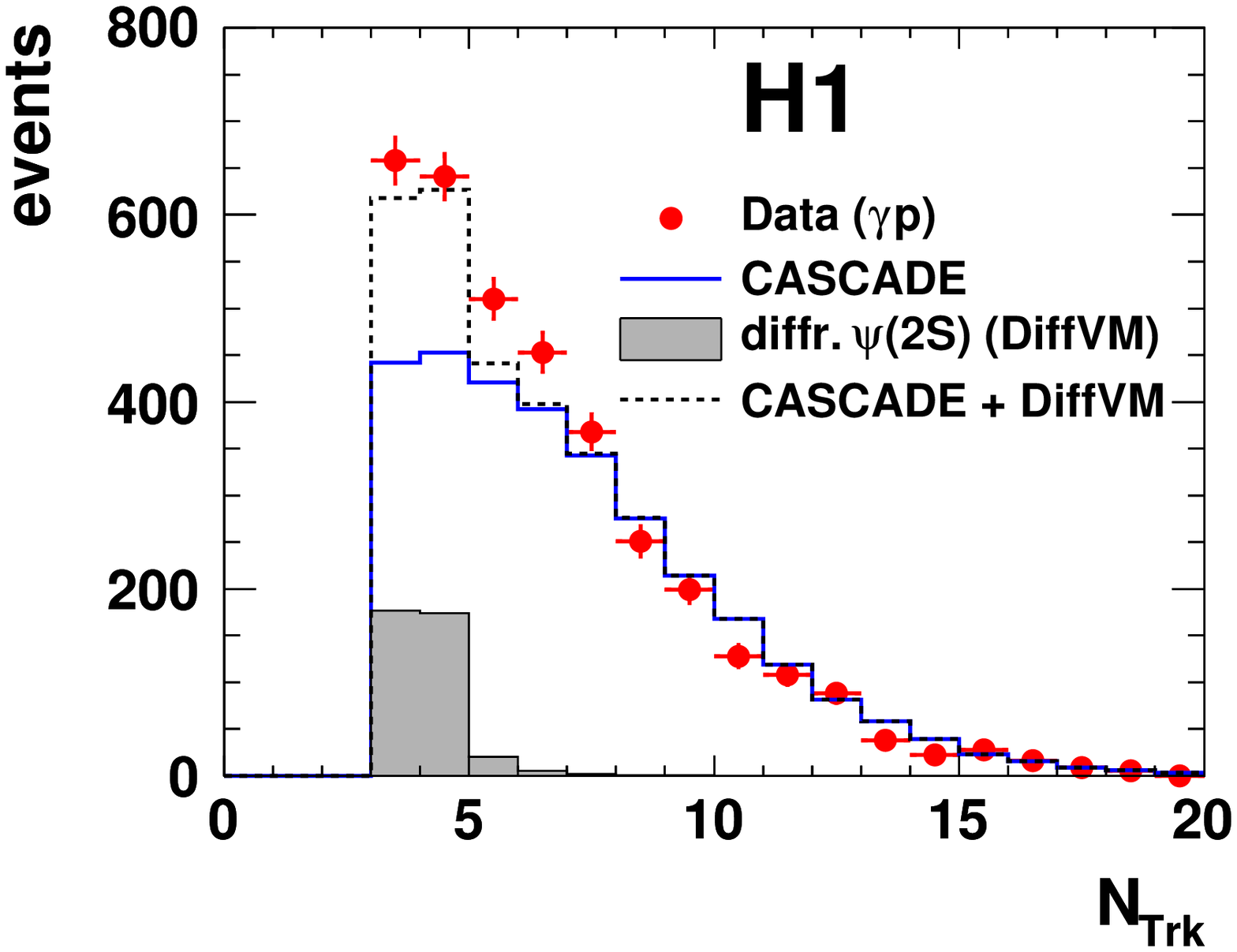}}%fin_control_Inel_3Trk.eps}}
\put(0. ,0.){\includegraphics*[width=7.5cm]{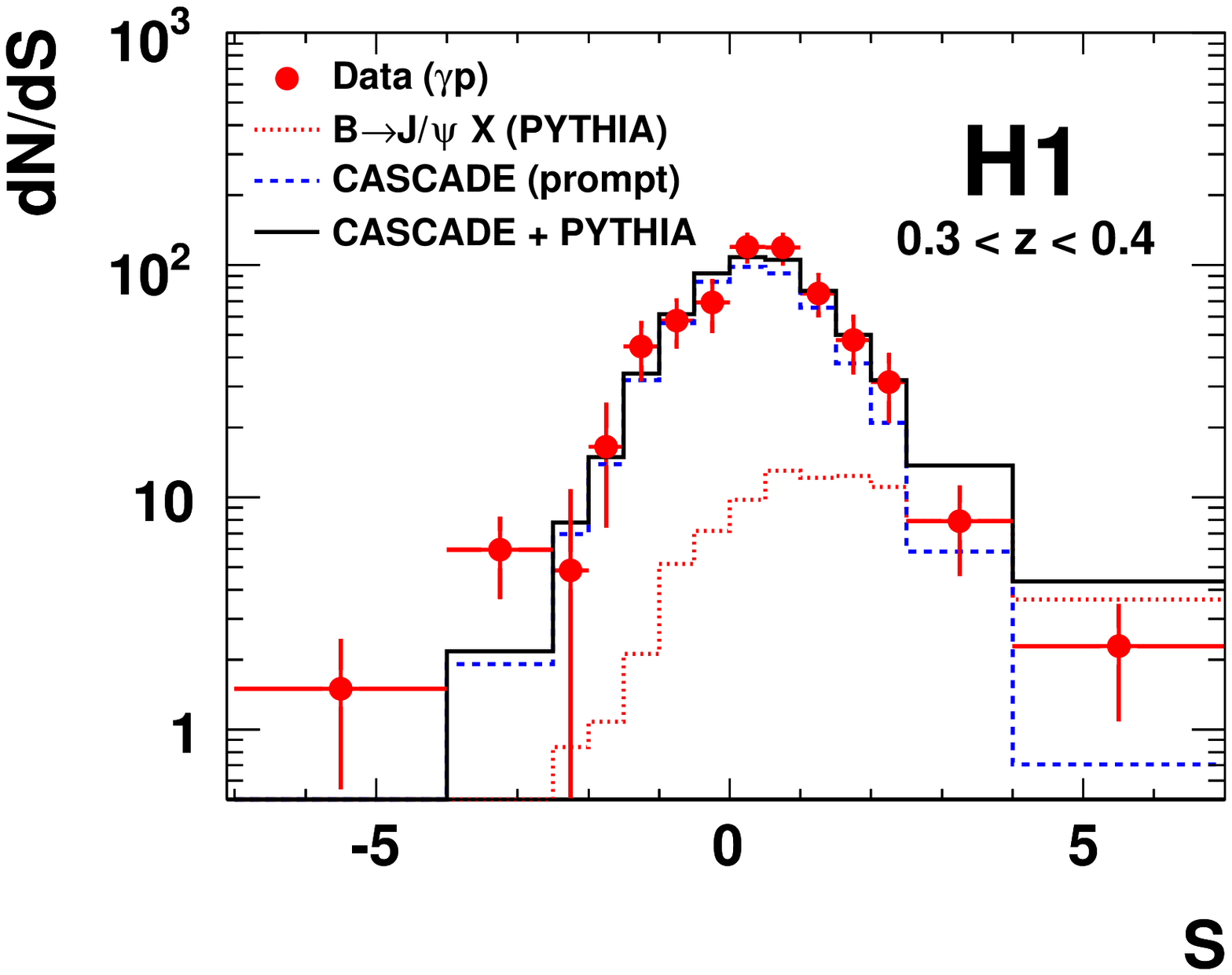}}%fin_lowZ_SIG_Z_3.eps}}
\put(8.5 ,0.){\includegraphics*[width=7.5cm]{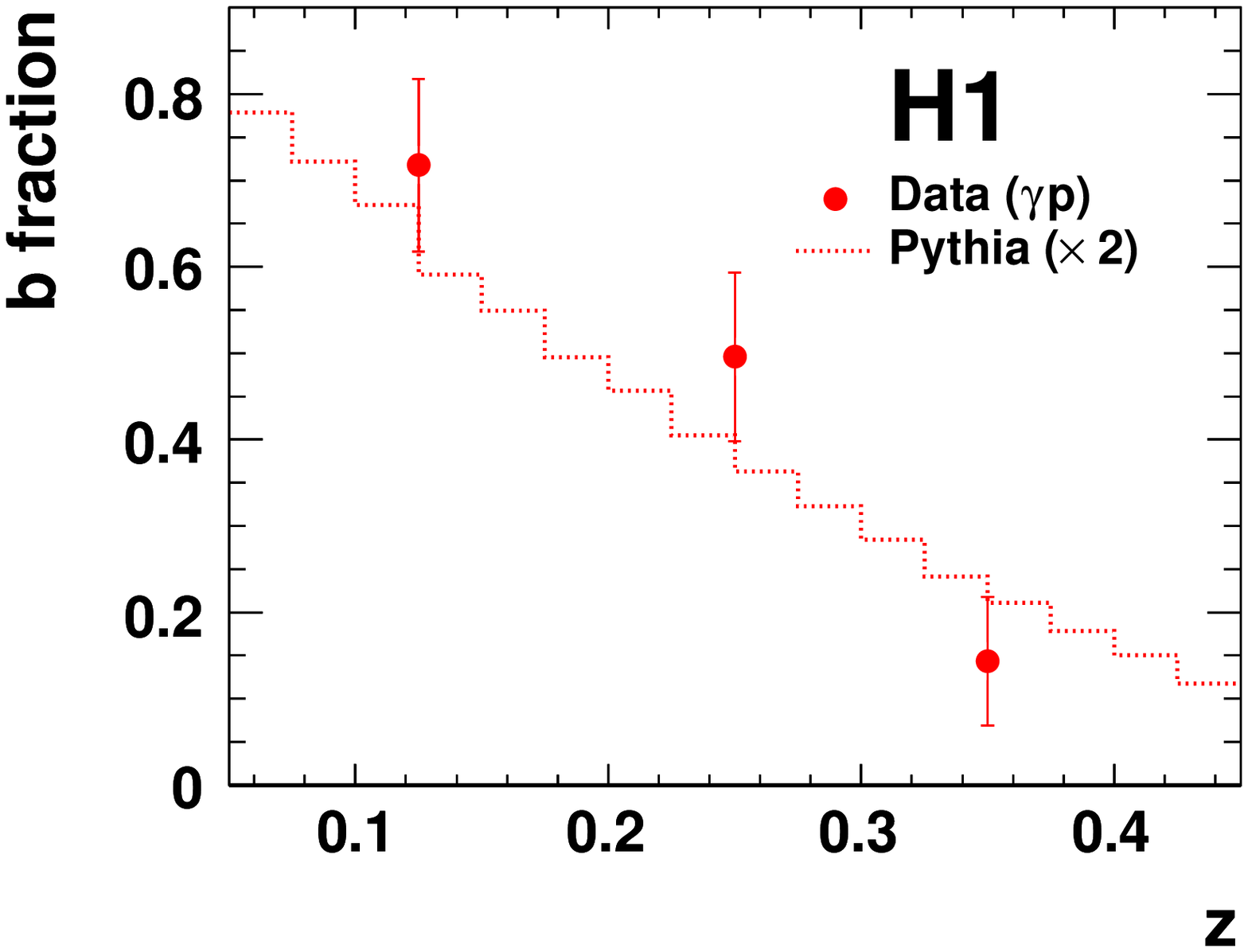}}%fin_lowZ_BFrac_Summary.eps}}

\put(4.7,12.8){\large \sf Inelastic \JPsi Photoproduction}

\put(4.9,12.35){a)}
\put(1.1,5.85){b)}
\put(9.6,5.85){c)}

\end{picture}

\caption{
a) Distribution of the multiplicity of tracks, $N_{Trk}$, in the central region of the detector 
($20^\circ < \ThetaTrk < 160^\circ$) for the photoproduction sample,
b) signed significance distribution $\mathcal{S}$ for the photoproduction sample at low elasticities 
($0.3 < \ZJPsi < 0.4$) and
c) measured contribution from $b$ hadron decays for three bins of the elasticity $z$
in comparison with the prediction based on \Pythia (scaled up by a factor of two) and \Cascade. }
\label{fig:ctrl:bgandcuts} 
\end{figure}

Diffractive production of \PsiPrime mesons contributes at large values of \ZJPsi by decays into a \JPsi meson and two charged pions. 
These events typically contain three or four reconstructed
charged tracks in the central detector ($20^\circ < \ThetaTrk < 160^\circ$). 
In figure~\ref{fig:ctrl:bgandcuts}a) the distribution of the charged track 
multiplicity measured in the central detector is shown for the photoproduction 
sample selected using all selection criteria given in table~\ref{tab:selection} except for the track multiplicity cut, which is relaxed to $\mathrm{N_{Trk}} \geq 3$.
The data are described by the sum 
of the \Cascade simulation and the prediction for diffractive \PsiPrime production, 
as simulated using the \DiffVM Monte Carlo generator. 
In the final selection remaining contributions from diffractive \PsiPrime meson production amount
to about 1.3\% in the total sample and to about 5\% in the highest elasticity bin, $0.75 < \ZJPsi < 0.9$.

The fraction of events arising from $b$ hadrons decaying into $\JPsi+X$ is 
estimated using the \Pythia simulation. The \Pythia prediction is scaled by
a factor of 2, based on results from previous measurements of beauty 
production at HERA~\cite{btomujj,btojj}. 
This scaled prediction by \Pythia amounts to 5\% in the total sample and 
about 20\% in the lowest \ZJPsi bin. It is confirmed within uncertainties by the following 
determination using data.
The fraction of events in the photoproduction sample containing $b$ hadrons
is estimated using the impact parameter of the decay muons to exploit the 
lifetime signature of $b$ hadrons. 
The impact parameter, $\delta$, of 
the decay muon tracks is defined as the distance of closest approach 
in the transverse plane to the reconstructed
primary vertex. The sign of the impact parameter is defined as positive if the 
angle between the decay muon and the \JPsi meson momentum direction is less than 
$\unit[90]{^\circ}$, and is defined as negative otherwise.
A signed significance $\mathcal{S} = \delta/\sigma(\delta)$ is reconstructed by weighting the reconstructed signed impact parameter with its uncertainty~\cite{CSTresol}.
Figure \ref{fig:ctrl:bgandcuts}b) shows the distribution of the signed 
significance for events in the interval $0.3 < \ZJPsi < 0.4$.
The histogram is filled with the signed significance of the decay muons 
for all events where both muon tracks have at least one hit in the CST.
The fraction of events coming from the decay of $b$ hadrons
is obtained from a fit of the significance distribution of \Cascade 
(simulating prompt \JPsi meson production) plus \Pythia 
(simulating \bbb events with subsequent decays into \JPsi + $X$) to that of the data.
The fit results are dominated by the region of small signed significances, $\mathcal{S} < 3$, due to large statistical uncertainties at larger values of $\mathcal{S}$.
The distribution of the data is corrected for non-resonant contributions using the side bands~\cite{msteder}.
The relative contribution from $b$ hadrons as resulting from the fits are 
shown in figure \ref{fig:ctrl:bgandcuts}c) for three bins of \ZJPsi. 
The scaled predictions from \Pythia are found to be in good agreement with the 
measured fractions, indicating that the background from $b$ hadrons is under control.

The contribution from $\chi_c$ production and decay was studied~\cite{Adloff:2002ex}
and found negligibly small in the present kinematic region, $0.3 < \ZJPsi < 0.9$.

%%%%%%%%%%%%%%%%%%%%%%%%%%%%%%%%%%%%%%%%%%%%%%%%%%%%%%%%%%%%
\section{Systematic Uncertainties}
\label{sec:syserr}

\begin{table}[tb] \centering
\begin{tabular}{lccc} 
\toprule
Source & \multicolumn{3}{c}{Uncertainty [$\%$]}\\
 \midrule  
 & Photoproduction & \multicolumn{2}{c}{Electroproduction} \\
 & $\JPsi \rightarrow \mu^+\mu^-$ & $\JPsi \rightarrow \mu^+\mu^-$ & $\JPsi \rightarrow e^+e^-$ \\
\midrule
Decay leptons reconstruction       &  $1$  & $1$ & $2$ \\
Decay leptons identification       &  $3$  & $3$ & $3$ \\
Number of signal events            &  $2$  & $2$ & $4$ \\
Trigger                            &  $3$  & $2$ & $2$ \\
Scattered electron energy scale    & ---   & $2$ & $2$ \\
Hadronic final state energy scale  &  $4$  & $3$ & $3$ \\
Integrated luminosity              &  $4$  & $3.2$ & $3.2$ \\
Model uncertainties                &  $5$  & $5$ & $5$ \\
Decay branching ratio              &  $1$  & $1$ & $1$ \\
\midrule
Sum                               &  $9.0$  & $8.2$ & $9.1$ \\
\bottomrule

\end{tabular}
\caption{Systematic uncertainties of the \JPsi meson production cross section. 
The total systematic uncertainty is the sum of the contributions 
added in quadrature.}
\label{tab:sys}
\end{table}
%%%%%%%%%%%%%%%%%%%%%%%%%%%%%%%%%%%%%%%%%%%%%%%%%%%%%

The sources of systematic uncertainties of the cross section measurement are listed in table~\ref{tab:sys} and are detailed in the following:

\begin{itemize}

\item 
The uncertainty on the cross section due to the track and vertex reconstruction efficiency has been 
determined to be $\unit[1]{\%}$ for $\JPsi \ra \mu\mu$ and $\unit[2]{\%}$ for $\JPsi \ra ee$.

\item
The efficiency for the identification of the leptons is determined using a high statistics sample of events of elastically produced \JPsi mesons~\cite{msteder}.
The detector simulation is reweighted to match the efficiency measured in the data as necessary. 
Remaining differences are smaller than $\unit[3]{\%}$ everywhere and are taken as systematic uncertainty. 

\item 
The systematic uncertainty on the determination of the number of 
signal events, obtained by a fit to the mass distributions in every analysis bin, is determined by a variation of the extraction method.
Comparing the number of signal events for binned and unbinned 
log-likelihood fits yields a systematic uncertainty of $\unit[0.5]{\%}$.
In addition, the result from the fit to background and signal is compared 
to the number of signal events above the fitted background function 
in the mass window between $2.95$ and $\unit[3.2]{GeV}$.
An uncertainty of $\unit[2]{\%}$ for the decay into muons and $\unit[4]{\%}$ for the electrons is found. 
The uncertainty for the electron is larger due to an additional uncertainty originating from the description of the radiative tail.

\item 
The trigger efficiencies are determined using independent trigger channels. 
For the electroproduction sample the trigger efficiency is measured to be $\unit[(97 \pm 2)]{\%}$.
In the photoproduction sample the trigger efficiency depends mainly on the identification of the decay muons in the central muon system.
The efficiency amounts to about $\unit[70]{\%}$ with a systematic uncertainty of $\unit[3]{\%}$.
A detailed description of the determination of the trigger efficiencies can be found in~\cite{msteder}.

\item For the electroproduction sample the measurement of the scattered 
electron energy is known with a scale uncertainty of $\unit[1]{\%}$. The uncertainty of the scattering angle is 1 mrad.
Both uncertainties combined lead to an uncertainty of the cross section 
measurement of $\unit[2]{\%}$ on average.

\item 
The hadronic energy scale uncertainty is $\unit[4]{\%}$ in the LAr and  $\unit[7]{\%}$ in the SpaCal.
This leads to an uncertainty on the cross sections measurement of  $\unit[3]{\%}$ for the electroproduction sample and $\unit[4]{\%}$ for the photoproduction sample.

\item 
The integrated luminosity is known to a precision of $\unit[3.2]{\%}$ for the electroproduction sample and $\unit[4.0]{\%}$ for the photoproduction sample. 

\item 
The dependence of the result on model assumptions made in the \Cascade
Monte Carlo simulation were investigated and found to amount to $\unit[5]{\%}$ in total.
The model uncertainty arising from the knowledge of the decay angular 
distributions, explained in section~\ref{sec:angles}, is determined 
by variation of the parameter $\alpha$  in the simulation by $\pm 0.3$.
This variation results in a change of the cross section of up to $\unit[4]{\%}$.
The systematic uncertainty originating from the uncertainty of the slope of the \PtJPsi(\PtStarJPsi) distribution 
in the simulation is determined by a variation of this distribution as described in~\cite{msteder}.
This variation results in a change of the cross section of up to $\unit[4]{\%}$.

\item
The branching ratios of the leptonic decay channels of the \JPsi meson are known with an accuracy of $\unit[1]{\%}$ \cite{PDG}.

\end{itemize}

The total systematic uncertainty is obtained by adding all the above contributions 
in quadrature. 
A total systematic uncertainty of $\unit[9]{\%}$ is determined for the photoproduction sample.
For the combined electroproduction cross section the total systematic uncertainty is $\unit[8.5]{\%}$
The same uncertainties are attributed to all bins of the cross section measurement. 
For the measurement of the helicity distributions only the uncorrelated systematic uncertainties
are taken into account. 
They amount to about $\unit[3.5]{\%}$ and are negligible compared to the statistical uncertainties.
%%%%%%%%%%%%%%%%%%%%%%%%%%%%%%%%%%%%%%%%%%%%%%%%%%%%%%%%%%%
\section{Cross Section Measurements}
\label{sec:xsec}

The cross section measurement is performed in the kinematic range $60 < \Wgp < \unit[240]{GeV}$, $0.3 < \ZJPsi < 0.9$ and $\PtPtStarJPsi > \unit[1]{GeV}$. The photon virtuality \Qsquared is limited in the electroproduction analysis to $3.6 < \Qsquared < \unit[100]{GeV^2}$ and for the photoproduction sample to $\Qsquared < \unit[2.5]{GeV^2}$.
 
For the measurement of differential cross sections the number of signal 
events in each bin is corrected for detector inefficiencies and acceptance
and normalised to integrated luminosity and branching ratio. 
They are not corrected for QED radiative effects.
The electroproduction cross sections, measured from $\JPsi \ra \mu\mu$ and $\JPsi \ra ee$, are combined~\cite{msteder}.
The differential cross sections are bin-centre corrected using MC simulations.
In order to avoid model dependencies, the measured cross sections are not corrected for contributions from backgrounds as described in section~\ref{sec:bg}.
All measured cross sections are listed in tables~\ref{tab:xsecs:gammap:pt2} -- \ref{tab:xsecs:dis:Z_PtStar} together with statistical and systematic uncertainties.

\begin{figure}[p]
\centering
\unitlength1cm
\begin{picture}(16,13)
\put(0. , 6.5){\includegraphics*[width=7.5cm]{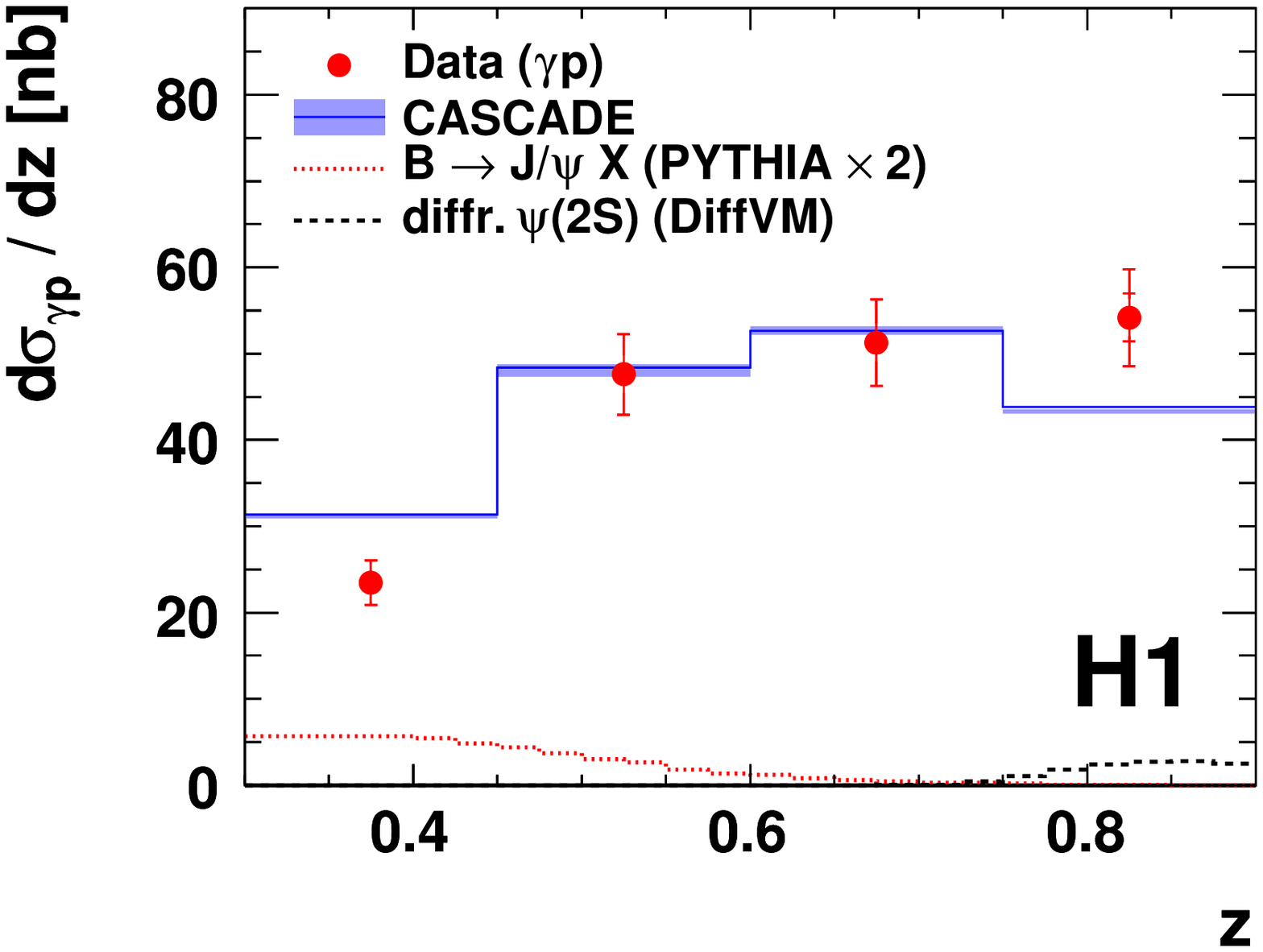}}%fin_xs_Z_Cascade.eps}} 
\put(8.5, 6.5){\includegraphics*[width=7.5cm]{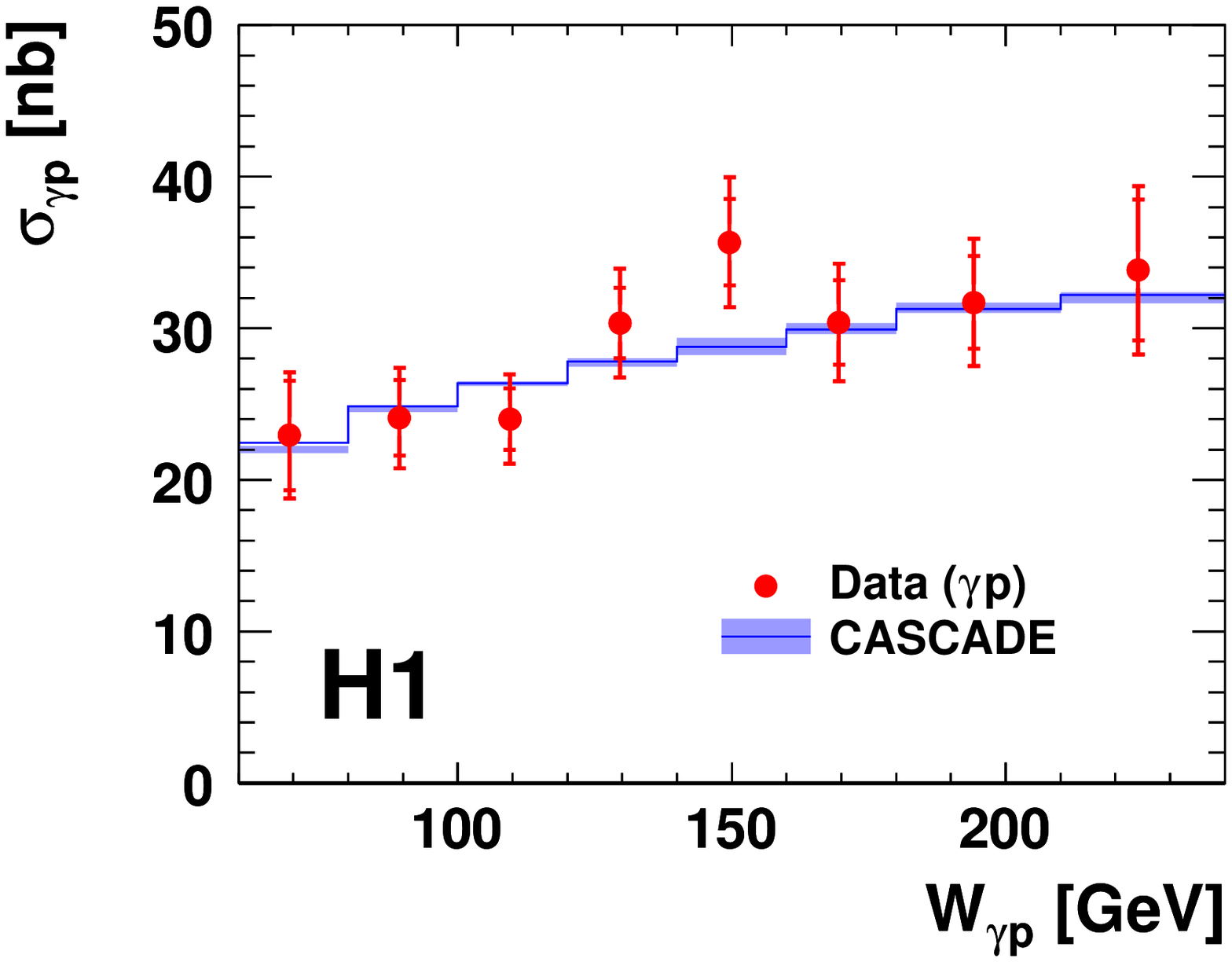}}%fin_xs_Wgp_Cascade.eps}} 
\put(4.25 , 0.){\includegraphics*[width=7.5cm]{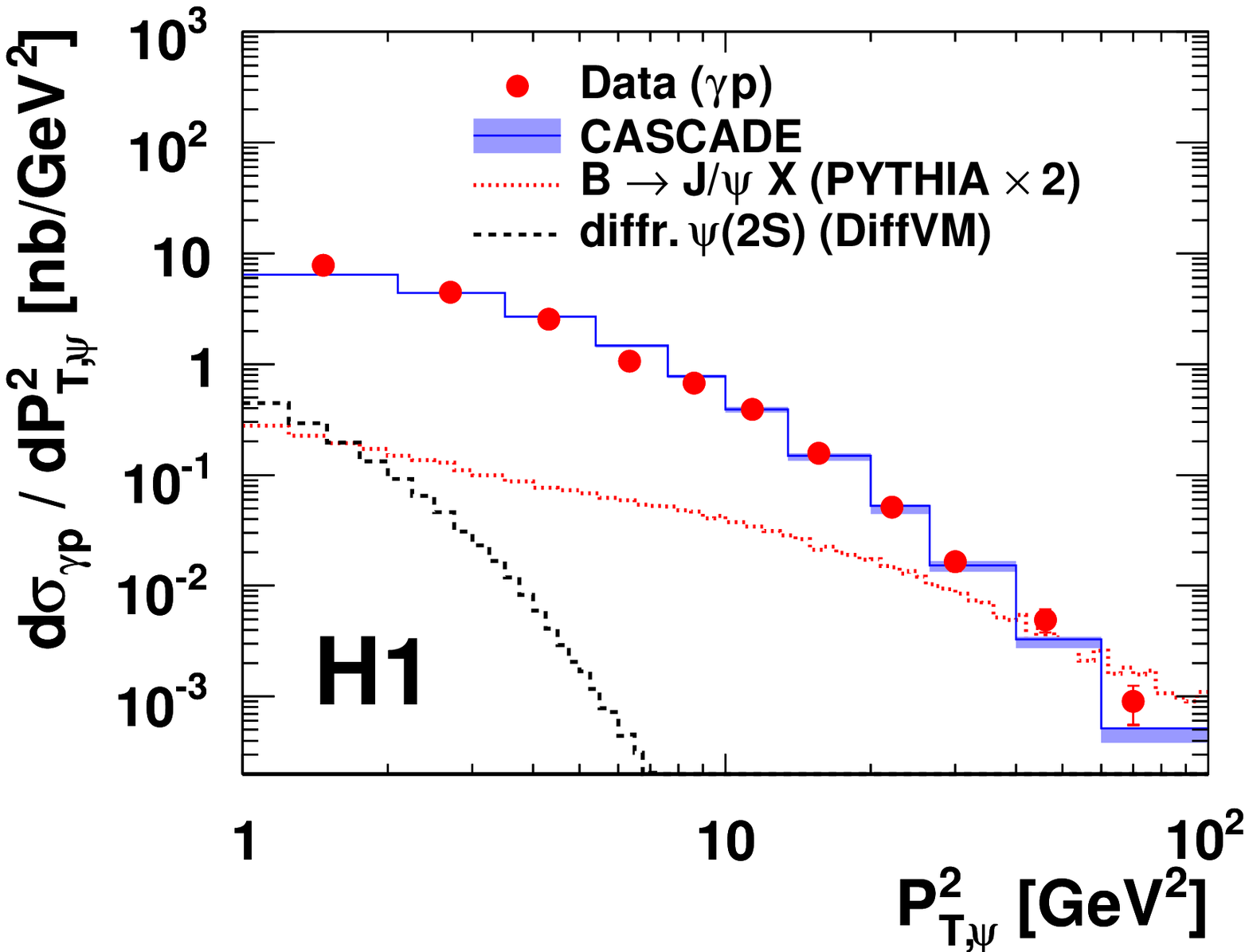}}%fin_xs_Pt2JPsi_Cascade.eps}} 
		
\put(4.7,13){\large \sf Inelastic \JPsi Photoproduction}

\put(1.4,12.3){a)}
\put(9.9,12.3){b)}
\put(5.6,5.8){c)}
\end{picture}
\caption{Differential \JPsi meson photoproduction cross sections for the kinematic range
$60 < \Wgp < \unit[240]{GeV}$, $0.3 < \ZJPsi < 0.9$ and $\PtJPsi > \unit[1]{GeV}$,
as functions of 
a) the elasticity \ZJPsi, 
b) the photon proton centre of mass energy \Wgp  and 
c) the squared transverse momentum of the \JPsi meson \PtJPsiSquare.
The inner error bar represents the statistical uncertainty and the outer error bar indicates the statistical and systematic uncertainties added in quadrature.
The data are compared to the predictions from \Cascade (solid line). 
The uncertainty band of the \Cascade prediction arises from a scale variation by a factor of two.
The dashed and dotted lines indicate the remaining background from diffractive \PsiPrime or $b$ hadron decays respectively as estimated using MC simulations.}
\label{fig:res:gp:xsecs:1d}
\end{figure}

\begin{figure}[p]
\centering
\unitlength1cm
\begin{picture}(16,13)
\put(3.8,6.5){\includegraphics*[width=7.5cm]{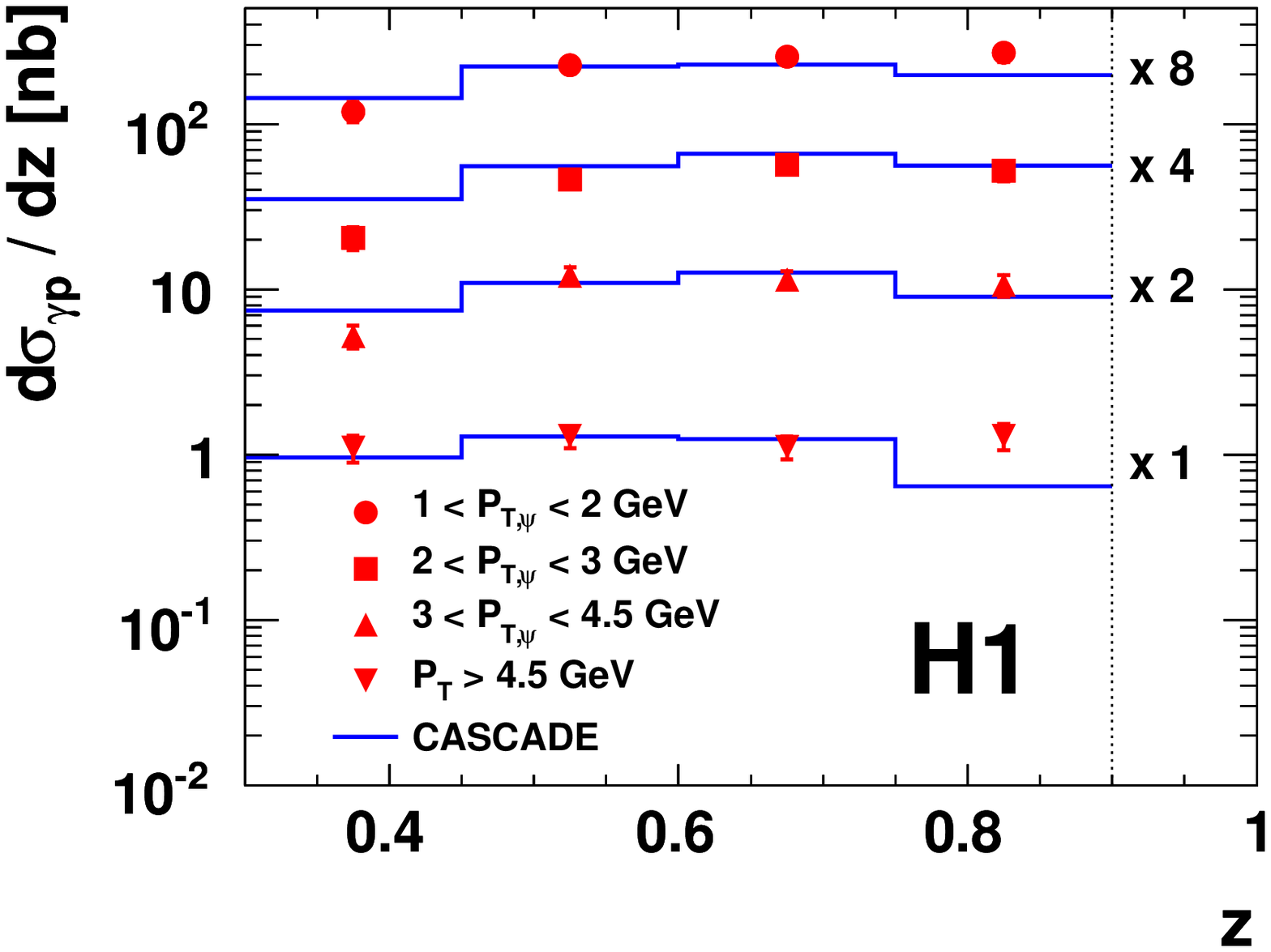}}%fin_xs_Z_PtBINS_Cascade.eps}}
\put(3.8,0.){\includegraphics*[width=7.5cm]{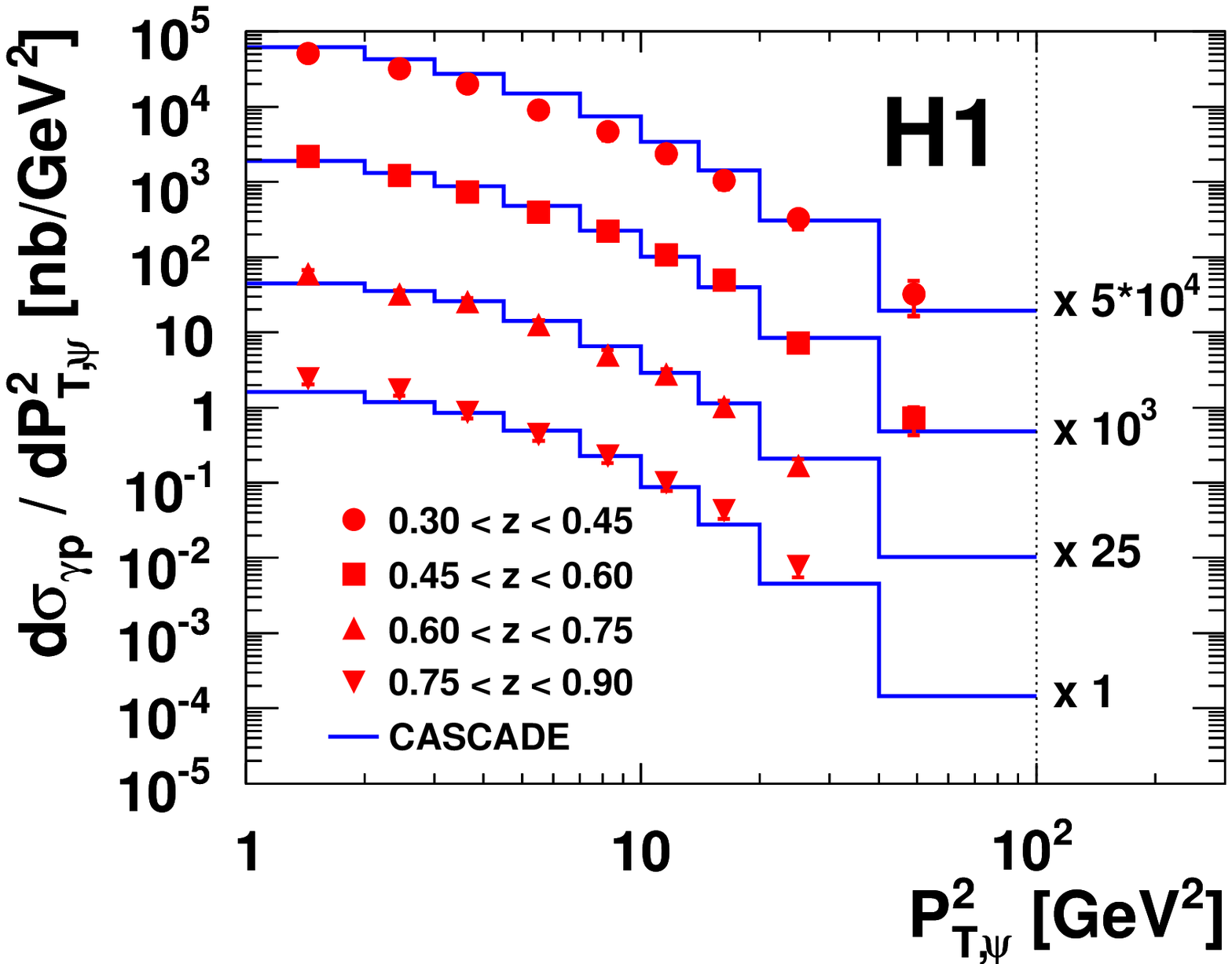}}%fin_xs_Pt2_ZBINS_Cascade.eps}}
\put(5.1,12.3){a)}
\put(5.1,5.8){b)}
\put(4.7,13.0){\large \sf Inelastic \JPsi Photoproduction}
\end{picture}
\caption{a) Differential \JPsi meson cross sections as a function 
of \ZJPsi in four bins of \PtJPsi and
b) differential \JPsi meson cross sections as a function 
of $\PtJPsiSquare$ in four bins of \ZJPsi. 
The inner error bar represents the statistical uncertainty and the outer error bar indicates the statistical and systematic uncertainties added in quadrature.
For visibility, the measured cross sections are scaled by the factors indicated in the figures. 
The data are compared to the predictions from \Cascade (lines).}
\label{fig:res:gp:xsecs:2d} 
\end{figure}

For the photoproduction sample the measured $ep$ cross sections are transformed to $\gamma p$ 
cross sections using the photon flux factors presented in
table~\ref{tab:fluxes}, calculated in the Weizs\"acker Williams approximation~\cite{WWA}.
The differential \JPsi meson photoproduction cross section is measured as function
of the elasticity \ZJPsi and the squared transverse momentum \PtJPsiSquare 
of the \JPsi meson. The total $\gamma p$ cross section is measured in bins of the 
photon proton centre of mass energy \Wgp.
The results are displayed in figure~\ref{fig:res:gp:xsecs:1d} and 
show a reasonable agreement with
the prediction from the \Cascade MC generator.
A variation of the renormalisation scale by a factor of two ($0.5 \mu_0 < \mu_\mr{r} < 2 \mu_0$) has little effect as shown by the band in the figures.
In addition to the \Cascade prediction, the remaining contributions from
diffractive \PsiPrime mesons and from $b$ hadrons are shown.
The distributions in \PtJPsiSquare and \ZJPsi are further investigated by 
dividing the sample into bins of \PtJPsiSquare and \ZJPsi, respectively as shown in figure~\ref{fig:res:gp:xsecs:2d}.
The \ZJPsi distribution tends to flatten off towards larger
values of \PtJPsi presented in figure~\ref{fig:res:gp:xsecs:2d}a).
It can be seen that differences between the data and the \Cascade prediction are localised at low elasticities and low transverse momenta of the \JPsi mesons, where \Cascade overshoots the data, and at large elasticities and large transverse momenta, where \Cascade is below the data.
Taking into account that the measured cross section in the lowest elasticity bin
includes a significant fraction of about $\unit[20]{\%}$ of events originating
from $b$ hadron decays, the difference to the \Cascade prediction is even more
significant.

\begin{figure}[p]
\centering
\unitlength1cm
\begin{picture}(16,13)
\put(0. ,6.5){\includegraphics*[width=7.5cm]{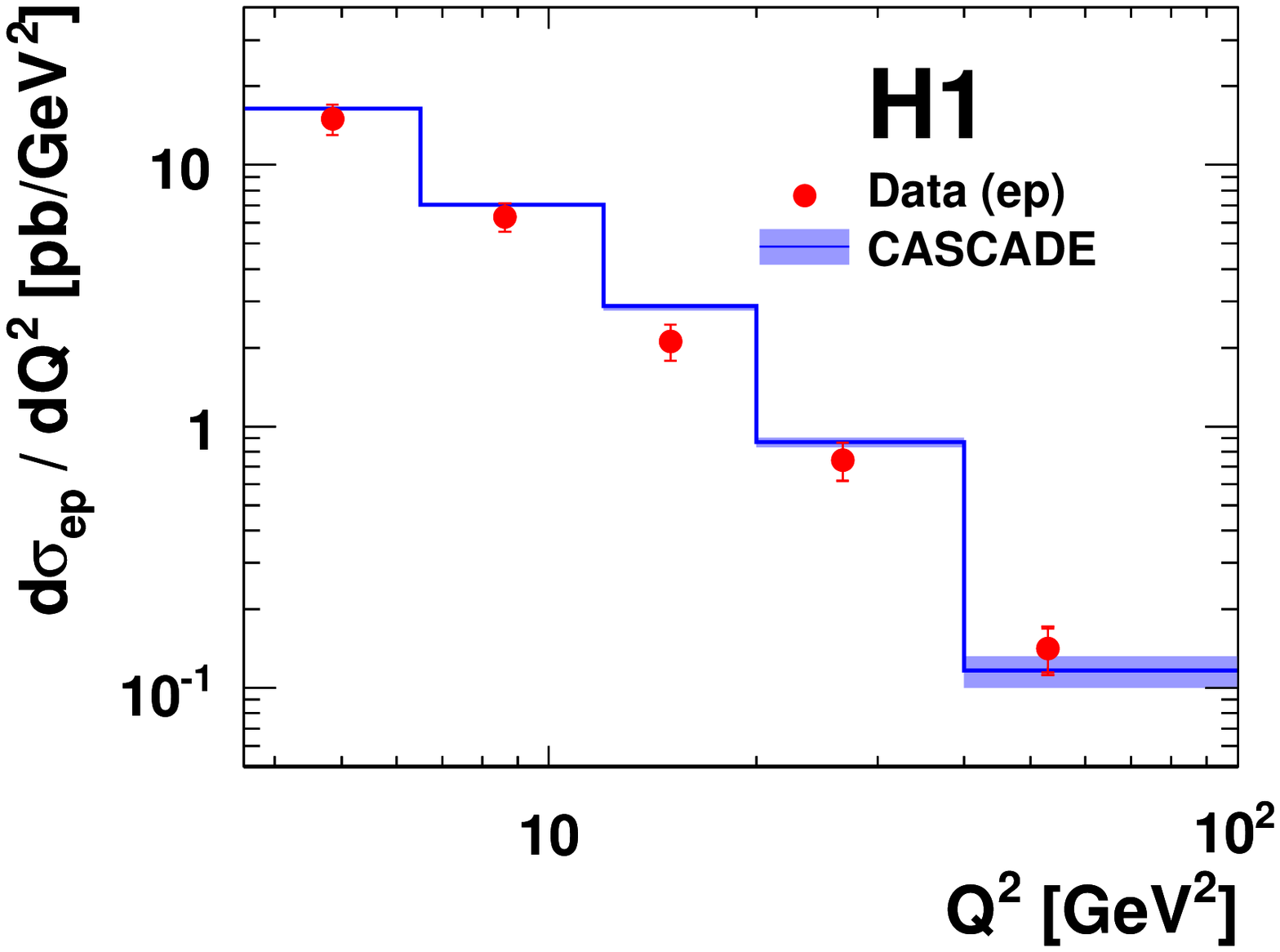}}%fin_xs_Q2e_Cascade.eps}}
\put(8.5,6.5){\includegraphics*[width=7.5cm]{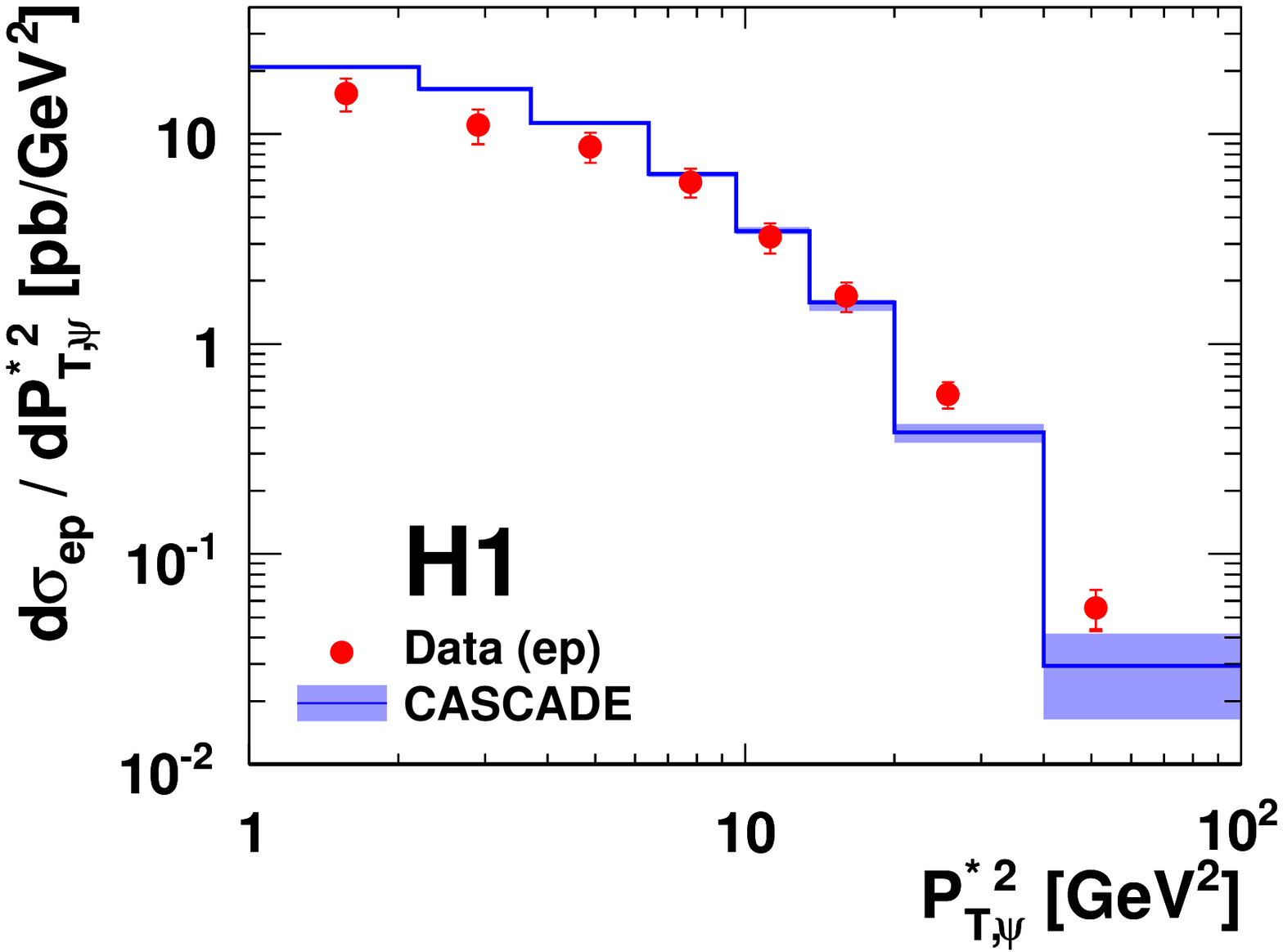}}%fin_xs_PtStar2_Cascade.eps}} 
\put(0., 0.){\includegraphics*[width=7.5cm]{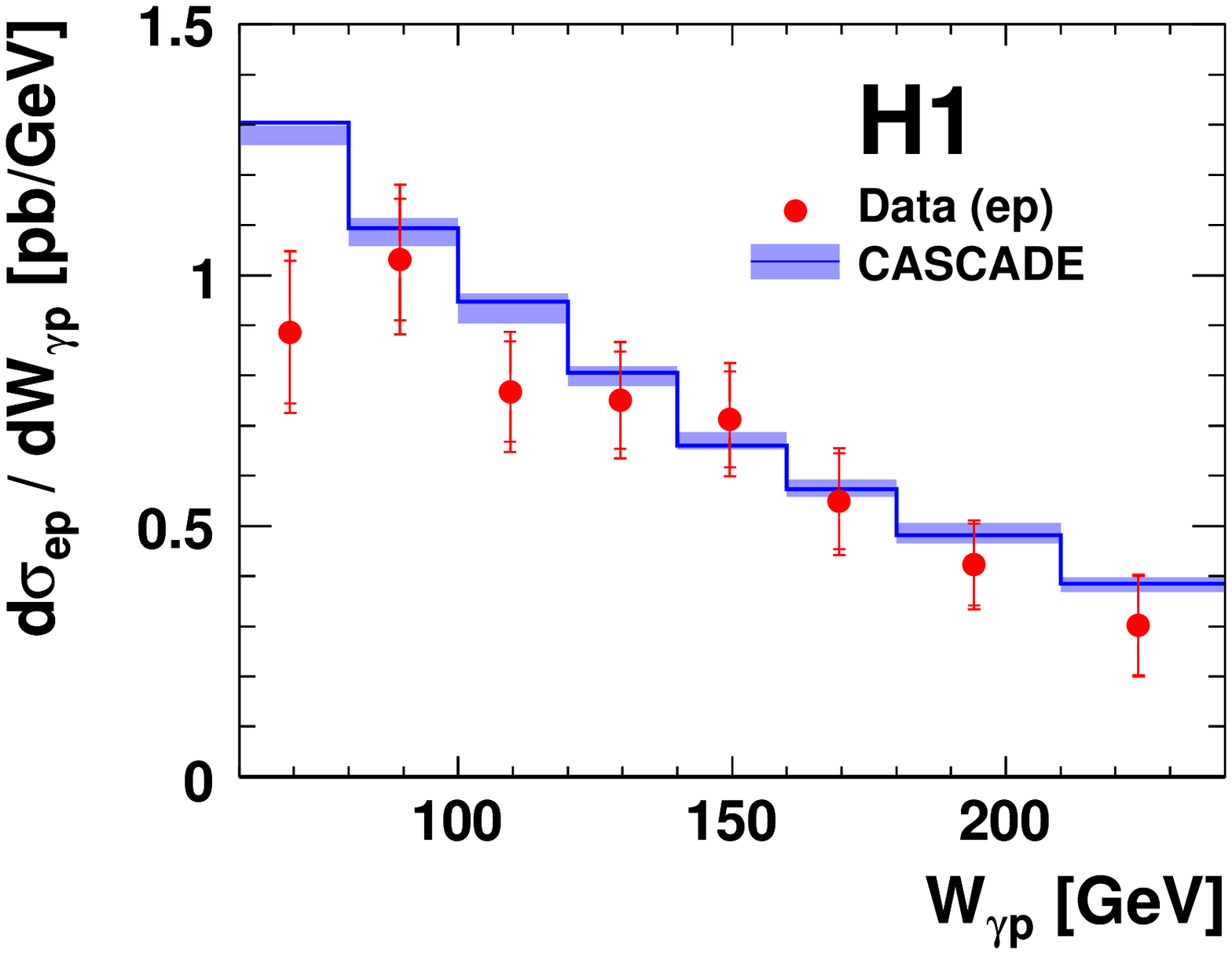}}%fin_xs_Wgp_Cascade.eps}} 
\put(8.5,0.){\includegraphics*[width=7.5cm]{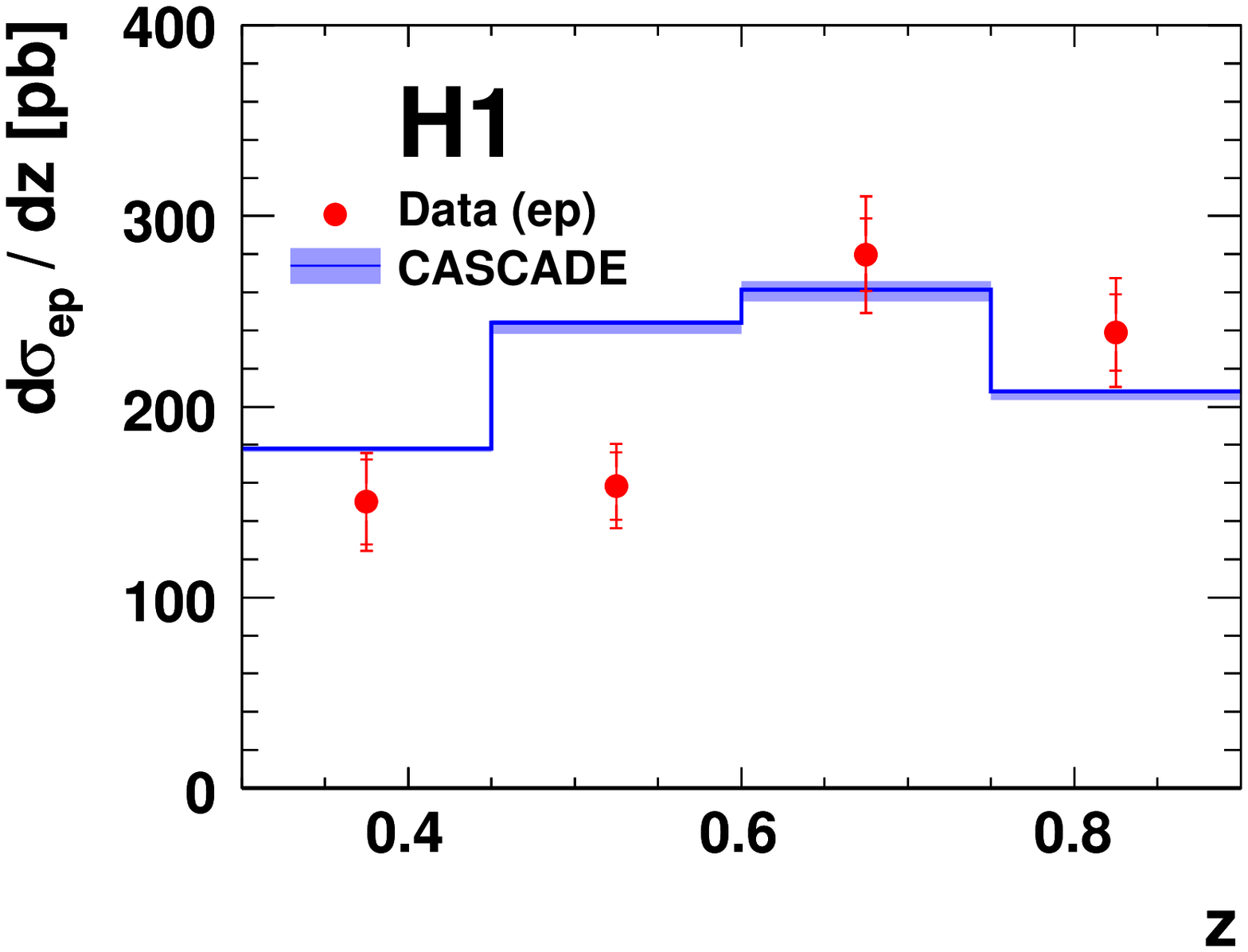}}%fin_xs_Z_Cascade.eps}} 
\put(4.55,13){\large \sf Inelastic \JPsi Electroproduction}
\put(1.3,12.3){a)}
\put(9.8,12.3){b)}
\put(1.3,5.8){c)}
\put(9.8,5.8){d)}
\end{picture}
\caption{Differential \JPsi meson cross sections for the kinematic range 
$3.6 < \Qsquared < \unit[100]{GeV^2}$, $60 < \Wgp < \unit[240]{GeV}$, $0.3 < \ZJPsi < 0.9$ and 
$\PtStarJPsiSquare > \unit[1]{GeV}$, as functions of 
a) the photon virtuality \Qsquared, 
b) the squared transverse momentum of the \JPsi meson in the photon proton rest 
frame \PtStarJPsiSquare, 
c) the energy in the photon proton rest frame \Wgp and
d) the elasticity \ZJPsi.
The inner error bar represents the statistical uncertainty and the outer error bar indicates the statistical and systematic uncertainties added in quadrature.
The data are compared to the predictions from \Cascade (solid line).
The uncertainty band of the \Cascade prediction arises from a scale variation by a factor of two.
}
\label{fig:dis:xsecs2}
\end{figure}

\begin{figure}[p]
\centering
\unitlength1cm
\begin{picture}(16,13.)
\put(3.8,6.5){\includegraphics*[width=7.5cm]{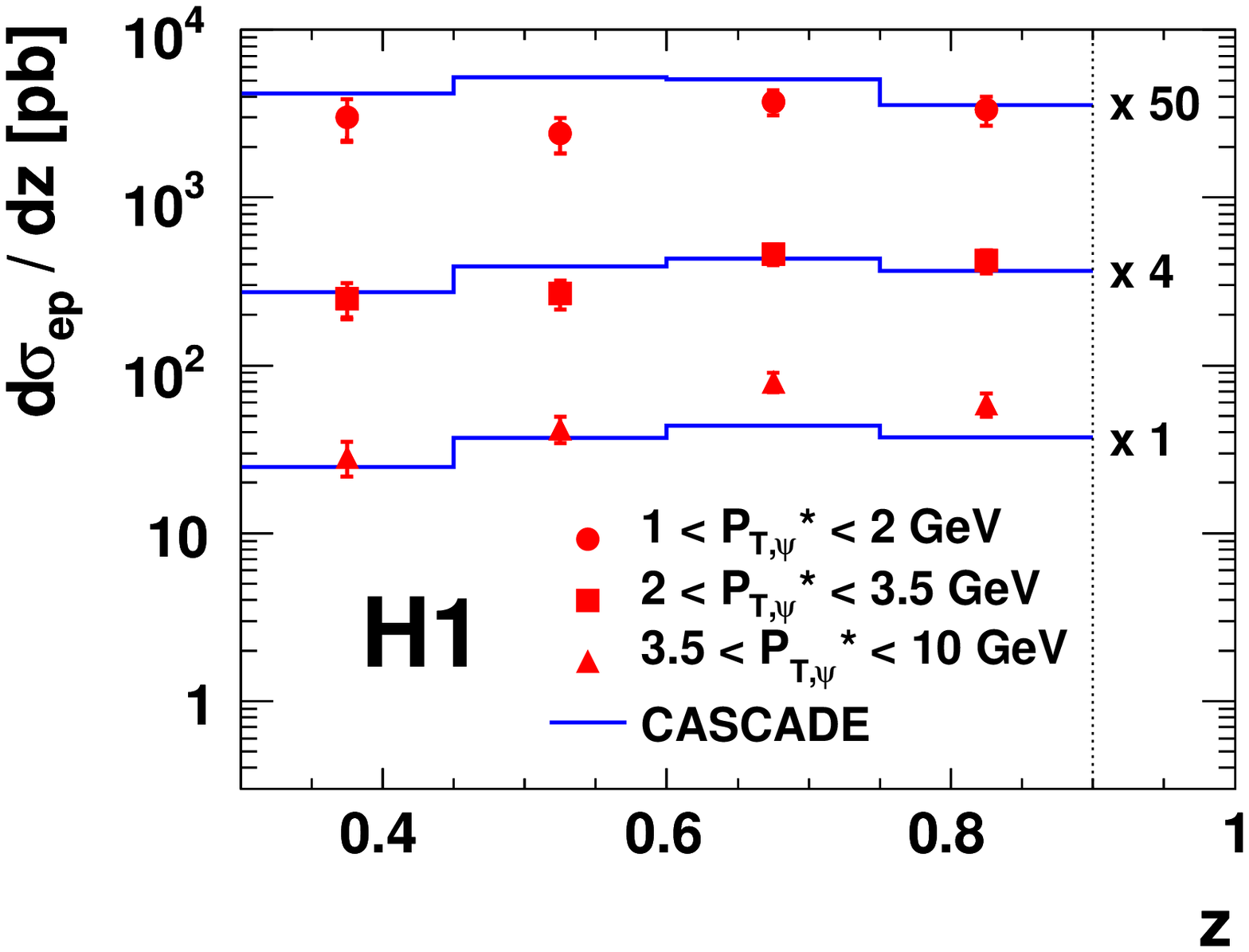}}%fin_xs_Z_PtStarBINS_Cascade.eps}}
\put(3.8,0.){\includegraphics*[width=7.5cm]{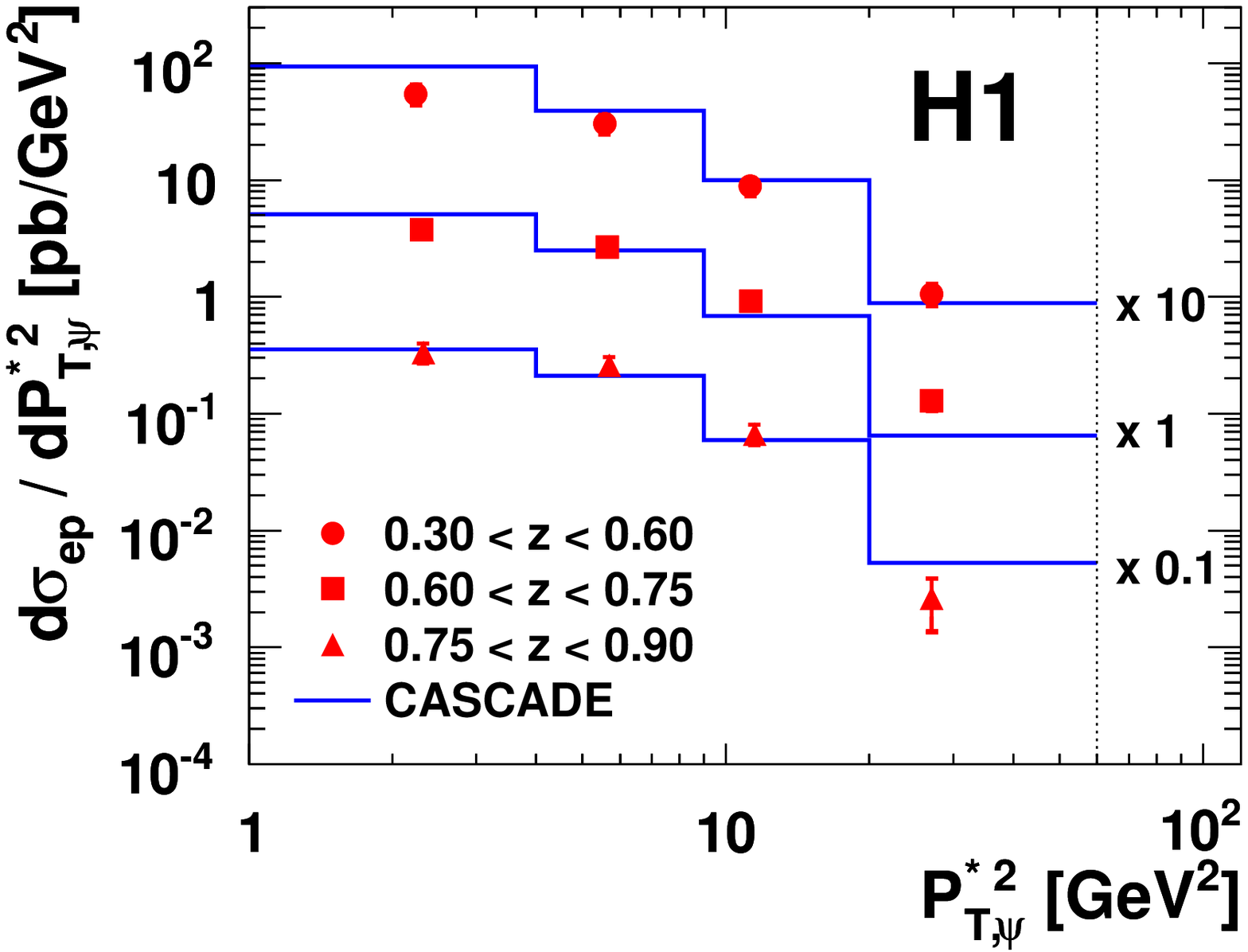}}%fin_xs_PtStar2_ZBINS_Cascade.eps}}
\put(5.15,12.3){a)}
\put(5.15,5.8){b)}
\put(4.55,13.0){\large \sf Inelastic \JPsi Electroproduction}
\end{picture}
\caption{a) Differential \JPsi meson cross sections as a function of \ZJPsi in three bins of \PtStarJPsi
and b) differential cross section as a function of \PtStarJPsiSquare in three bins of \ZJPsi.
For visibility, the measured cross sections are scaled by the factors indicated in the figures. 
Predictions from \Cascade are shown as solid line.}
\label{fig:res:xsecsptz} 
\end{figure}

Results for electroproduction are shown in figure~\ref{fig:dis:xsecs2} and figure~\ref{fig:res:xsecsptz}.
Differential $ep$ cross sections are measured as functions of the photon virtuality \Qsquared, the squared transverse momentum of the \JPsi meson in the photon proton rest frame \PtStarJPsiSquare, the energy \Wgp and the elasticity \ZJPsi. 
Figure~\ref{fig:res:xsecsptz} shows differential cross sections as a function of the elasticity \ZJPsi in bins of \PtStarJPsi and as a function of \PtStarJPsiSquare in bins of \ZJPsi.
A comparison of the electroproduction data with predictions from the Monte 
Carlo generator \Cascade reveals in general a reasonable agreement with the data. 
Differences in shape can be seen in the differential cross section as a function of \PtStarJPsiSquare.

\begin{figure}[p]
\centering
\unitlength1cm
\begin{picture}(16,13.)
\put(0.,6.5){\includegraphics*[width=7.5cm]{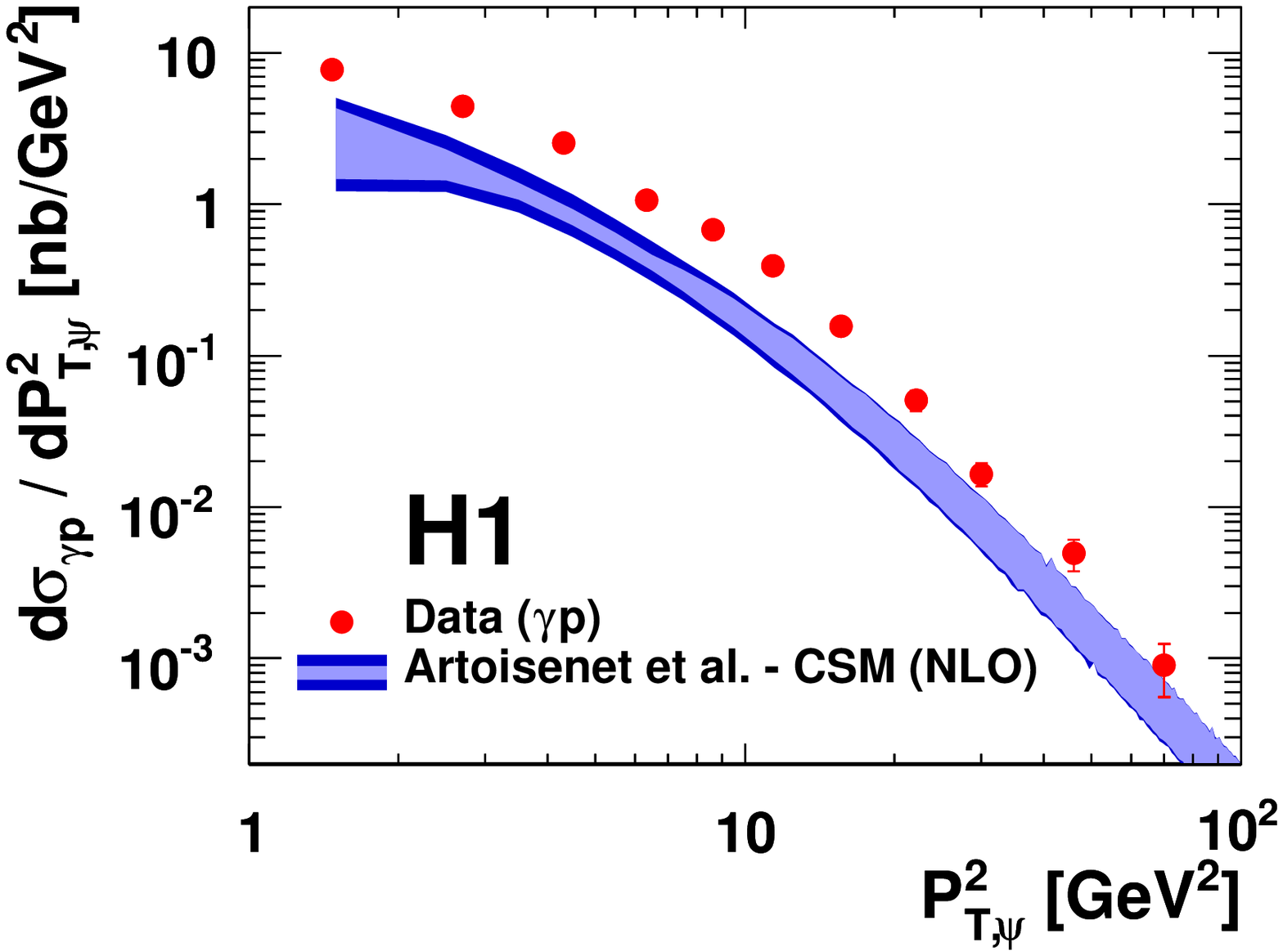}}%fin_xs_Pt2JPsi_CSMNLO.eps}}
\put(8.5,6.5){\includegraphics*[width=7.5cm]{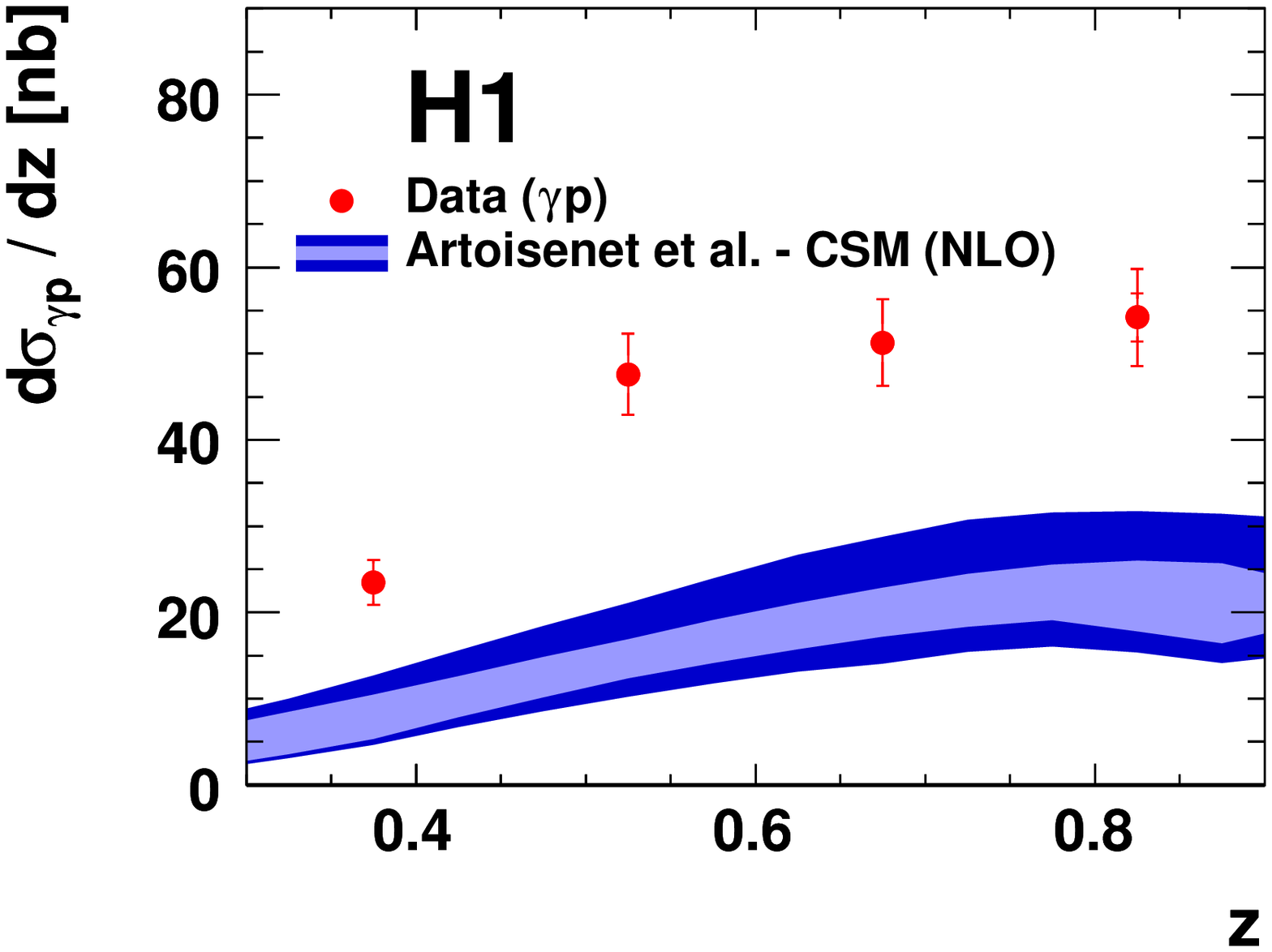}}%fin_xs_Z_CSMNLO.eps}}
\put(0.,0.){\includegraphics*[width=7.5cm]{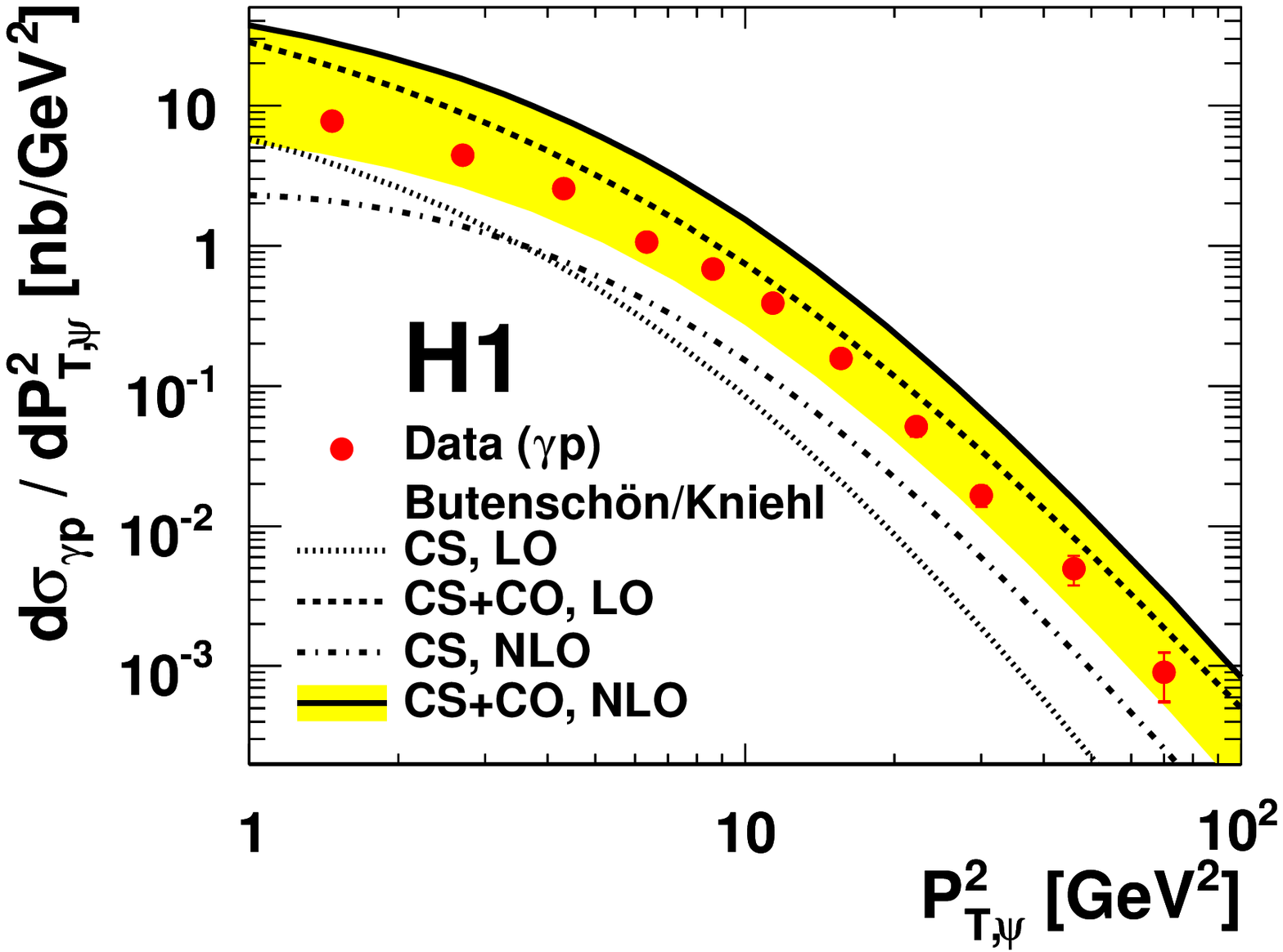}}%fin_xs_Pt2JPsi_NRQCD.eps}}
\put(8.5,0.){\includegraphics*[width=7.5cm]{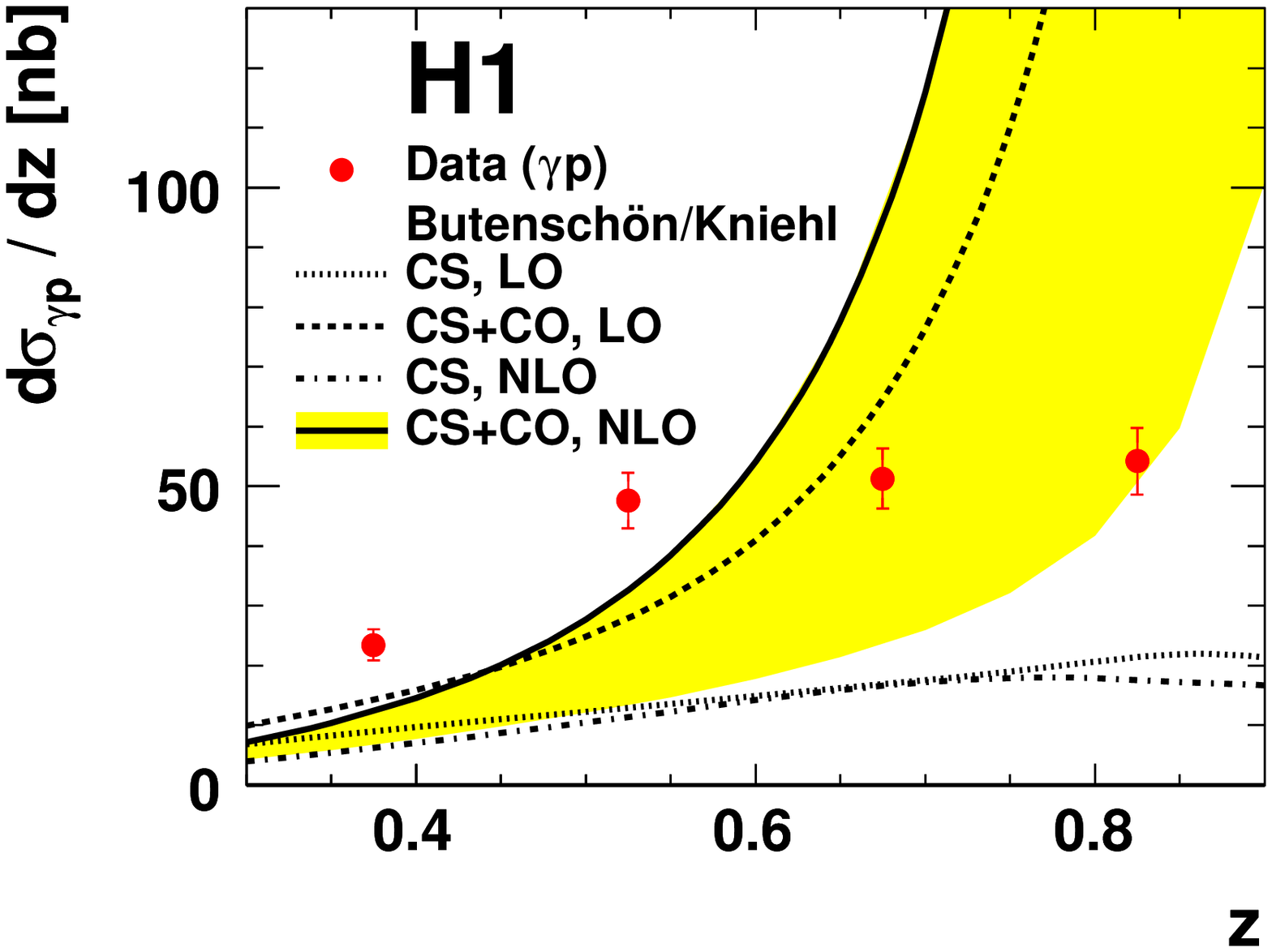}}%fin_xs_Z_NRQCD.eps}}
\put(1.2,12.3){a)}
\put(9.8,12.3){b)}
\put(1.2,5.8){c)}
\put(9.8,5.8){d)}
\put(4.7,13.){\large \sf Inelastic \JPsi Photoproduction}
\end{picture}
\caption{Differential \JPsi meson photoproduction cross sections for the kinematic range
$60 < \Wgp < \unit[240]{GeV}$, $0.3 < \ZJPsi < 0.9$ and $\PtJPsi > \unit[1]{GeV}$
as functions of the squared transverse momentum of the \JPsi meson \PtJPsiSquare (a) and c)) and the elasticity \ZJPsi (b) and d)).
The inner error bar represents the statistical uncertainty and the outer error bar indicates the statistical and systematic uncertainties added in quadrature.
The data are compared with calculations to next-to-leading order:
a,b) a colour singlet model (CSM) calculation~\cite{Maltoni} and 
c,d) a NRQCD calculation including contributions from colour octet states (CS + CO)~\cite{Kniehl}. 
The colour singlet component (CS) of the latter calculation is shown separately in addition.}
\label{fig:res:gp:xsecs:1d:CSMNLO}
\end{figure}

For photoproduction, several theory calculations to next-to-leading order have 
been performed and are compared with the data in figure~\ref{fig:res:gp:xsecs:1d:CSMNLO}.
A calculation in the CSM at NLO~\cite{Kraemer}
	was repeated using up-to-date sets of scale parameters~\cite{Maltoni, Kniehl}, 
yielding predictions as shown in figure~\ref{fig:res:gp:xsecs:1d:CSMNLO}a)-b). 
The shapes of the data are reasonably described, whereas the normalisation of the prediction 
is about a factor three below the data, with large uncertainties, 
indicating that corrections beyond next-to-leading order are necessary in order to describe the data.
Estimates of the NNLO contribution for charmonium production at the Tevatron~\cite{maltoni1,maltoni2} indicate that these contributions can be large indeed. 

The calculation to next-to-leading order has been extended to include 
colour octet contributions resulting in a larger cross section~\cite{Kniehl}.
A comparison of this prediction with the data is shown in figure~\ref{fig:res:gp:xsecs:1d:CSMNLO}c)-d).
The dominant uncertainty arises from the difference in the predicted cross section when using LO colour octet LDMEs or higher order improved LDMEs~\cite{Kniehl}.
The NRQCD prediction fails however in describing the shape of the differential cross section as a function of the elasticity \ZJPsi,
even within the presently large uncertainties of the calculation.

%%%%%%%%%%%%%%%%%%%%%%%%%%%%%%%%%%%%%%%%%%%%%%%%%%%%%%%%%%%%
\section{Polarisation Measurement}
\label{sec:angles}

The measurement of the \JPsi meson helicity distributions provides an independent method to distinguish between different production mechanisms.
The measurement is performed for the photoproduction data sample.
The \JPsi meson polarisation is measured by analysing the decay angle
distributions of the \JPsi meson, and their dependence on \PtJPsi and \ZJPsi, in
two complementary 
frames~\cite{polarization}: the helicity frame and the Collins-Soper frame.
In the helicity frame the polarisation axis $z$ in the \JPsi meson rest frame 
is defined by the flight direction of the \JPsi meson in the $\gamma p$ 
rest frame, whereas the polarisation in the Collins-Soper frame is 
measured with respect to the bisector of proton ($-\vec{p_p}$) and 
photon ($\vec{p_\gamma}$) in the \JPsi meson rest frame \cite{beneke}.
Subsequently, the frame-dependent polarisation axis is taken as $z$
axis of a right handed coordinate system, where the $x$ and $z$ axis 
lie in a plane spanned by the photon and proton directions.
The $y$ axis is perpendicular to this plane and is the same in both reference frames.
The polar ($\Theta^*$) and azimuthal ($\phi^*$) angles of the positive decay muons are used.

The parametrisation of the measured decay angle
distributions as function of \CosThetaStar and \PhiStar is given by~\cite{beneke}:
\be
\frac{\mr{d}\sigma}{\mr{d}\cos{\Theta^*}}&\propto&1+\alpha\cos^2{\Theta^*}\, ;
\label{eq:theta}\\[.3em]
\frac{\mr{d}\sigma}{\mr{d}\phi^*}&\propto&1+\frac{\alpha}{3}+\frac{\nu}{3}\,\cos{2\phi^*}.
\label{eq:phi}
\ee
The polarisation variables $\alpha$ and $\nu$ can be related to elements of the spin density matrix for the \JPsi meson. Moreover, $\alpha = +1$ and $-1$  corresponds to fully transverse and longitudinal polarisation of the \JPsi meson, respectively. 

\begin{figure}[p]
\centering
\unitlength1cm
\begin{picture}(16,13)
\put(0. , 6.5){\includegraphics*[width=7.5cm]{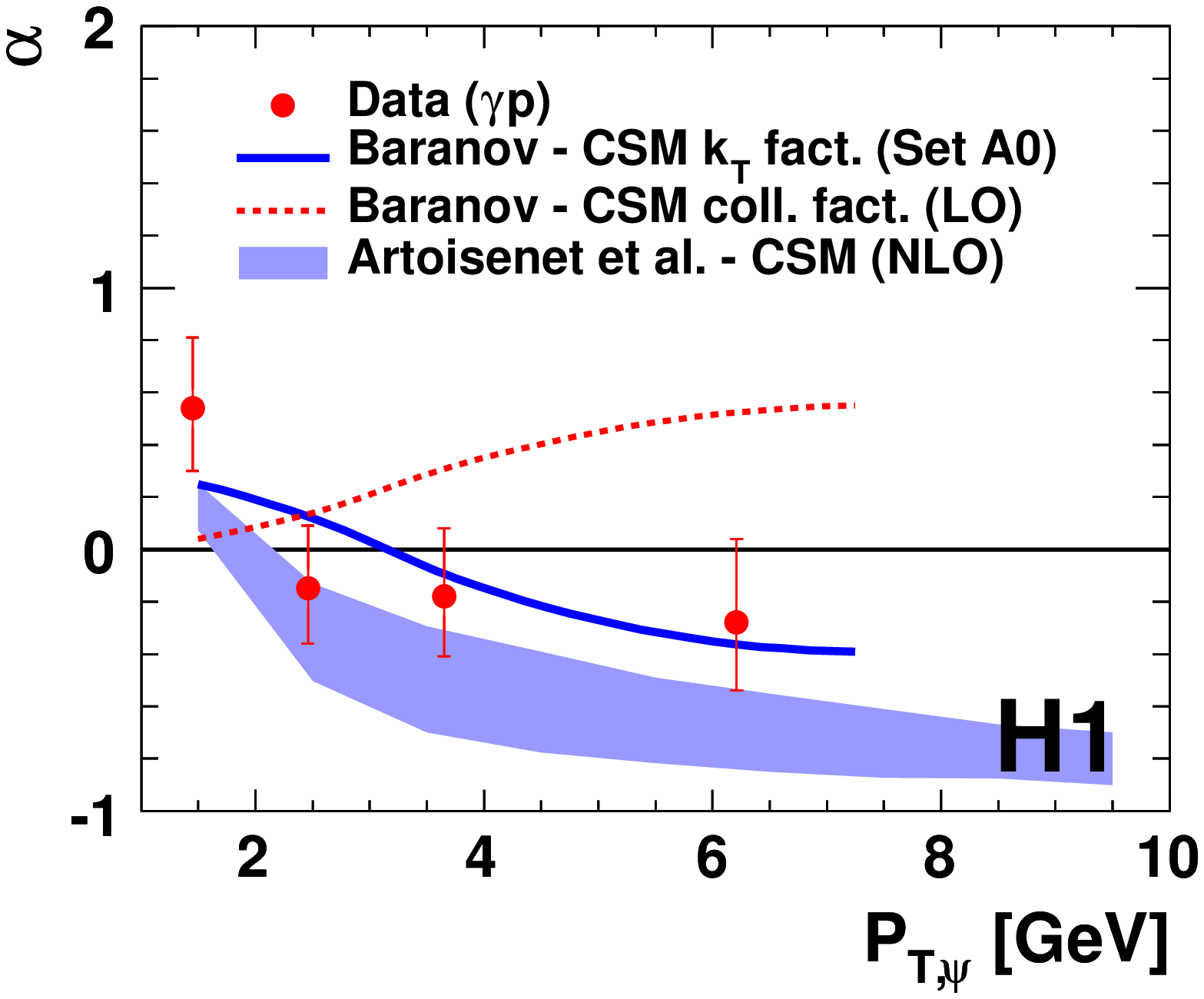}}%fin_Polarisation_Alpha_Pt_kT_CSMNLO.eps}}
\put(8.5, 6.5){\includegraphics*[width=7.5cm]{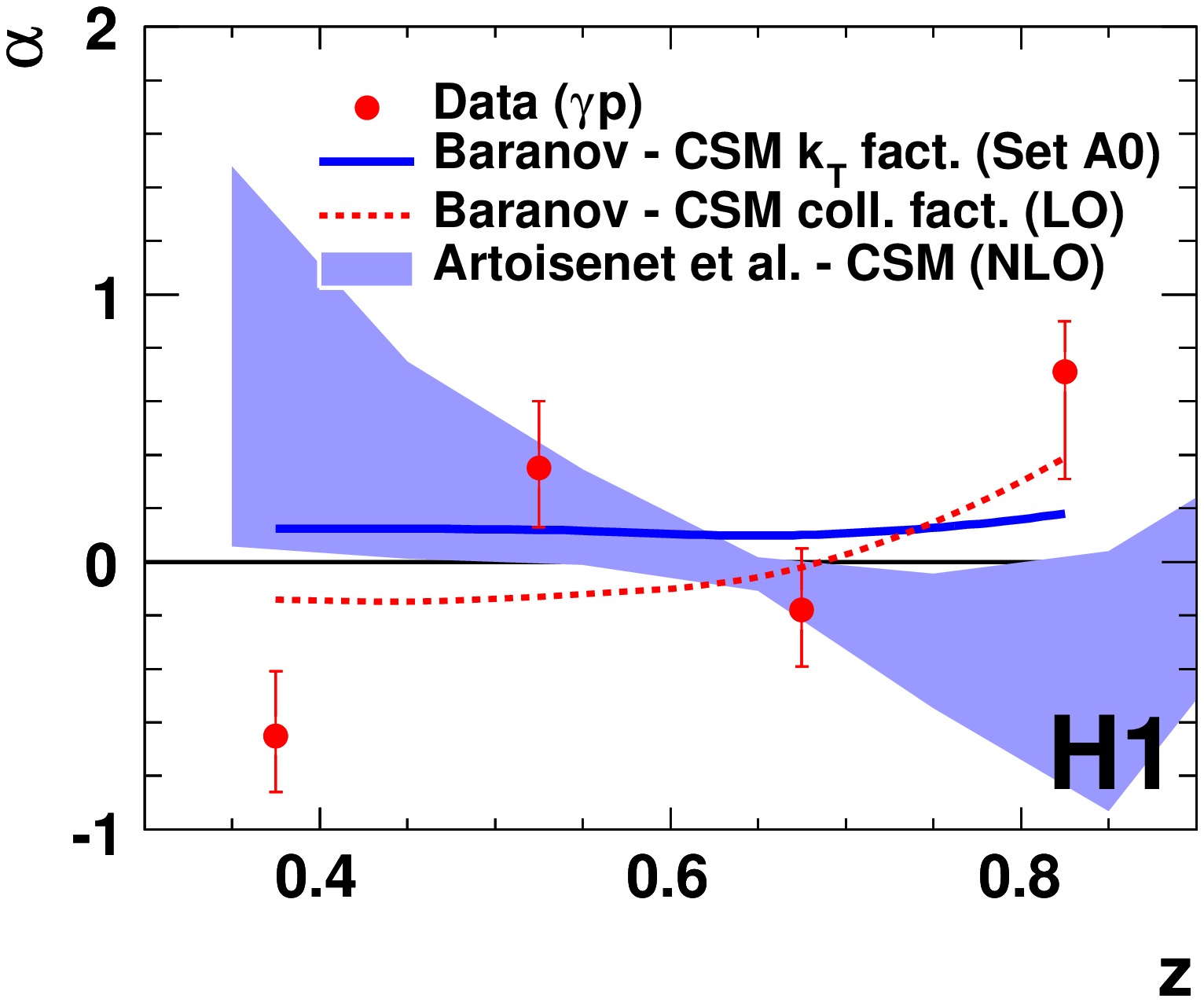}}%fin_Polarisation_Alpha_Z_kT_CSMNLO.eps}}
\put(0. , 0.){\includegraphics*[width=7.5cm]{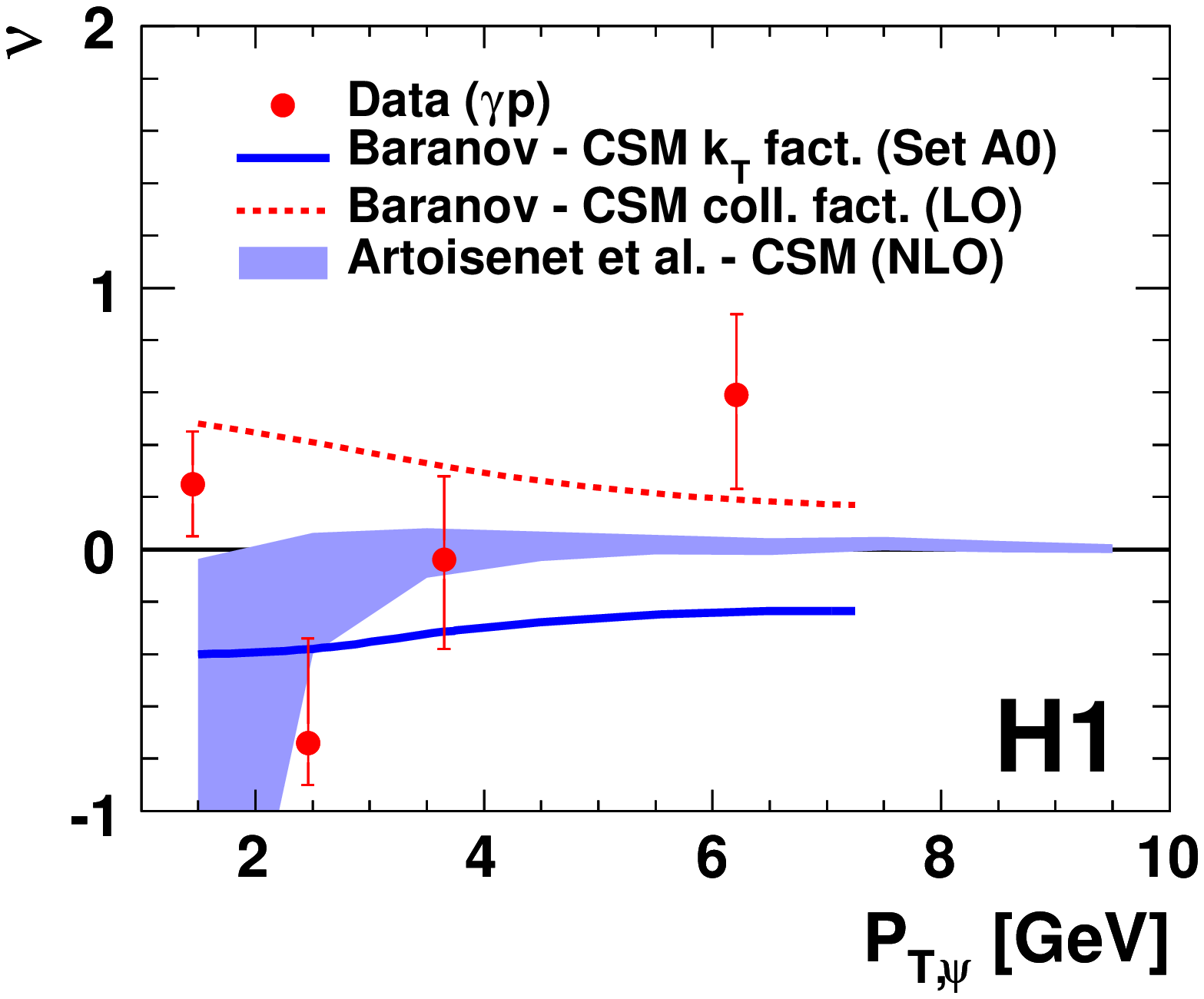}}%fin_Polarisation_Nu_Pt_kT_CSMNLO.eps}}
\put(8.5, 0.){\includegraphics*[width=7.5cm]{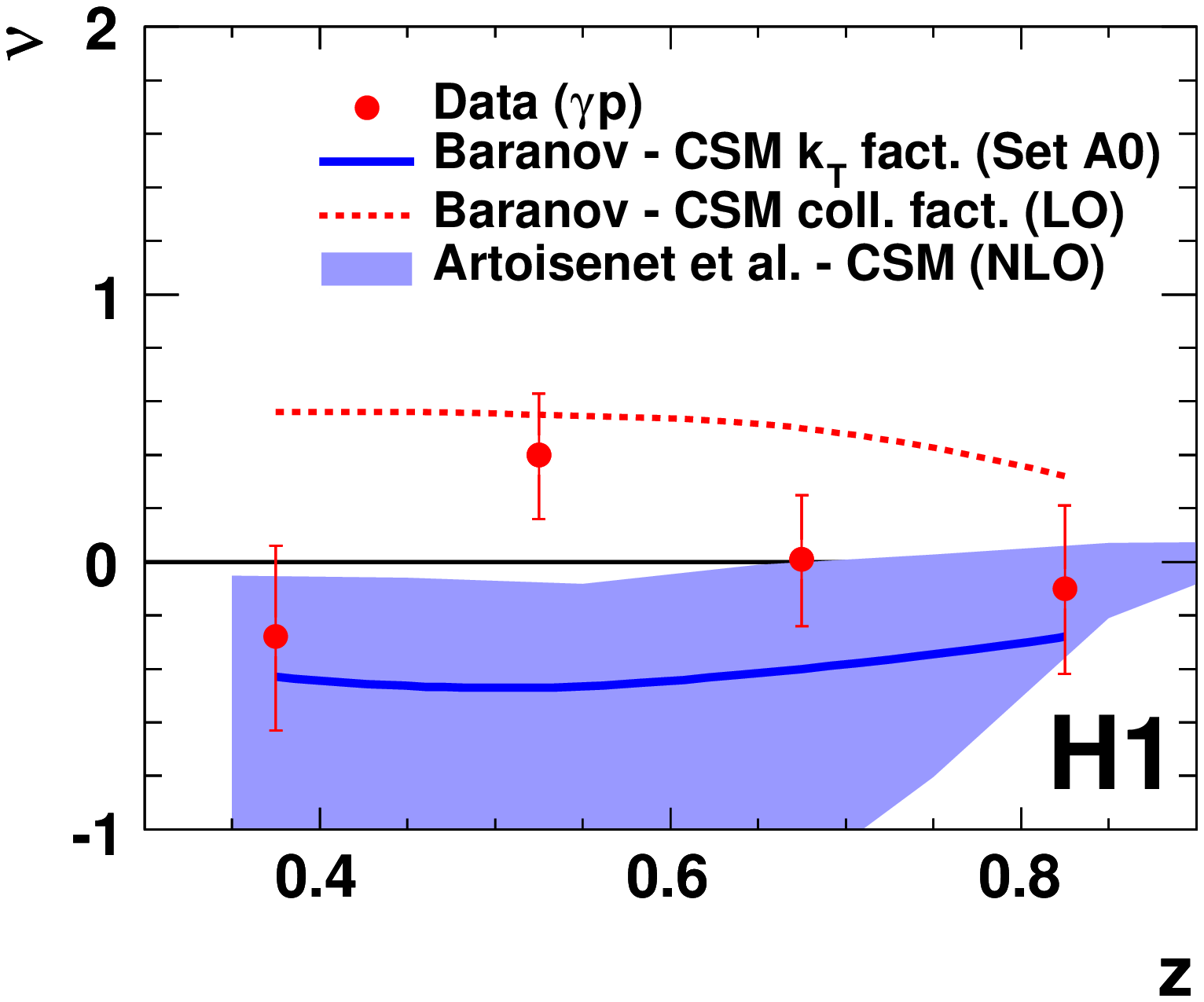}}%fin_Polarisation_Nu_Z_kT_CSMNLO.eps}}

\put(5.2,14){\large \sf Inelastic \JPsi Polarisation }
\put(6.425,13.4){\large \sf Helicity Frame}

\put(1.1,12.8){a)}
\put(9.6,12.8){b)}
\put(1.1,6.3){c)}
\put(9.6,6.3){d)}
\end{picture}
\caption{Polarisation parameters $\alpha$ and $\nu$ measured in the helicity frame 
for the kinematic range $60 < \Wgp < \unit[240]{GeV}$, $0.3 < \ZJPsi < 0.9$ and $\PtJPsi > \unit[1]{GeV}$, 
as a function of \ZJPsi and \PtJPsi. The measurement is compared with predictions calculated 
in a \kT factorisation ansatz ~\cite{Baranov} 
and with calculations in CSM (collinear 
factorisation) at leading~\cite{Baranov} and next-to-leading order~\cite{Maltoni}.}
\label{fig:gp:Polarisation}
\end{figure}

\begin{figure}[p]
\centering
\unitlength1cm
\begin{picture}(16,13)
\put(0. , 6.5){\includegraphics*[width=7.5cm]{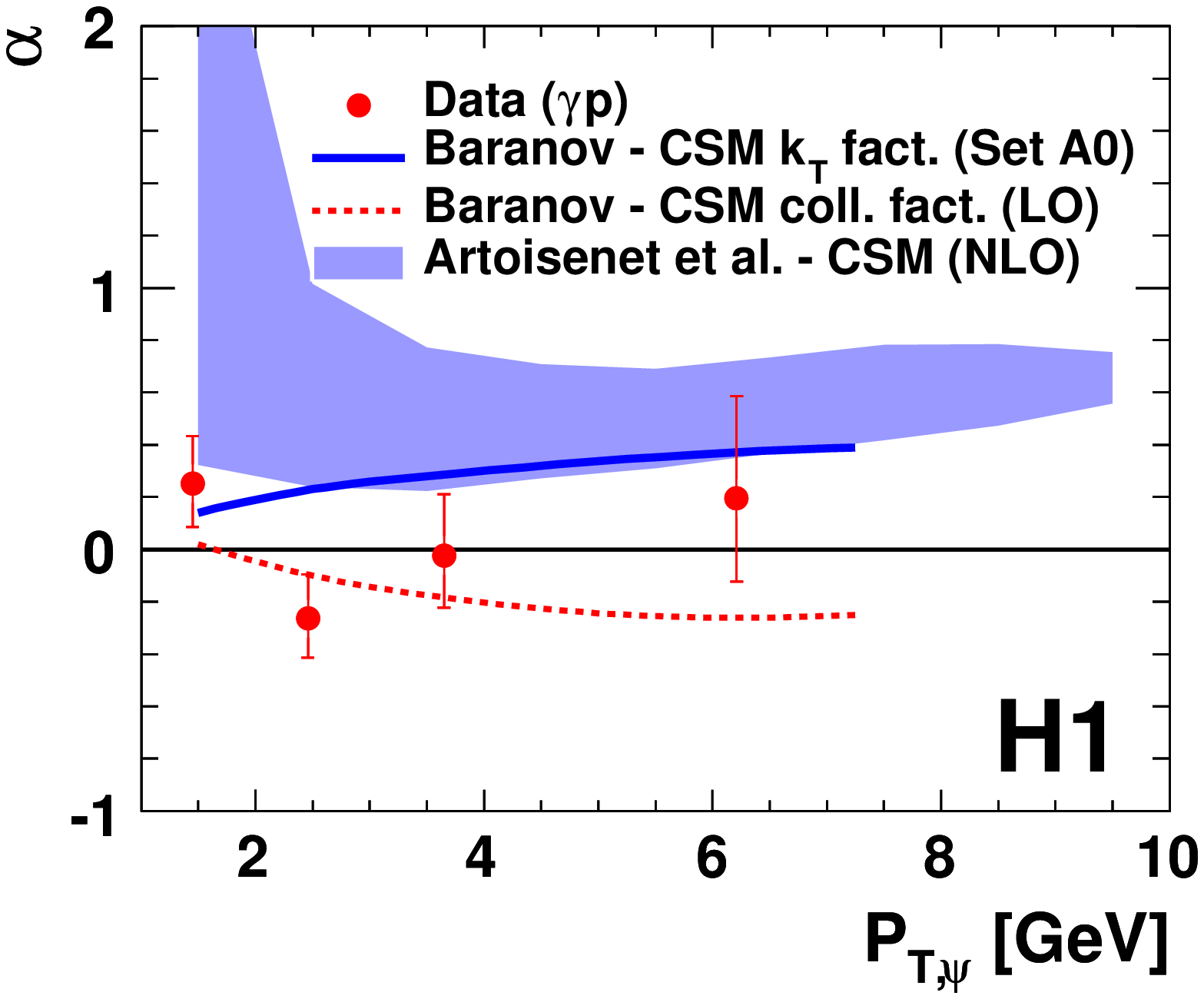}}%fin_Polarisation_Alpha_Pt_kT_CSMNLO_CS.eps}}
\put(8.5, 6.5){\includegraphics*[width=7.5cm]{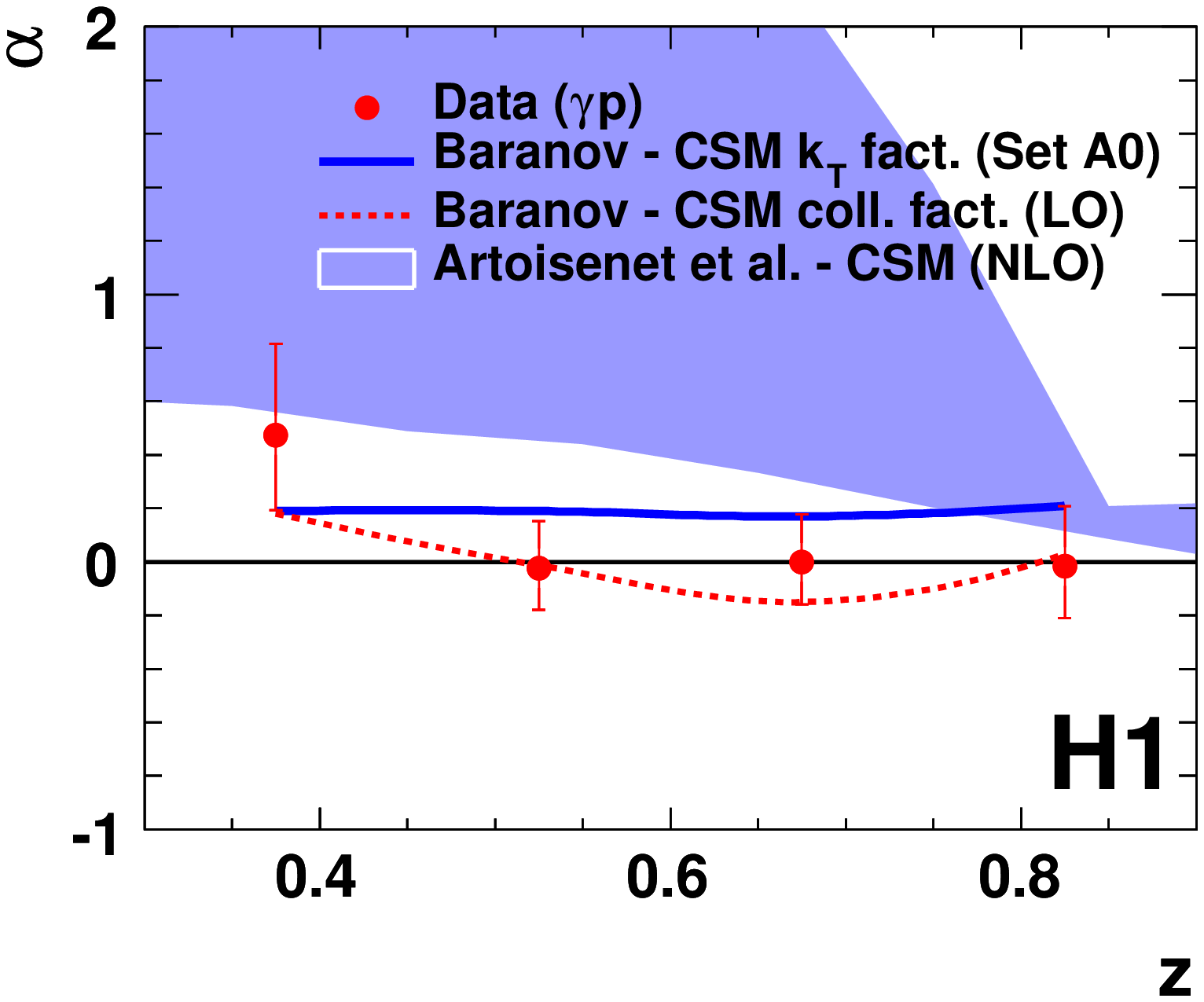}}%fin_Polarisation_Alpha_Z_kT_CSMNLO_CS.eps}}
\put(0. , 0.){\includegraphics*[width=7.5cm]{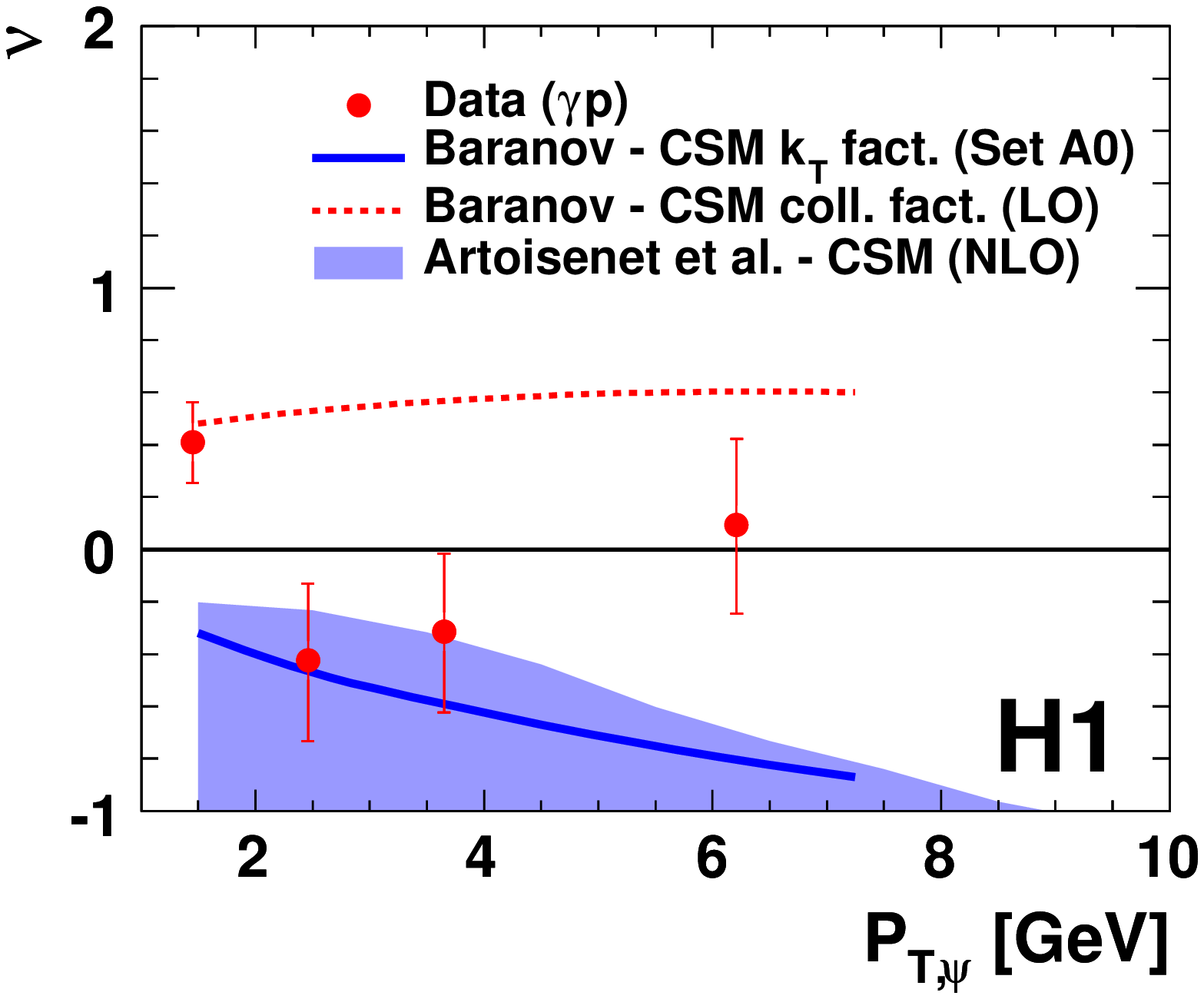}}%fin_Polarisation_Nu_Pt_kT_CSMNLO_CS.eps}}
\put(8.5, 0.){\includegraphics*[width=7.5cm]{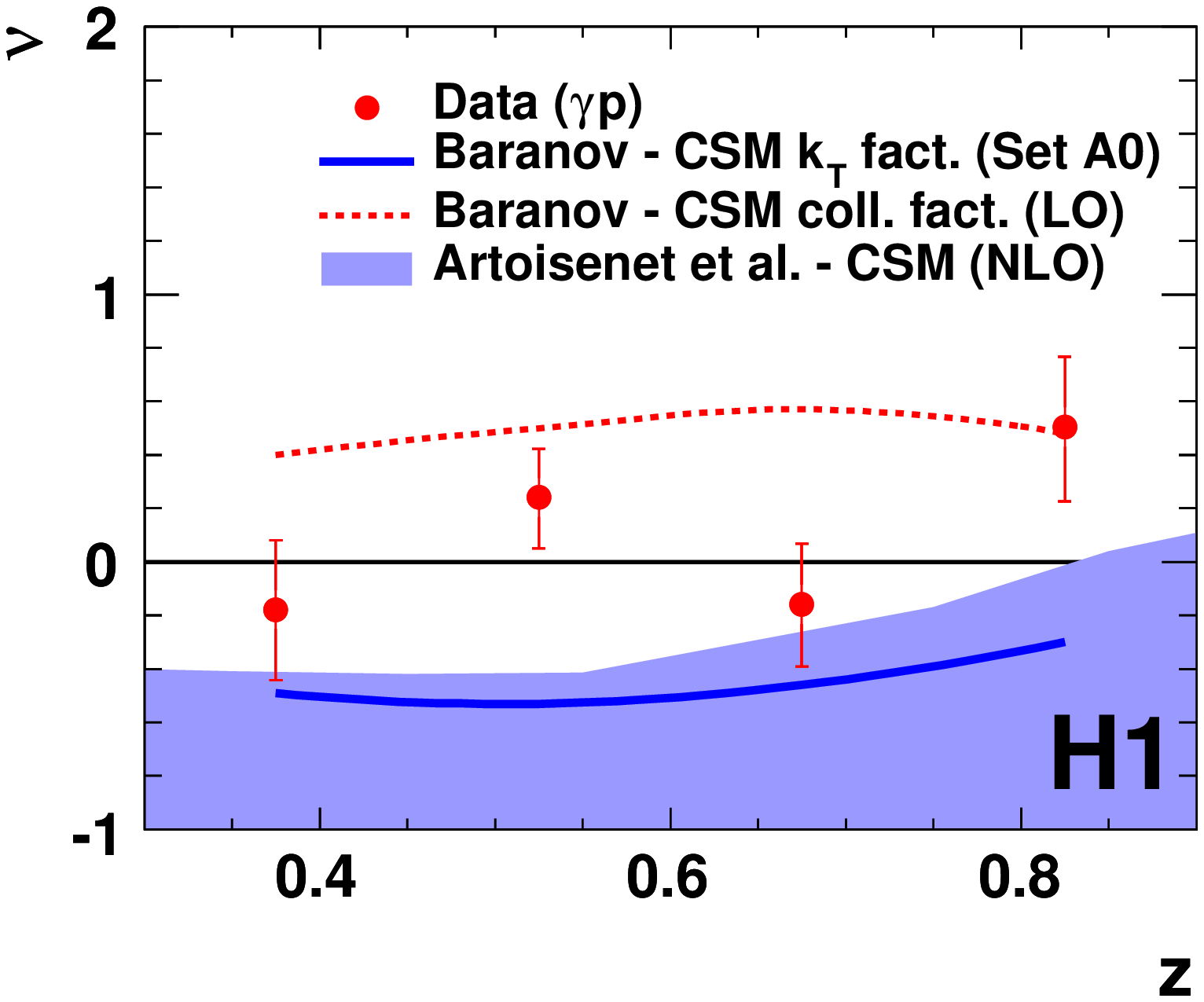}}%fin_Polarisation_Nu_Z_kT_CSMNLO_CS.eps}}

\put(5.2,14){\large \sf Inelastic \JPsi Polarisation }
\put(5.7,13.4){\large \sf Collins-Soper Frame}

\put(1.1,12.8){a)}
\put(9.6,12.8){b)}
\put(1.1,6.3){c)}
\put(9.6,6.3){d)}
\end{picture}
\caption{Polarisation parameters $\alpha$ and $\nu$ in the Collins-Soper frame 
for the kinematic range $60 < \Wgp < \unit[240]{GeV}$, $0.3 < \ZJPsi < 0.9$ and $\PtJPsi > \unit[1]{GeV}$, 
as a function of 
\ZJPsi and \PtJPsi. The measurement is compared with predictions calculated in a \kT factorisation 
ansatz~\cite{Baranov} 
and with calculations in CSM (collinear factorisation) at leading~\cite{Baranov} and 
next-to-leading order~\cite{Maltoni}.}
\label{fig:gp:Polarisation:CS}
\end{figure}

A $\chi^2$ fit is performed in each bin of the polarisation measurement, 
comparing data to Monte Carlo samples on reconstruction level probing values for $\alpha$ and $\nu$ between $-1$ and $+1$.
Systematic uncertainties on this measurement are negligible compared to rather large statistical uncertainties.
The results for $\alpha$ and $\nu$ as a function of \PtJPsi and \ZJPsi are presented for the helicity frame  in figure~\ref{fig:gp:Polarisation} and in figure~\ref{fig:gp:Polarisation:CS} for the Collins-Soper frame.
The values for the polarisation parameters in both frames are listed in
table~\ref{tab:pol:heli}.

Within uncertainties the \JPsi mesons produced inelastically at HERA are unpolarised.
The measurements are compared to predictions using 
a \kT factorisation ansatz~\cite{Baranov} and to calculations in the CSM in collinear 
factorisation at leading order~\cite{Baranov} and next-to-leading order~\cite{Maltoni}.
The predictions in the \kT factorisation ansatz describe the data. 
The NLO calculations show a similar
trend within large uncertainties.
In contrast, the leading order CSM calculation predicts larger values for the polarisation variables than the measured ones for many bins and is disfavoured by the measurement.
A similar measurement was published by the ZEUS collaboration in a different kinematic range~\cite{ZEUS_Pol}.

%%%%%%%%%%%%%%%%%%%%%%%%%%%%%%%%%%%%%%%%%%%%%%%%%%%%%%%%%%%%
\section{Conclusions}
\label{sec:conclusions}
A measurement of inelastic \JPsi meson production is performed. 
Differential cross sections with improved statistical and systematic uncertainties are presented for both electroproduction
and photoproduction. Polarisation parameters for the photoproduction of \JPsi mesons are measured in
two different reference frames, the helicity frame and the Collins-Soper frame.  

The data are compared to a number of recent theory predictions.
It is found that predictions based on \kT factorisation in the colour singlet model are able to describe the cross sections and the helicity distributions well. 
Calculations based on collinear factorisation in the colour singlet model at next-to-leading order
produce a reasonable description of the shape of the measured cross sections, but are lower in normalisation.
They give an acceptable description of the polarisation parameter measurements
within the large uncertainties.
The failure to describe the cross section measurements and the strong sensitivity to scale variations indicate that calculations beyond next-to-leading order are necessary.
Moreover contributions from colour octet states may be significant. 

%%%%%%%%%%%%%%%%%%%%%%%%%%%%%%%%%%%%%%%%%%%%%%%%%%%%%%%%%%%%
\section*{Acknowledgements}

We are grateful to the HERA machine group whose outstanding
efforts have made this experiment possible.
We thank the engineers and technicians for their work in constructing and
maintaining the H1 detector, our funding agencies for
financial support, the DESY technical staff for continual assistance
and the DESY directorate for support and for the
hospitality which they extend to the non-DESY
members of the collaboration.
We would like to thank 
Pierre Artoisenet, Sergey Baranov, Mathias Butensch\"on, Bernd Kniehl, Michael Kr\"amer and Fabio Maltoni 
for providing theory calculations for this paper as well as for helpful discussions.

\clearpage

%%%%%%%%%%%%%%%%%%%%%%%%%%%%%%%%%%%%%%%%%%%%%%%%%%%%%%%%%%%%

%%%%%%%%%%%%%%%%%%%%%%%%%%%%%%%%%%%%%%%%%%%%%%%%%%%%%%%%%%%

%%%%%%%%%%%%%%%%%%%%%%%%%%%%%%%%%%%%%%%%%%%%%%%%%%%%%%%%%%%

%%%%%%%%%%%%%%%%%%%%%%%%%%%%%%%%%%%%%%%%%%%%%%%%%%%%%%%%%%%

%%%%%%%%%%%%%%%%%%%%%%%%%%%%%%%%%%%%%%%%%%%%%%%%%%%%%%%%%%%

%%%%%%%%%%%%%%%%%%%%%%%%%%%%%%%%%%%%%%%%%%%%%%%%%%%%%%%%%%%%

%%%%%%%%%%%%%%%%%%%%%%%%%%%%%%%%%%%%%%%%%%%%%%%%%%%%%%%%%%%

%%%%%%%%%%%%%%%%%%%%%%%%%%%%%%%%%%%%%%%%%%%%%%%%%%%%%%%%%%%

%%%%%%%%%%%%%%%%%%%%%%%%%%%%%%%%%%%%%%%%%%%%%%%%%%%%%%%%%%%

%%%%%%%%%%%%%%%%%%%%%%%%%%%%%%%%%%%%%%%%%%%%%%%%%%%%%%%%%%%
\begin{figure}[p]
\centering
\unitlength1cm
\begin{picture}(16,13)
\put(3.8 , 6.5){\includegraphics*[width=7.5cm]{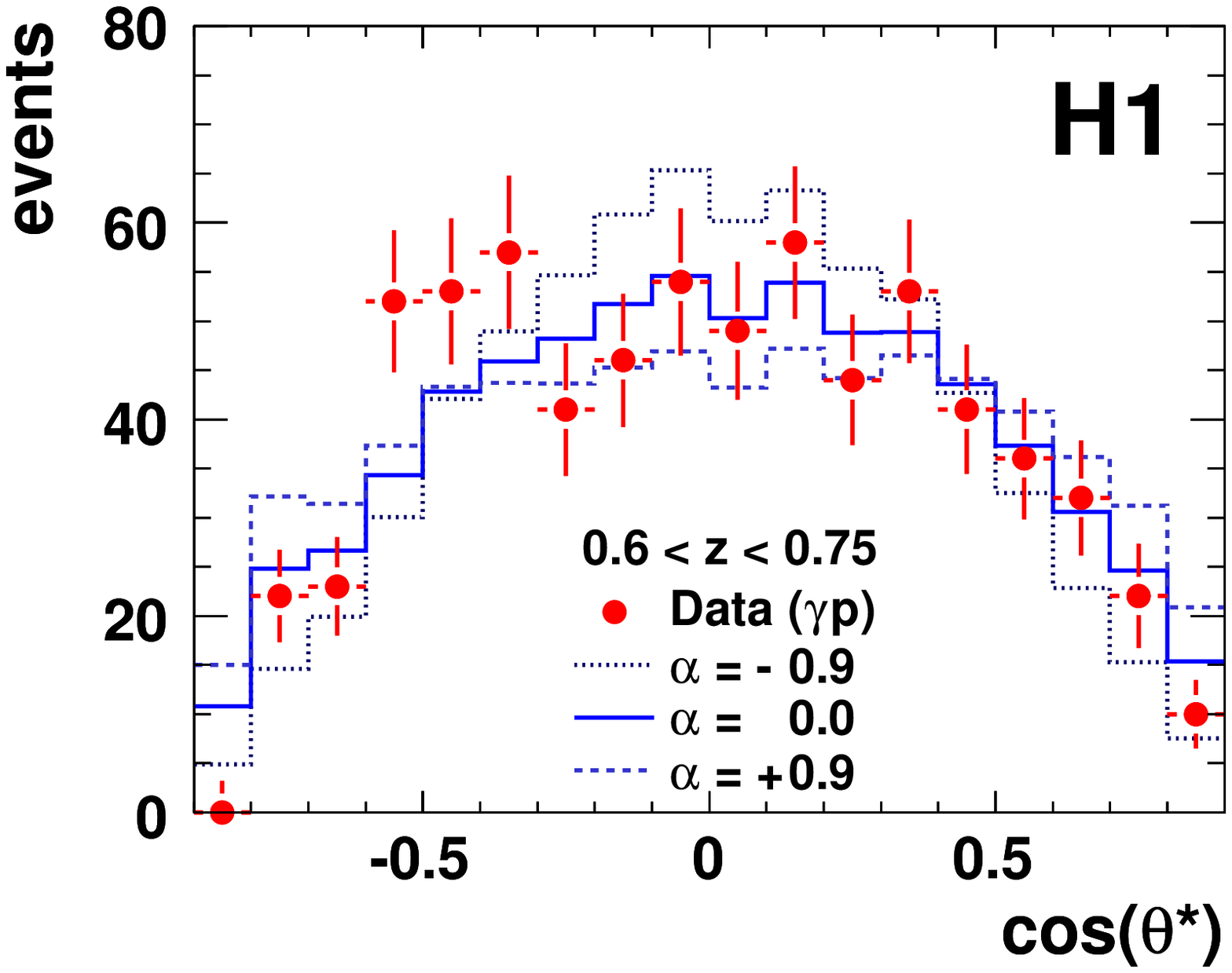}}%fin_Polarisation_con_CosTheta_Z_3.eps}}
\put(3.8 , 0.){\includegraphics*[width=7.5cm]{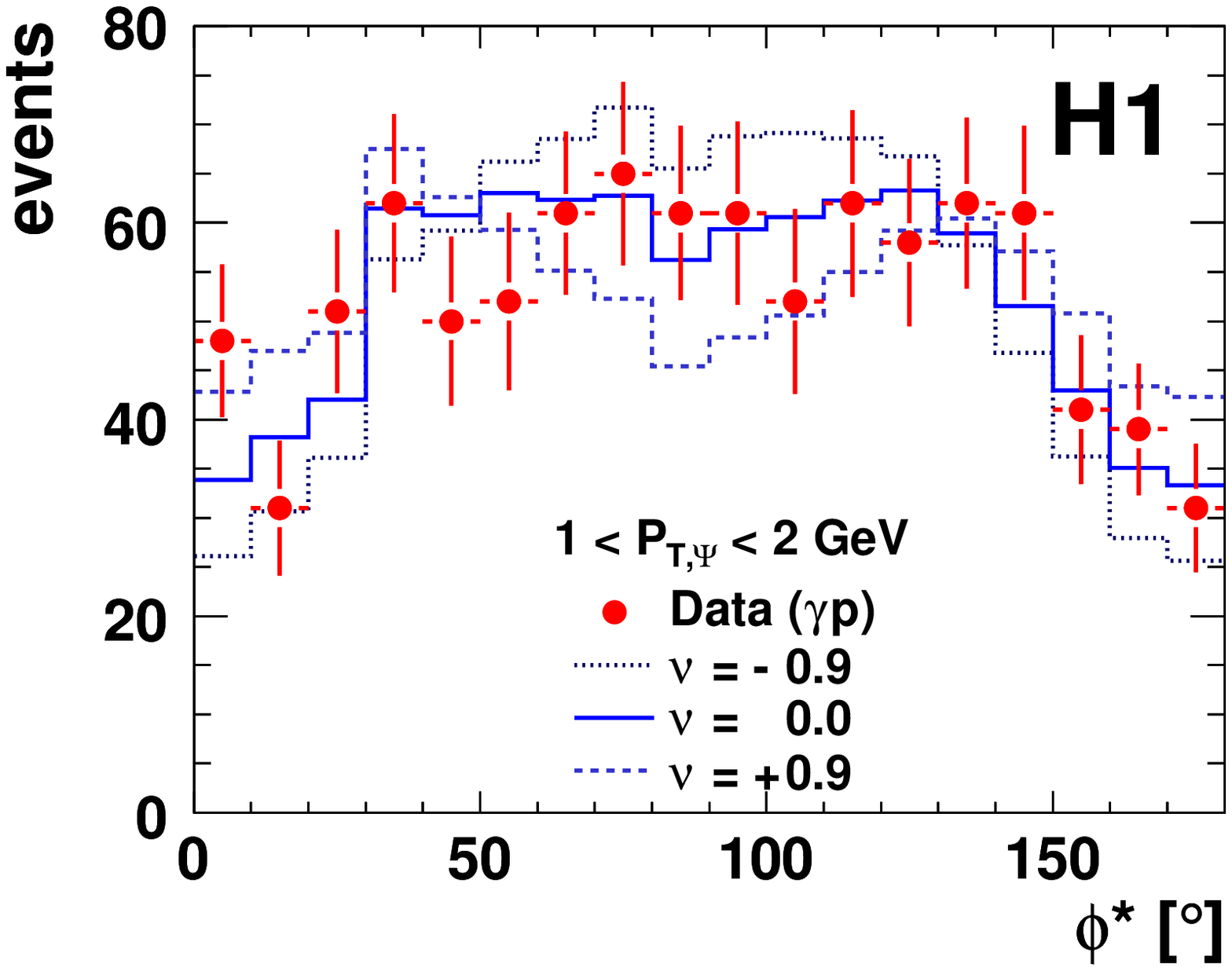}}%fin_Polarisation_con_Phi_Pt_1.eps}}
	
\put(5.1,12.35){a)}
\put(5.1,5.85){b)}
\put(5.2,13.5){\large \sf Inelastic \JPsi Polarisation }
\put(6.425,12.9){\large \sf Helicity Frame}

\end{picture}
\caption{Distributions for a) \CosThetaStar in the range $0.6 < \ZJPsi < 0.75$ and b) \PhiStar in the range $1 < \PtJPsi < \unit[2]{GeV}$. The data are compared to the corrected \Cascade simulation with three different assumptions for the polarisation variables $\alpha$ or $\nu$.}
\label{fig:gp:con:Polarisation}
\end{figure}

%%%%%%%%%%%%%%%%%%%%%%%%%%%%%%%%%%%%%%%%%%%%%%%%%%%%%%%%%%%

%%%%%%%%%%%%%%%%%%%%%%%%%%%%%%%%%%%%%%%%%%%%%%%%%%%%%%%%%%%

%%%%%%%%%%%%%%%%%%%%%%%%%%%%%%%%%%%%%%%%%%%%%%%%%%%%%%%%%%%
%%%%%%%%%%%%%%%%%%%%%%%%%%%%%%%%%%%%%%%%%%%%%%%%%%%%%%%%%%%
%%%%%%%%%%%%%%%%%%%%%%%%%%%%%%%%%%%%%%%%%%%%%%%%%%%%%%%%%%%
%%%%%%%%%%%%%%%%%%%%%%%%%%%%%%%%%%%%%%%%%%%%%%%%%%%%%%%%%%%

\clearpage

\begin{table}[ptb]
\begin{center}
%\begin{tabular}{D{.}{.}{4.0} l D{.}{.}{4.1} D{.}{.}{2.1} l c l c l}
\begin{tabular}{r c r c l p{0.1cm} l p{0.1cm} l}
%\hline 
\multicolumn{9}{ c }{Inelastic \JPsi Photoproduction} \\
%\multicolumn{5}{ c }{ $0.3 < \ZJPsi < 0.9, 60 < \Wgp < \unit[240]{GeV}$} \\
\toprule
\multicolumn{3}{ c }{$\PtJPsiSquare~[\mathrm{GeV}^2]$} & \multicolumn{1}{ c }{$\left< \PtJPsiSquare \right>~[\mathrm{GeV}^2]$} & \multicolumn{5}{ c }{$\mathrm{d}\sigma_\mathrm{\gamma p}/\mathrm{d}\PtJPsiSquare~\mathrm{[nb/GeV^2]}$}  \\
\midrule
 $1.0$ &$\div$& $  2.1$ & $~~1.5$  & $7.75$   & $\pm$ & $ 0.82    $ & $\pm$ & $ 0.70    $ \\
 $2.1$ &$\div$& $  3.5$ & $~~2.7$  & $4.43$   & $\pm$ & $ 0.48    $ & $\pm$ & $ 0.40    $ \\
 $3.5$ &$\div$& $  5.4$ & $~~4.3$  & $2.55$   & $\pm$ & $ 0.28    $ & $\pm$ & $ 0.23    $ \\
 $5.4$ &$\div$& $  7.6$ & $~~6.3$  & $1.06$   & $\pm$ & $ 0.13    $ & $\pm$ & $ 0.10   $ \\
 $7.6$ &$\div$& $ 10.0$ & $~~8.6$  & $0.677$  & $\pm$ & $ 0.084   $ & $\pm$ & $ 0.061   $ \\
$10.0$ &$\div$& $ 13.5$ & $11.4$  & $0.391$   & $\pm$ & $ 0.048   $ & $\pm$ & $ 0.035   $ \\
$13.5$ &$\div$& $ 20.0$ & $15.6$  & $0.156$   & $\pm$ & $ 0.020   $ & $\pm$ & $ 0.014   $ \\
$20.0$ &$\div$& $ 26.5$ & $22.1$  & $0.0509$  & $\pm$ & $ 0.0078  $ & $\pm$ & $ 0.0046  $ \\
$26.5$ &$\div$& $ 40.0$ & $30.0$  & $0.0175$  & $\pm$ & $ 0.0029  $ & $\pm$ & $ 0.0015  $ \\
$40.0$ &$\div$& $ 60.0$ & $46.0$  & $0.0049 $ & $\pm$ & $ 0.0012  $ & $\pm$ & $ 0.0004 $ \\
$60.0$ &$\div$& $100.0$ & $70.0$  & $0.00090$ & $\pm$ & $ 0.00035 $ & $\pm$ & $ 0.00008 $ \\
\specialrule{\heavyrulewidth}{\cmidrulesep}{\cmidrulesep}
\multicolumn{3}{ c }{\ZJPsi} & \multicolumn{1}{ c }{$\left< \ZJPsi \right>$} & \multicolumn{5}{ c }{$ \mathrm{d}\sigma_\mathrm{\gamma p}/\mathrm{d}\ZJPsi~\mathrm{[nb]}$}  \\
\midrule
$0.30$ &$\div$& $0.45$ & $0.375$  & $ 23.4 $ & $\pm$ & $ 2.6 $ & $\pm$ & $ 2.1 $ \\
$0.45$ &$\div$& $0.60$ & $0.525$  & $ 47.6 $ & $\pm$ & $ 4.7 $ & $\pm$ & $ 4.3 $ \\
$0.60$ &$\div$& $0.75$ & $0.675$  & $ 51.3 $ & $\pm$ & $ 5.0 $ & $\pm$ & $ 4.6 $ \\
$0.75$ &$\div$& $0.90$ & $0.825$  & $ 54.2 $ & $\pm$ & $ 5.6 $ & $\pm$ & $ 4.9 $ \\
\bottomrule
\end{tabular}
\end{center}
\caption{Measured differential photoproduction cross sections in the kinematic range $0.3 < \ZJPsi < 0.9, \PtJPsi > \unit[1]{GeV}$ and $60 < \Wgp < \unit[240]{GeV}$ as function of the squared transverse momentum \PtJPsiSquare and the elasticity \ZJPsi of the \JPsi meson. The bin centre values, $\left< \PtJPsiSquare \right>$ and $\left< \ZJPsi \right>$, are also given in the table.}
\label{tab:xsecs:gammap:z}
\label{tab:xsecs:gammap:pt2}
\end{table}

\begin{table}[ptb]
\begin{center}
\begin{tabular}{ r c r c l l p{0.1cm} l p{0.1cm} l }
\multicolumn{10}{ c }{Inelastic \JPsi Photoproduction} \\
\toprule
\multicolumn{3}{ c }{$\Wgp~\mathrm{[GeV]}$} & \multicolumn{1}{ c }{$\left< \Wgp \right>~\mathrm{[GeV]}$} & \multicolumn{1}{ c }{$\Phi_\gamma$} & \multicolumn{5}{ c }{$ \sigma_\mathrm{\gamma p}~[nb]$}  \\
\midrule
 $60$ &$\div$&  $80$ & $~~69$  & $0.0269 $  & $ 22.9 $ & $\pm $ & $ 4.1 $ & $\pm $ & $ 2.1 $ \\
 $80$ &$\div$& $100$ & $~~89$  & $0.0192 $  & $ 24.1 $ & $\pm $ & $ 3.3 $ & $\pm $ & $ 2.2 $ \\
$100$ &$\div$& $120$ & $110$   & $0.0145 $  & $ 24.0 $ & $\pm $ & $ 3.0 $ & $\pm $ & $ 2.2 $ \\
$120$ &$\div$& $140$ & $130$   & $0.0112 $  & $ 30.3 $ & $\pm $ & $ 3.6 $ & $\pm $ & $ 2.7 $ \\
$140$ &$\div$& $160$ & $150$   & $0.00891$  & $ 35.7 $ & $\pm $ & $ 4.3 $ & $\pm $ & $ 3.2 $ \\
$160$ &$\div$& $180$ & $170$   & $0.00716$  & $ 30.4 $ & $\pm $ & $ 3.9 $ & $\pm $ & $ 2.7 $ \\
$180$ &$\div$& $210$ & $194$   & $0.00832$  & $ 31.7 $ & $\pm $ & $ 4.2 $ & $\pm $ & $ 2.9 $ \\
$210$ &$\div$& $240$ & $224$   & $0.00621$  & $ 33.8 $ & $\pm $ & $ 5.6 $ & $\pm $ & $ 3.0 $ \\
\bottomrule
\end{tabular}
\end{center}
\caption{Measured photoproduction cross sections in the kinematic range $\PtJPsi > \unit[1]{GeV}$ and $0.3 < \ZJPsi < 0.9$ in bins of the photon proton centre-of-mass energy \Wgp. The bin centre values $\left< \Wgp \right>$ are also given in the table. $\Phi_\gamma$ denotes the photon flux factors~\cite{WWA} employed in the photoproduction analysis using an upper \Qsquared boundary of $\Qsquared = \unit[2.5]{GeV^2}$. For the range $60 < \Wgp < \unit[240]{GeV}$ a photon flux factor of $\Phi_\gamma = 0.1024$ is calculated.}
\label{tab:xsecs:gammap:wgp}
\label{tab:fluxes}
\end{table}

\begin{table}[ptb]
\begin{center}
\begin{tabular}{r c r c l p{0.1cm} l p{0.1cm} l}
\multicolumn{9}{ c }{Inelastic \JPsi Photoproduction} \\
\toprule
\multicolumn{3}{ c }{$\PtJPsiSquare~[\mathrm{GeV}^2]$} & \multicolumn{1}{ c }{$\left< \PtJPsiSquare \right>~[\mathrm{GeV}^2]$} & \multicolumn{5}{ c }{$ \mathrm{d}\sigma_\mathrm{\gamma p}/\mathrm{d}\PtJPsiSquare~\mathrm{[nb/GeV^2]}$ } \\
\specialrule{\lightrulewidth}{\cmidrulesep}{0pt}
\multicolumn{9}{ c }{ $0.30 < \ZJPsi < 0.45$} \\
\specialrule{\lightrulewidth}{0pt}{\cmidrulesep}
 $1.0$ &$\div$& $2.0 $  & $~~1.4$  & $ 1.02    $ & $\pm$ & $ 0.20    $ & $\pm$ & $ 0.09 $ \\
 $2.0$ &$\div$& $3.0 $  & $~~2.5$  & $ 0.64    $ & $\pm$ & $ 0.13    $ & $\pm$ & $ 0.06 $ \\
 $3.0$ &$\div$& $4.5 $  & $~~3.6$  & $ 0.402   $ & $\pm$ & $ 0.077   $ & $\pm$ & $ 0.036 $ \\
 $4.5$ &$\div$& $7.0 $  & $~~5.5$  & $ 0.180   $ & $\pm$ & $ 0.036   $ & $\pm$ & $ 0.016 $ \\
 $7.0$ &$\div$& $10.0$  & $~~8.2$  & $ 0.093   $ & $\pm$ & $ 0.021   $ & $\pm$ & $ 0.008 $ \\
$10.0$ &$\div$& $14.0$  & $11.6$   & $ 0.047   $ & $\pm$ & $ 0.011   $ & $\pm$ & $ 0.004 $ \\
$14.0$ &$\div$& $20.0$  & $16.2$   & $ 0.0210  $ & $\pm$ & $ 0.0052  $ & $\pm$ & $ 0.0019 $ \\
$20.0$ &$\div$& $40.0$  & $25.0$   & $ 0.0065  $ & $\pm$ & $ 0.0018  $ & $\pm$ & $ 0.0006 $ \\
$40.0$ &$\div$& $100.0$ & $49.0$   & $ 0.00065 $ & $\pm$ & $ 0.00032 $ & $\pm$ & $ 0.00006 $ \\
\specialrule{\lightrulewidth}{\cmidrulesep}{0pt}
\multicolumn{9}{ c }{ $0.45 < \ZJPsi < 0.60$} \\
\specialrule{\lightrulewidth}{0pt}{\cmidrulesep}
$1.0 $ &$\div$& $2.0  $ & $~~1.4$   & $ 2.17    $ & $\pm$ & $ 0.29    $ & $\pm$ & $ 0.19 $ \\
$2.0 $ &$\div$& $3.0  $ & $~~2.5$   & $ 1.21    $ & $\pm$ & $ 0.18    $ & $\pm$ & $ 0.11 $ \\
$3.0 $ &$\div$& $4.5  $ & $~~3.6$   & $ 0.74    $ & $\pm$ & $ 0.11    $ & $\pm$ & $ 0.07 $ \\
$4.5 $ &$\div$& $7.0  $ & $~~5.5$   & $ 0.392   $ & $\pm$ & $ 0.057   $ & $\pm$ & $ 0.035 $ \\
$7.0 $ &$\div$& $10.0 $ & $~~8.2$   & $ 0.219   $ & $\pm$ & $ 0.033   $ & $\pm$ & $ 0.020 $ \\
$10.0$ &$\div$& $14.0 $ & $11.6$    & $ 0.107   $ & $\pm$ & $ 0.014   $ & $\pm$ & $ 0.010 $ \\
$14.0$ &$\div$& $20.0 $ & $16.2$    & $ 0.0497  $ & $\pm$ & $ 0.0084  $ & $\pm$ & $ 0.0045 $ \\
$20.0$ &$\div$& $40.0 $ & $25.0$    & $ 0.0072  $ & $\pm$ & $ 0.0015  $ & $\pm$ & $ 0.0007 $ \\
$40.0$ &$\div$& $100.0$ & $49.0$    & $ 0.00072 $ & $\pm$ & $ 0.00030 $ & $\pm$ & $ 0.00007 $ \\
\specialrule{\lightrulewidth}{\cmidrulesep}{0pt}
\multicolumn{9}{ c }{ $0.60 < \ZJPsi < 0.75$} \\
\specialrule{\lightrulewidth}{0pt}{\cmidrulesep}
$1.0 $ &$\div$& $2.0 $ & $~~1.4$   & $ 2.40   $ & $\pm$ & $ 0.31   $ & $\pm$ & $ 0.22 $ \\
$2.0 $ &$\div$& $3.0 $ & $~~2.5$   & $ 1.79   $ & $\pm$ & $ 0.18   $ & $\pm$ & $ 0.11 $ \\
$3.0 $ &$\div$& $4.5 $ & $~~3.6$   & $ 1.01   $ & $\pm$ & $ 0.13   $ & $\pm$ & $ 0.09 $ \\
$4.5 $ &$\div$& $7.0 $ & $~~5.5$   & $ 0.506  $ & $\pm$ & $ 0.070  $ & $\pm$ & $ 0.046 $ \\
$7.0 $ &$\div$& $10.0$ & $~~8.2$   & $ 0.200  $ & $\pm$ & $ 0.032  $ & $\pm$ & $ 0.018 $ \\
$10.0$ &$\div$& $14.0$ & $11.6$    & $ 0.112  $ & $\pm$ & $ 0.018  $ & $\pm$ & $ 0.010 $ \\
$14.0$ &$\div$& $20.0$ & $16.2$    & $ 0.0413 $ & $\pm$ & $ 0.0076 $ & $\pm$ & $ 0.0037 $ \\
$20.0$ &$\div$& $40.0$ & $25.0$    & $ 0.0068 $ & $\pm$ & $ 0.0014 $ & $\pm$ & $ 0.0006 $ \\
\specialrule{\lightrulewidth}{\cmidrulesep}{0pt}
\multicolumn{9}{ c }{ $0.75 < \ZJPsi < 0.90$} \\
\specialrule{\lightrulewidth}{0pt}{\cmidrulesep}
$1.0 $ &$\div$& $2.0 $ & $~~1.4$  & $ 2.40   $ & $\pm$ & $ 0.36   $ & $\pm$ & $ 0.22 $ \\
$2.0 $ &$\div$& $3.0 $ & $~~2.5$  & $ 1.69   $ & $\pm$ & $ 0.27   $ & $\pm$ & $ 0.15 $ \\
$3.0 $ &$\div$& $4.5 $ & $~~3.6$  & $ 0.86   $ & $\pm$ & $ 0.15   $ & $\pm$ & $ 0.08 $ \\
$4.5 $ &$\div$& $7.0 $ & $~~5.5$  & $ 0.437  $ & $\pm$ & $ 0.076  $ & $\pm$ & $ 0.039 $ \\
$7.0 $ &$\div$& $10.0$ & $~~8.2$  & $ 0.226  $ & $\pm$ & $ 0.042  $ & $\pm$ & $ 0.020 $ \\
$10.0$ &$\div$&$ 14.0$ & $11.6$   & $ 0.099  $ & $\pm$ & $ 0.022  $ & $\pm$ & $ 0.009 $ \\
$14.0$ &$\div$&$ 20.0$ & $16.2$   & $ 0.0428 $ & $\pm$ & $ 0.0098 $ & $\pm$ & $ 0.0039 $ \\
$20.0$ &$\div$&$ 40.0$ & $25.0$   & $ 0.0076 $ & $\pm$ & $ 0.0021 $ & $\pm$ & $ 0.0007 $ \\
\bottomrule
\end{tabular}
\end{center}
\caption{Measured differential photoproduction cross sections in the kinematic
range $0.3 < \ZJPsi < 0.9$ and $60 < \Wgp < \unit[240]{GeV}$ as a function of the squared transverse momentum of the \JPsi meson in bins of the elasticity \ZJPsi. The bin centre values $\left< \PtJPsiSquare \right>$ are also given in the table.} 
\label{tab:xsecs:gammap:pt2_Z}
\end{table}

\begin{table}[ptb]
\begin{center}
\begin{tabular}{r c r p{0.5cm} c p{0.5cm} l p{0.1cm} l p{0.1cm} l}
\multicolumn{11}{ c }{Inelastic \JPsi Photoproduction} \\
\toprule
\multicolumn{3}{ c }{$\ZJPsi$} & & \multicolumn{1}{ c }{$\left< \ZJPsi \right>$} & & \multicolumn{5}{ c }{$ \mathrm{d}\sigma_\mathrm{\gamma p}/\mathrm{d}\ZJPsi~\mathrm{[nb]}$}  \\
\specialrule{\lightrulewidth}{\cmidrulesep}{0pt}
\multicolumn{11}{ c }{ $1.0 < \PtJPsi < \unit[2.0]{GeV}$} \\
\specialrule{\lightrulewidth}{0pt}{\cmidrulesep}
$0.30$ &$\div$& $0.45$ & & $0.375$  & & $ 14.9 $ & $\pm$ & $ 2.1 $ & $\pm$ & $ 1.3 $ \\
$0.45$ &$\div$& $0.60$ & & $0.525$  & & $ 28.3 $ & $\pm$ & $ 3.1 $ & $\pm$ & $ 2.5 $ \\
$0.60$ &$\div$& $0.75$ & & $0.675$  & & $ 31.8 $ & $\pm$ & $ 3.4 $ & $\pm$ & $ 2.9 $ \\
$0.75$ &$\div$& $0.90$ & & $0.825$  & & $ 33.6 $ & $\pm$ & $ 4.0 $ & $\pm$ & $ 3.0 $ \\
\specialrule{\lightrulewidth}{\cmidrulesep}{0pt}
\multicolumn{11}{ c }{ $2.0 < \PtJPsi < \unit[3.0]{GeV}$} \\
\specialrule{\lightrulewidth}{0pt}{\cmidrulesep}
$0.30$ &$\div$& $0.45$ & & $0.375$  & & $~~5.1 $ & $\pm$ & $ 0.8 $ & $\pm$ & $ 0.5 $ \\
$0.45$ &$\div$& $0.60$ & & $0.525$  & & $ 11.6 $ & $\pm$ & $ 1.4 $ & $\pm$ & $ 1.0 $ \\
$0.60$ &$\div$& $0.75$ & & $0.675$  & & $ 14.1 $ & $\pm$ & $ 1.6 $ & $\pm$ & $ 1.3 $ \\
$0.75$ &$\div$& $0.90$ & & $0.825$  & & $ 13.1 $ & $\pm$ & $ 1.8 $ & $\pm$ & $ 1.2 $ \\
\specialrule{\lightrulewidth}{\cmidrulesep}{0pt}
\multicolumn{11}{ c }{ $3.0 < \PtJPsi < \unit[4.5]{GeV}$} \\
\specialrule{\lightrulewidth}{0pt}{\cmidrulesep}
$0.30$ &$\div$& $0.45$ & & $0.375$  & & $~~2.60 $ & $\pm$ & $ 0.42 $ & $\pm$ & $ 0.23 $ \\
$0.45$ &$\div$& $0.60$ & & $0.525$  & & $~~6.05 $ & $\pm$ & $ 0.73 $ & $\pm$ & $ 0.54 $ \\
$0.60$ &$\div$& $0.75$ & & $0.675$  & & $~~5.71 $ & $\pm$ & $ 0.71 $ & $\pm$ & $ 0.51 $ \\
$0.75$ &$\div$& $0.90$ & & $0.825$  & & $~~5.32 $ & $\pm$ & $ 0.80 $ & $\pm$ & $ 0.48 $ \\
\specialrule{\lightrulewidth}{\cmidrulesep}{0pt}
\multicolumn{11}{ c }{ $\PtJPsi > \unit[4.5]{GeV}$} \\
\specialrule{\lightrulewidth}{0pt}{\cmidrulesep}
$0.30$ &$\div$& $0.45$ & & $0.375$  & & $~~1.10 $ & $\pm$ & $ 0.20 $ & $\pm$ & $ 0.1 $ \\
$0.45$ &$\div$& $0.60$ & & $0.525$  & & $~~1.30 $ & $\pm$ & $ 0.20 $ & $\pm$ & $ 0.1 $ \\
$0.60$ &$\div$& $0.75$ & & $0.675$  & & $~~1.11 $ & $\pm$ & $ 0.17 $ & $\pm$ & $ 0.1 $ \\
$0.75$ &$\div$& $0.90$ & & $0.825$  & & $~~1.30 $ & $\pm$ & $ 0.24 $ & $\pm$ & $ 0.1 $ \\
\bottomrule
\end{tabular}
\end{center}
\caption{Measured differential photoproduction cross sections in the kinematic
range $\PtJPsi > \unit[1]{GeV}$ and $60 < \Wgp < \unit[240]{GeV}$ as a function of the elasticity \ZJPsi in bins of the transverse momentum of the \JPsi meson.}
\label{tab:xsecs:gammap:Z_Pt}
\end{table}

%%%%%%%%%%%%%%%%%%%%%%%%%%%%%%%%%%%%%%%%%%%%%%%%%%%%%%%%%%%
\clearpage

\begin{table}[ptb]
\begin{center}
\begin{tabular}{r c r p{0.5cm} c p{0.5cm}  l p{0.1cm} l p{0.1cm} l}
\multicolumn{11}{ c }{Inelastic \JPsi Electroproduction} \\
\toprule
\multicolumn{3}{ c }{$\Qsquared~\mathrm{[GeV^2]}$} & \multicolumn{3}{ c }{$\left< \Qsquared \right>~\mathrm{[GeV^2]}$} & \multicolumn{5}{ c }{$ \mathrm{d}\sigma_\mathrm{ep}/\mathrm{d}\Qsquared~\mathrm{[pb/GeV^2]}$}  \\
\midrule
$~~3.6 $&$\div$& $~~6.5$ & & $~~4.9$  & & $  14.98 $ & $\pm$ & $ 1.97  $ & $\pm$ & $ 1.27   $ \\
$~~6.5 $&$\div$& $12.0 $ & & $~~8.6$  & & $ ~~6.33 $ & $\pm$ & $ 0.75  $ & $\pm$ & $ 0.54  $ \\
$12.0$  &$\div$& $20.0 $ & & $15.0$   & & $ ~~2.11 $ & $\pm$ & $ 0.33  $ & $\pm$ & $ 0.18  $ \\
$20.0$  &$\div$& $40.0 $ & & $26.7$   & & $ ~~0.74 $ & $\pm$ & $ 0.12  $ & $\pm$ & $ 0.06 $ \\
$40.0$  &$\div$& $100.0$ & & $53.0$   & & $ ~~0.141$ & $\pm$ & $ 0.029 $ & $\pm$ & $ 0.012 $ \\
\toprule
\multicolumn{3}{ c }{$\PtStarJPsiSquare~\mathrm{[GeV^2]}$} &  \multicolumn{3}{ c }{$\left< \PtStarJPsiSquare \right>~\mathrm{[GeV^2]}$} & \multicolumn{5}{ c }{$ \mathrm{d}\sigma_\mathrm{ep}/\mathrm{d}\PtStarJPsiSquare~\mathrm{[pb/GeV^2]}$}  \\
\midrule
$1.0 $ &$\div$& $  2.2$ & & $~~1.6$   & & $ 15.5   $ & $\pm$ & $ 2.7    $ & $\pm$ & $ 1.3 $ \\
$2.2 $ &$\div$& $  3.7$ & & $~~2.9$   & & $ 11.0   $ & $\pm$ & $ 2.1    $ & $\pm$ & $ 0.9 $ \\
$3.7 $ &$\div$& $  6.4$ & & $~~4.9$   & & $ ~~8.7  $ & $\pm$ & $ 1.4    $ & $\pm$ & $ 0.7 $ \\
$6.4 $ &$\div$& $  9.6$ & & $~~7.8$   & & $ ~~5.90 $ & $\pm$ & $ 0.92   $ & $\pm$ & $ 0.50 $ \\
$9.6 $ &$\div$& $ 13.5$ & & $11.2$  & & $ ~~3.23   $ & $\pm$ & $ 0.53   $ & $\pm$ & $ 0.27 $ \\
$13.5$ &$\div$& $ 20.0$ & & $16.0$  & & $ ~~1.69   $ & $\pm$ & $ 0.27   $ & $\pm$ & $ 0.14 $ \\
$20.0$ &$\div$& $ 40.0$ & & $25.7$  & & $ ~~0.576  $ & $\pm$ & $ 0.083  $ & $\pm$ & $ 0.049 $ \\
$40.0$ &$\div$& $100.0$ & & $51.0$  & & $ ~~0.055  $ & $\pm$ & $ 0.012  $ & $\pm$ & $ 0.005 $ \\
\toprule
\multicolumn{3}{ c }{\ZJPsi} & \multicolumn{3}{ c }{$\left< \ZJPsi \right>$}  & \multicolumn{5}{ c }{$ \mathrm{d}\sigma_\mathrm{ep}/\mathrm{d}\ZJPsi~\mathrm{[pb]}$}  \\
\midrule
$0.30$ &$\div$& $0.45$ & & $0.375$ & & $ 150  $ & $\pm$ & $ 26  $ & $\pm$ & $ 13 $ \\
$0.45$ &$\div$& $0.60$ & & $0.525$ & & $ 158  $ & $\pm$ & $ 22  $ & $\pm$ & $ 14 $ \\
$0.60$ &$\div$& $0.75$ & & $0.675$ & & $ 280  $ & $\pm$ & $ 31  $ & $\pm$ & $ 24 $ \\
$0.75$ &$\div$& $0.90$ & & $0.825$ & & $ 239  $ & $\pm$ & $ 29  $ & $\pm$ & $ 20 $ \\
\toprule
\multicolumn{3}{ c }{$\Wgp~\mathrm{[GeV]}$} & \multicolumn{3}{ c }{$\left< \Wgp \right>~\mathrm{[GeV]}$} & \multicolumn{5}{ c }{$ \mathrm{d}\sigma_\mathrm{ep}/\mathrm{d}\Wgp~\mathrm{[pb/GeV]}$ } \\
\midrule
$~~60$&$\div$& $~~80$ & & $~~69$     & & $ 0.89  $ & $\pm$ & $ 0.16  $ & $\pm$ & $ 0.08 $ \\
$~~80$&$\div$& $100$  & & $~~89$    & & $ 1.03  $ & $\pm$ & $ 0.15  $ & $\pm$ & $ 0.09 $ \\
$100$ &$\div$&$ 120$  & & $110$  & & $ 0.77  $ & $\pm$ & $ 0.12  $ & $\pm$ & $ 0.007 $ \\
$120$ &$\div$&$ 140$  & & $130$  & & $ 0.75  $ & $\pm$ & $ 0.11  $ & $\pm$ & $ 0.06 $ \\
$140$ &$\div$&$ 160$  & & $150$  & & $ 0.71  $ & $\pm$ & $ 0.11  $ & $\pm$ & $ 0.06 $ \\
$160$ &$\div$&$ 180$  & & $170$  & & $ 0.55  $ & $\pm$ & $ 0.10  $ & $\pm$ & $ 0.05 $ \\
$180$ &$\div$&$ 210$  & & $194$  & & $ 0.42  $ & $\pm$ & $ 0.09  $ & $\pm$ & $ 0.04 $ \\
$210$ &$\div$&$ 240$  & & $224$  & & $ 0.30  $ & $\pm$ & $ 0.10  $ & $\pm$ & $ 0.03 $ \\
\bottomrule
\end{tabular}
\end{center}
\caption{Measured differential electroproduction cross sections in the kinematic
range $3.6 < \Qsquared < \unit[100]{GeV^2}, \PtStarJPsi > \unit[1]{GeV}$ and $0.3 < \ZJPsi < 0.9$ as function of the four momentum transfer \Qsquared, the squared transverse momentum of the \JPsi meson in the photon proton rest frame \PtStarJPsiSquare, the elasticity \ZJPsi and the photon proton centre-of-mass energy \Wgp.}
\label{tab:xsecs:dis:Q2e}
\label{tab:xsecs:dis:PtStar2}
\label{tab:xsecs:dis:z}
\label{tab:xsecs:dis:Wgp}
\end{table}

\begin{table}[ptb]
\begin{center}
\begin{tabular}{r c r p{0.5cm} c p{0.5cm} l p{0.1cm} l p{0.1cm} l}
\multicolumn{11}{ c }{Inelastic \JPsi Electroproduction} \\
\toprule
\multicolumn{3}{ c }{$\PtStarJPsiSquare~\mathrm{[GeV^2]}$} & \multicolumn{3}{ c }{$\left< \PtStarJPsiSquare \right>\mathrm{[GeV^2]}$} & \multicolumn{5}{ c }{$ \mathrm{d}\sigma_\mathrm{ep}/\mathrm{d}\PtStarJPsiSquare~\mathrm{[nb/GeV^2]}$}  \\
\midrule
\multicolumn{11}{ c }{ $0.30 < \ZJPsi < 0.60$} \\
\midrule
$ 1.0$ &$\div$& $ 4.0$ & & $ 2.2$  & & $ 5.5   $ & $\pm$ & $ 1.1   $ & $\pm$ & $ 0.5 $ \\
$ 4.0$ &$\div$& $ 9.0$ & & $ 5.6$  & & $ 3.0   $ & $\pm$ & $ 0.6   $ & $\pm$ & $ 0.3 $ \\
$ 9.0$ &$\div$& $20.0$ & & $11.3$  & & $ 0.89  $ & $\pm$ & $ 0.17  $ & $\pm$ & $ 0.08 $ \\
$20.0$ &$\div$& $60.0$ & & $27.0$  & & $ 0.11  $ & $\pm$ & $ 0.02  $ & $\pm$ & $ 0.01 $ \\
\midrule
\multicolumn{11}{ c }{ $0.60 < \ZJPsi < 0.75$} \\
\midrule
$ 1.0$ &$\div$& $ 4.0$ & & $ 2.3$  & & $ 3.7   $ & $\pm$ & $ 0.7   $ & $\pm$ & $ 0.3 $ \\
$ 4.0$ &$\div$& $ 9.0$ & & $ 5.7$  & & $ 2.7   $ & $\pm$ & $ 0.4   $ & $\pm$ & $ 0.2 $ \\
$ 9.0$ &$\div$& $20.0$ & & $11.3$  & & $ 0.92  $ & $\pm$ & $ 0.15  $ & $\pm$ & $ 0.08 $ \\
$20.0$ &$\div$& $60.0$ & & $27.0$  & & $ 0.13  $ & $\pm$ & $ 0.03  $ & $\pm$ & $ 0.01 $ \\
\midrule
\multicolumn{11}{ c }{ $0.75 < \ZJPsi < 0.90$} \\
\midrule
$ 1.0$ &$\div$& $ 4.0$ & & $ 2.3$  & & $ 3.3    $ & $\pm$ & $ 0.7  $ & $\pm$ & $ 0.3 $ \\
$ 4.0$ &$\div$& $ 9.0$ & & $ 5.7$  & & $ 2.6    $ & $\pm$ & $ 0.5  $ & $\pm$ & $ 0.2 $ \\
$ 9.0$ &$\div$& $20.0$ & & $11.5$  & & $ 0.67   $ & $\pm$ & $ 0.13  $ & $\pm$ & $ 0.06 $ \\
$20.0$ &$\div$& $60.0$ & & $27.0$  & & $ 0.026  $ & $\pm$ & $ 0.013  $ & $\pm$ & $ 0.002 $ \\
\toprule
\multicolumn{3}{ c }{$\ZJPsi$} & \multicolumn{3}{ c }{$\left< \ZJPsi \right>$} & \multicolumn{5}{ c }{$\mathrm{d}\sigma_\mathrm{ep}/\mathrm{d}\ZJPsi~\mathrm{[nb]}$}  \\
\midrule
\multicolumn{11}{ c }{ $1.0 < \PtStarJPsi < \unit[2.0]{GeV}$} \\
\midrule
$0.30$ &$\div$& $0.45$ & & $0.375$  & & $ 60.0  $ & $\pm$ & $ 17.0  $ & $\pm$ & $ 5.1 $ \\
$0.45$ &$\div$& $0.60$ & & $0.525$  & & $ 48.0  $ & $\pm$ & $ 11.4  $ & $\pm$ & $ 4.1 $ \\
$0.60$ &$\div$& $0.75$ & & $0.675$  & & $ 74.6  $ & $\pm$ & $ 12.8  $ & $\pm$ & $ 6.3 $ \\
$0.75$ &$\div$& $0.90$ & & $0.825$  & & $ 66.8  $ & $\pm$ & $ 12.9  $ & $\pm$ & $ 5.7 $ \\
\midrule
\multicolumn{11}{ c }{ $2.0 < \PtStarJPsi < \unit[3.5]{GeV}$} \\
\midrule
$0.30$ &$\div$& $0.45$ & & $0.375$  & & $ 62.4  $ & $\pm$ & $ 15.1  $ & $\pm$ & $ 5.3 $ \\
$0.45$ &$\div$& $0.60$ & & $0.525$  & & $ 67.1  $ & $\pm$ & $ 13.1  $ & $\pm$ & $ 5.7 $ \\
$0.60$ &$\div$& $0.75$ & & $0.675$  & & $ 115.3  $ & $\pm$ & $ 16.2  $ & $\pm$ & $ 9.8 $ \\
$0.75$ &$\div$& $0.90$ & & $0.825$  & & $ 105.0  $ & $\pm$ & $ 16.7  $ & $\pm$ & $ 8.9 $ \\
\midrule
\multicolumn{11}{ c }{ $3.5 < \PtStarJPsi < \unit[10.]{GeV}$} \\
\midrule
$0.30$ &$\div$& $0.45$ & & $0.375$  & & $ 28.4  $ & $\pm$ & $ 6.7  $ & $\pm$ & $ 2.4 $ \\
$0.45$ &$\div$& $0.60$ & & $0.525$  & & $ 41.9  $ & $\pm$ & $ 7.5  $ & $\pm$ & $ 3.6 $ \\
$0.60$ &$\div$& $0.75$ & & $0.675$  & & $ 79.6  $ & $\pm$ & $ 10.6  $ & $\pm$ & $ 6.8 $ \\
$0.75$ &$\div$& $0.90$ & & $0.825$  & & $ 58.9  $ & $\pm$ & $ 9.5  $ & $\pm$ & $ 5.0 $ \\
\bottomrule
\end{tabular}
\end{center}
\caption{Measured differential electroproduction cross sections in the kinematic
range $3.6 < \Qsquared < \unit[100]{GeV^2}, \PtStarJPsi > \unit[1]{GeV}$ and $60 < \Wgp < \unit[240]{GeV}$ as a function of the squared transverse momentum in the photon proton rest frame \PtStarJPsiSquare in bins of the elasticity \ZJPsi and the elasticity \ZJPsi in bins of the transverse momentum in the photon proton rest frame \PtStarJPsi.}
\label{tab:xsecs:dis:ptstar2_Z}
\label{tab:xsecs:dis:Z_PtStar}
\end{table}

%%%%%%%%%%%%%%%%%%%%%%%%%%%%%%%%%%%%%%%%%%%%%%%%%%%%%%%%%%%
%%%%%%%%%%%%%%%%%%%%%%%%%%%%%%%%%%%%%%%%%%%%%%%%%%%%%%%%%%%

\clearpage

\begin{table}[htbp]
\vspace{3.5cm}
\begin{center}
\begin{tabular}{r c l c l l}
\multicolumn{6}{c}{Inelastic \JPsi Photoproduction} \\
\multicolumn{6}{c}{{Helicity Frame}} \\
\toprule 
\multicolumn{3}{c}{\PtJPsi [GeV]} & $\left< \PtJPsi \right>$ [GeV] & \multicolumn{1}{c}{$\alpha$} & \multicolumn{1}{c}{$\nu$}\\
\midrule
$1.0$ &$\div$& $2.0 $ & $1.45$ & $ +0.54~^{+ 0.27}_{- 0.24}$ & $ +0.25~^{+ 0.20}_{- 0.20}$\\
$2.0$ &$\div$& $3.0 $ & $2.46$ & $ -0.15~^{+ 0.24}_{- 0.21}$ & $ -0.74~^{+ 0.40}_{- 0.16}$\\
$3.0$ &$\div$& $4.5 $ & $3.65$ & $ -0.18~^{+ 0.26}_{- 0.23}$ & $ -0.04~^{+ 0.32}_{- 0.34}$\\
$4.5$ &$\div$& $10.0$ & $6.21$ & $ -0.28~^{+ 0.32}_{- 0.26}$ & $ +0.59~^{+ 0.31}_{- 0.36}$\\
\midrule
\multicolumn{3}{c}{\ZJPsi} & $\left< \ZJPsi \right>$ & \multicolumn{1}{c}{$\alpha$} & \multicolumn{1}{c}{$\nu$} \\
\hline
$0.30$ &$\div$& $0.45$ & $0.375$  & $ -0.65~^{+ 0.24}_{- 0.21}$ & $ -0.28~^{+ 0.34}_{- 0.35}$\\
$0.45$ &$\div$& $0.60$ & $0.525$  & $ +0.35~^{+ 0.25}_{- 0.22}$ & $ +0.40~^{+ 0.23}_{- 0.24}$\\
$0.60$ &$\div$& $0.75$ & $0.675$  & $ -0.18~^{+ 0.23}_{- 0.21}$ & $ +0.01~^{+ 0.24}_{- 0.25}$\\
$0.75$ &$\div$& $0.90$ & $0.825$  & $ +0.71~^{+ 0.19}_{- 0.40}$ & $ -0.10~^{+ 0.31}_{- 0.32}$\\
\toprule
\multicolumn{6}{c}{{Collins-Soper Frame}} \\
\midrule 
\multicolumn{3}{c}{\PtJPsi [GeV]} & $\left< \PtJPsi \right>$ [GeV] & \multicolumn{1}{c}{$\alpha$} & \multicolumn{1}{c}{$\nu$} \\
\midrule
$1.0$ &$\div$& $2.0 $ & $1.45$  & $ +0.25~^{+ 0.18}_{- 0.17}$ & $ +0.41~^{+ 0.15}_{- 0.16}$ \\
$2.0$ &$\div$& $3.0 $ & $2.46$  & $ -0.26~^{+ 0.17}_{- 0.15}$ & $ -0.42~^{+ 0.29}_{- 0.31}$\\
$3.0$ &$\div$& $4.5 $ & $3.65$  & $ -0.02~^{+ 0.23}_{- 0.20}$ & $ -0.31~^{+ 0.30}_{- 0.31}$\\
$4.5$ &$\div$& $10.0$ & $6.21$  & $ +0.19~^{+ 0.39}_{- 0.32}$ & $ +0.09~^{+ 0.33}_{- 0.34}$\\

\midrule
\multicolumn{3}{c}{\ZJPsi} & $\left< \ZJPsi \right>$ & \multicolumn{1}{c}{$\alpha$} & \multicolumn{1}{c}{$\nu$}\\
\midrule

$0.30 $&$\div$& $0.45$ & $0.375$  & $ +0.47~^{+ 0.34}_{- 0.28}$ & $ -0.18~^{+ 0.26}_{- 0.26}$\\
$0.45$ &$\div$& $0.60$ & $0.525$  & $ -0.02~^{+ 0.18}_{- 0.16}$ & $ +0.24~^{+ 0.18}_{- 0.19}$\\
$0.60 $&$\div$& $0.75$ & $0.675$  & $ -0.00~^{+ 0.18}_{- 0.16}$ & $ -0.16~^{+ 0.23}_{- 0.23}$\\
$0.75$ &$\div$& $0.90$ & $0.825$  & $ -0.02~^{+ 0.23}_{- 0.19}$ & $ +0.50~^{+ 0.26}_{- 0.28}$\\

\bottomrule
\end{tabular}
\end{center}
\caption{Measured polarisation parameters in the helicity and the Collins-Soper
frame as function of \PtJPsi and \ZJPsi in the kinematic range $\PtJPsi >
\unit[1]{GeV}, 60 < \Wgp < \unit[240]{GeV}$ and $0.3 < \ZJPsi < 0.9$.}
\label{tab:pol:heli}
\label{tab:pol:cs}
\end{table}
\renewcommand{\arraystretch}{1.0}

\end{document}